\documentclass[a4paper,11pt]{article}
\usepackage{jheppub}
\usepackage[utf8]{inputenc}
\usepackage[T1]{fontenc}
\usepackage{mathrsfs}
\usepackage{amsmath,amssymb,graphicx}
\usepackage{xcolor}
\hypersetup{colorlinks=true,linkcolor=blue,citecolor=magenta}

\newcommand{\dd}{\mathrm{d}}

\title{Black hole free energy landscapes from constrained instantons}
\author[a]{Shan-Ping Wu,}
\emailAdd{wushanping@zsit.edu.cn}
\author[b,c]{Jia-Zhou Liu,}
\emailAdd{liujzh2025@lzu.edu.cn}
\author[b,c]{Yu-Xiao Liu,}
\emailAdd{liuyx@lzu.edu.cn}
\author[b,c,*]{Shao-Wen Wei\note[*]{Corresponding author.}}
\emailAdd{weishw@lzu.edu.cn}
\affiliation[a]{School of Electronics and Information Engineering, Shaoxing Institute of Technology, Shaoxing, 312000, China}
\affiliation[b]{Lanzhou Center for Theoretical Physics, Key Laboratory of Theoretical Physics of Gansu Province, and Key Laboratory of Quantum Theory and Applications of MoE, Lanzhou University, Lanzhou, Gansu 730000, China}
\affiliation[c]{Institute of Theoretical Physics \& Research Center of Gravitation, Lanzhou University, Lanzhou 730000, China}

\abstract{We apply gravitational path integral factorization to provide a semiclassical interpretation of generalized black hole free energy landscapes at fixed external temperature and angular velocities. Boundary integral transforms express the formal partition function as an integral over constrained sectors labeled by complete local partial-Neumann data on an auxiliary hypersurface. Independent variation establishes paired regular subregion solutions as constrained instantons, with Lagrange multipliers supporting lapse and tangential shift jumps that accommodate thermal and rotational mismatches. For selected nonextremal stationary families with an asymptotic boundary, the constrained saddle action takes the form $I_{\rm constr}=\beta\mathcal F_{\rm gen}=\beta(E_{\mathcal B}-\Omega_{\mathcal B}^{I}J_{\mathcal B,I})-S$, with $\beta=T^{-1}$. Here $E_{\mathcal B}$ and $J_{\mathcal B,I}$ are renormalized Hamiltonian charges and $S$ is the black hole entropy, whose contribution to the action comes from the regular horizon cap. Interior stationary points of the landscape correspond to smooth equilibrium saddles when the retained variations independently span the entropy and angular momentum directions. We extend the construction to local, parity-even $f(\mathrm{Riemann})$ theories without curvature derivatives by retaining the complete auxiliary interface data. The same thermodynamic form follows with Wald entropy, without a small-deficit expansion. Schwarzschild and Kerr black holes illustrate the Einstein construction, while five-dimensional static AdS black holes in Einstein--Gauss--Bonnet gravity and the rotating BTZ branch of new massive gravity provide explicit checks of the higher-curvature extension.}
\begin{document}
\maketitle
\flushbottom
\section{Introduction}
\label{sec:introduction}
Black hole thermodynamics connects gravity with quantum theory and statistical mechanics through the laws of black hole mechanics, black hole entropy, and Hawking radiation~\cite{Bekenstein:1973ur,Bardeen:1973gs,Hawking:1975vcx}. In the Euclidean formulation, regularity fixes the thermal period of a smooth nonextremal black hole saddle~\cite{Hartle:1976tp,Gibbons:1976ue,Hawking:1995fd,Witten:2024upt}. This framework describes equilibrium phase structure, including the Hawking--Page transition~\cite{Hawking:1982dh,Witten:1998zw} and van der Waals--type behavior in charged AdS black holes~\cite{Chamblin:1999tk,Chamblin:1999hg,Kubiznak:2012wp}. This equilibrium description can be extended by allowing collective variables, such as the horizon radius, to vary at fixed external thermodynamic sources.

\begin{figure}[t]
        \centering
        \includegraphics[width=0.6\textwidth]{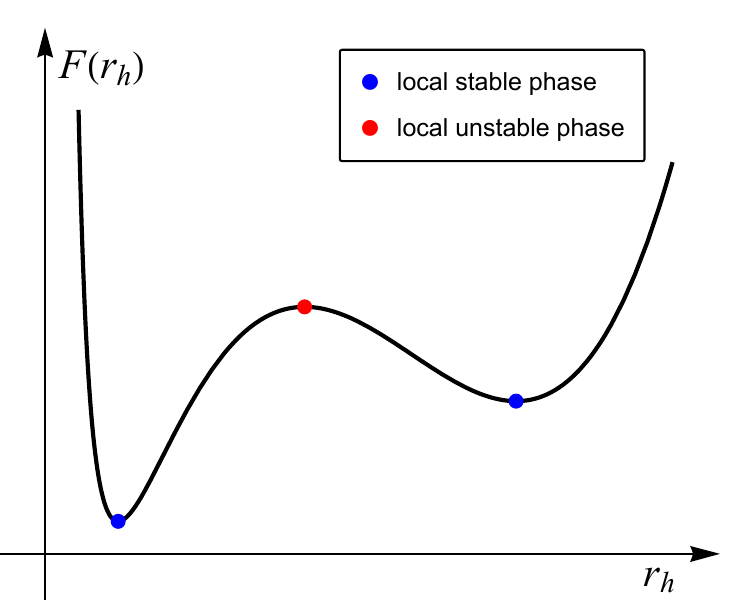}
        \caption{Schematic free energy landscape at fixed external thermodynamic data. Along the retained collective direction, local minima (blue) represent stable or metastable equilibrium black holes, local maxima (red) represent unstable equilibrium black holes, and generic points on the curve are off shell with respect to these data.}
        \label{fig:frh}
    \end{figure}

Early reduced action and Hamiltonian treatments expressed gravitational partition functions in terms of collective parameters~\cite{York:1986it,Whiting:1988qr,Whiting:1988ge,Louko:1994tv}. Microcanonical densities of states at fixed local boundary data were also related to thermal ensembles by integral transforms~\cite{Brown:1992bq}. Free energy landscape methods retain a continuous family of black hole configurations~\cite{Li:2020khm,Li:2020nsy}. In the canonical ensemble at fixed angular momentum $J$ and charge $Q$, the usual generalized free energy landscape is
\begin{equation}
    \mathcal F_{\rm gen}(r_+;T)=M(r_+)-T S(r_+),
    \label{eq:intro_naive_landscape}
\end{equation}
where $T$ is the temperature of the external heat bath, prescribed independently of the horizon radius $r_+$ and not required to coincide with the Hawking temperature $T_H(r_+)$. At fixed remaining external parameters, the first law implies that interior stationary points satisfy $T=T_H$ when $\partial S/\partial r_+\neq0$. Their stability concerns the retained collective direction, as illustrated in Fig.~\ref{fig:frh}.

Free energy landscape methods have been used to study stochastic phase transition dynamics, including transition rates and characteristic timescales~\cite{Li:2021vdp,Li:2021zep,Liu:2021lmr,Li:2022ylz,Li:2022yti,Wei:2020rcd,Yang:2021ljn,Li:2023ppc,Liu:2023sbf,Fairoos:2024yrh,Yue:2024qms}, and to develop thermodynamic topological classifications~\cite{Wei:2022dzw,Yerra:2022coh,Wu:2022whe,Fang:2022rsb,Wei:2024gfz,Wu:2024asq,Wei:2026upy}. Reduced ensembles have been used to compute averaged thermodynamic quantities for static and rotating black holes, as well as AdS black holes in Einstein--Gauss--Bonnet (EGB) gravity~\cite{Cheng:2024hxh,Cheng:2024efw,Ali:2024adt}. Collective variable path integrals also yield subleading effective thermodynamics~\cite{Liu:2025iei}.

A Euclidean realization of the generalized free energy in Eq.~\eqref{eq:intro_naive_landscape} uses geometries with conical singularities~\cite{Grumiller:2012rt,Eune:2013qs,Li:2022oup}, building on earlier treatments of off-shell black holes and conical geometry~\cite{Carlip:1993sa,Susskind:1993ws,Banados:1993qp,Susskind:1994sm,Teitelboim:1994is,Solodukhin:1994yz,Fursaev:1995ef,Frolov:1995xe,Mann:1996bi}. A prescribed period $\beta=T^{-1}$ differing from the regularity period $\beta_H$ produces a conical defect at the bolt. In Einstein gravity, including the defect contribution and the appropriate physical boundary terms gives $I_{\rm E}=\beta M-S_{\rm BH}$ in the fixed-$(J,Q)$ ensemble above. The conical construction has also been extended to AdS black holes in EGB gravity, yielding generalized free energies determined by the mass and Wald entropy~\cite{Li:2023men}.

Motivated by this conical realization, we seek a complementary path integral derivation of generalized free energy in which the black hole parameters can vary at fixed ensemble temperature without imposing $T=T_H$. Our aim is to determine the corresponding leading semiclassical weights and their relation to the free energy landscape. A further motivation arises in more general $f(\mathrm{Riemann})$ theories: although the conical expansion around a stationary black hole reproduces the Wald entropy~\cite{Wald:1993nt,Iyer:1994ys,Iyer:1995kg} contribution at linear order, higher-order terms in the deficit can depend on the smoothing prescription~\cite{Fursaev:1995ef,Wu:2025jes}.

Gravitational path integral factorization provides a framework for such a derivation. In Schwarzschild--de Sitter spacetime, generally unequal horizon temperatures prevent a single Euclidean period from regularizing both horizons. Dividing the geometry at fixed microcanonical interface data yields two regular caps with different lapse normalizations and a constrained action equal to minus the sum of the horizon entropies~\cite{Draper:2022xzl}, consistent with earlier results~\cite{Wu:2004db}. A related study further explored the interpretation of Schwarzschild--de Sitter geometries as constrained instantons at fixed mass and obtained the same action~\cite{Morvan:2022ybp}. The factorization framework has also been used to extend gravitational partition functions at fixed spatial volume~\cite{Jacobson:2022jir} to include mass-constrained instantons~\cite{Wu:2025rbx}. The same manifold decomposition approach extends beyond Einstein gravity to $f(\mathrm{Riemann})$ effective actions, yielding the total Wald entropy in a four-dimensional Schwarzschild--de Sitter example with curvature-squared corrections~\cite{Draper:2023bhg}. The feature we adopt is factorization at fixed common interface data, with the lapse allowed to jump across the interface.

Within this canonical factorization framework, we use Arnowitt--Deser--Misner (ADM) variables~\cite{Arnowitt:1962hi,Regge:1974zd} to specify local data on a codimension-one auxiliary interface. In Einstein gravity, a boundary Legendre transform selects partial-Neumann (pN) data: the densitized Brown--York energy and momentum responses $(\mathcal E,\mathcal J_A)$ and the cross-section metric $\sigma_{AB}$~\cite{Brown:1992br,Brown:1992bq}. Fixing these data leaves the lapse and tangential shift unfixed at the interface. For selected nonextremal stationary axisymmetric families, this freedom accommodates thermal and rotational mismatches while keeping the horizon cap regular and the physical boundary sources fixed. The resulting lapse and shift may jump across the auxiliary interface, and these jumps require a variational interpretation.

The method of constrained instantons provides this interpretation~\cite{Affleck:1980mp,Cotler:2020lxj}. A constraint $C[\phi]=\zeta$ defines a sector whose stationary configurations are found with $\zeta$ held fixed; they may contribute semiclassically even when they are not stationary under unrestricted variations. In our construction, boundary integral transforms relate the Dirichlet and pN representations, and delta functionals select the fixed common interface data. Independent variation determines the constraint multipliers that balance the lapse and shift jumps, establishing the paired subregion solutions as constrained instantons. Integrating these fixed-sector amplitudes over the common data reconstructs the formal thermal partition function.

Restricting these amplitudes to a selected nonextremal stationary black hole family labeled by collective parameters $\lambda^m$ defines a finite-dimensional landscape. For a regular horizon cap paired with an exterior reaching the physical asymptotic boundary $\mathcal B$, the leading semiclassical exponential weight is $\exp(-\beta\mathcal F_{\rm gen})$, with
\begin{equation}
    \mathcal F_{\rm gen}(\lambda^m;T,\Omega_{\mathcal B}^I)
    =E_{\mathcal B}(\lambda^m)-\Omega_{\mathcal B}^I J_{\mathcal B,I}(\lambda^m)
    -T S(\lambda^m).
    \label{eq:intro_constrained_free_energy}
\end{equation}
Here $\Omega_{\mathcal B}^I$ are the imposed angular velocities, and $E_{\mathcal B}$ and $J_{\mathcal B,I}$ are renormalized Hamiltonian charges in the chosen reference frame. The regular cap supplies the entropy $S$, which equals the Bekenstein--Hawking entropy $S_{\rm BH}$ in Einstein gravity. Using the auxiliary field Hamiltonian formulation~\cite{Deruelle:2009zk,Draper:2023bhg}, we extend this construction to local, parity-even $f(\mathrm{Riemann})$ theories without curvature derivatives. With complete canonical interface data and the stationary sewing conditions specified below, the same form holds with the corresponding generalized charges and Wald entropy, without a small-deficit expansion. Explicit checks of the sewing data and charge normalization are provided by static five-dimensional AdS black holes in EGB gravity and the rotating, locally AdS$_3$ Ba\~nados--Teitelboim--Zanelli (BTZ) black hole branch of new massive gravity (NMG). These results establish a semiclassical path integral construction of black hole free energy landscapes from constrained geometries assembled across a codimension-one interface, applicable to Einstein gravity and the higher-curvature theories considered here.

Related uses of constrained sectors include path integrals at fixed area or geometric entropy~\cite{Dong:2018seb,Dong:2019piw}, corner formulations with a fixed Jacobson--Myers functional in Lovelock gravity~\cite{Kastikainen:2023omj}, and Lorentzian and Euclidean constructions of black hole thermal weights from constrained saddles with codimension-two singularities~\cite{Marolf:2022ybi,Chen:2025leq}. A comparison with these approaches is given in Sec.~\ref{sec:conclusions}.

This paper is organized as follows. Sections~\ref{sec:adm_factorization} and~\ref{sec:quasilocal_constrained_instantons} establish action additivity, sewing, and constrained variation. Section~\ref{sec:free_energy_landscape} derives the landscape, and Sec.~\ref{sec:bh_solutions} treats the Einstein examples, including compact two-cap de Sitter configurations. Section~\ref{sec:modified_gravity} develops the higher-curvature extension and its EGB and NMG applications; Sec.~\ref{sec:conclusions} concludes. Appendices~\ref{app:ci_proof} and~\ref{app:fr_technical_details} contain the variational proofs and canonical calculations.
\section{Manifold Decomposition and Additivity of the Gravitational Action}
\label{sec:adm_factorization}
When the gravitational system is factorized across an auxiliary hypersurface, the boundary terms in each subregion action must be chosen to make the variational problem well posed for the specified boundary data. In this section, we first specify the geometric setup, orientation conventions, and ADM canonical variables. We then discuss the Dirichlet action, for which the induced metric is held fixed, and the pN action, for which a subset of the Brown--York responses is fixed.

These two variational principles select different sets of interface data to be fixed and identified in the factorized description. It is therefore important to distinguish the geometric cutting of a smooth covariant configuration from the boundary-value problems defined on the field space after the cut. Here we restrict attention to the actions and their variations; the corresponding construction at the level of the gravitational path integral is deferred to Sec.~\ref{sec:quasilocal_constrained_instantons}. Throughout, we work in units $G=\hbar=c=k_{\rm B}=1$.

\subsection{Manifold Decomposition and Geometric Conventions}
\label{subsec:adm_geometry}

Consider a $D=d+2$ dimensional Euclidean manifold $\mathcal M$, foliated away from any bolts (defined below) by constant-$\tau$ hypersurfaces $\Sigma_\tau$, with $0\leq\tau<\beta$. The initial and final slices are glued according to the prescribed thermal identification; in the rotating case, this identification is supplemented by the corresponding angular identification. Introducing a radial coordinate $\rho$, we define an auxiliary hypersurface $\mathcal T_r$ at $\rho=r$. The portion of each time slice on its inner side and the corresponding interface cross-section are
\begin{equation}
\begin{aligned}
    \Sigma_\tau^{(r)}
       &=\{p\in\Sigma_\tau\mid\rho(p)\le r\},\\
    \mathcal S_\tau^{(r)}
       &=\Sigma_\tau^{(r)}\cap\mathcal T_r
        =\Sigma_\tau\cap\mathcal T_r.
\end{aligned}
    \label{eq:truncated_time_slice}
\end{equation}
The cross-section $\mathcal S_\tau^{(r)}$ is the boundary component of $\Sigma_\tau^{(r)}$ that lies on $\mathcal T_r$. If the truncated slice has no other boundary components, then $\partial\Sigma_\tau^{(r)}=\mathcal S_\tau^{(r)}$.

In the factorization construction, $\mathcal T_r$ is an internal auxiliary interface at fixed finite $r$, and its interface data are distinct from the physical boundary conditions imposed on the outer boundary $\mathcal B$, when present. We denote an outer radial surface at $\rho=R>r$ by $\mathcal T_R$. Its cross-section on a constant-$\tau$ slice is $\mathcal S_\tau^{(R)}=\Sigma_\tau\cap\mathcal T_R$. If the physical boundary is at this finite radius, we identify $\mathcal B=\mathcal T_R$ and impose the physical boundary conditions directly there, keeping $R$ finite. When $\mathcal B$ lies at asymptotic infinity, $\mathcal T_R$ serves as the outer boundary of the regulated geometry. Its data are chosen to implement the prescribed asymptotic conditions at $\mathcal B$ as $R\to\infty$. Any required rescalings, reference subtraction, or counterterm contributions are treated according to the chosen asymptotic boundary prescription. The local geometric definitions below apply on both radial surfaces.

We denote the induced metrics on $\Sigma_\tau$, $\mathcal T_r$, and $\mathcal S_\tau^{(r)}$ by $h_{ab}$, $\gamma_{ij}$, and $\sigma_{AB}$, respectively, with $h_{ab}$ restricted to $\Sigma_\tau^{(r)}$ when appropriate. We write $h=\det(h_{ab})$, $\gamma=\det(\gamma_{ij})$, and $\sigma=\det(\sigma_{AB})$. Greek indices refer to bulk components, while $a,b,\ldots$, $i,j,\ldots$, and $A,B,\ldots$ denote components intrinsic to $\Sigma_\tau$, $\mathcal T_r$, and $\mathcal S_\tau^{(r)}$, respectively. The covariant derivatives compatible with $h_{ab}$, $\gamma_{ij}$, and $\sigma_{AB}$ are denoted by $D_a$, $\mathcal D_i$, and $\mathscr D_A$, respectively.

We now impose two simplifying geometric assumptions. First, the
cross-sections $\mathcal S_\tau^{(r)}$ are taken to be compact and without boundary, $\partial\mathcal S_\tau^{(r)}=\varnothing$. Second, the temporal foliation is assumed to be orthogonal to the radial hypersurfaces, $u^\mu n_\mu=0$, where $u$ and $n$ denote the temporal and radial unit normals, respectively. We choose adapted coordinates in which their dual one-forms take the form
\begin{equation}
    u=N\,\dd\tau,\qquad n=P\,\dd\rho.
\end{equation}
Here $N$ and $P$ are the temporal and radial lapses, respectively, and $n$ is oriented toward increasing $\rho$. The thermal identification removes the initial and final time slices as physical boundary components, while the closed cross-sections allow tangential integration by parts without boundary contributions. Orthogonality further simplifies the ADM boundary decomposition. On the other hand, bolts are an important feature of nonextremal Euclidean black hole geometries. The temporal foliation degenerates there as the lapse $N\to0$, and the associated contribution to the action is evaluated by the limiting procedure described below.

\begin{figure}[!htbp]
    \centering
    \includegraphics[width=0.58\textwidth]{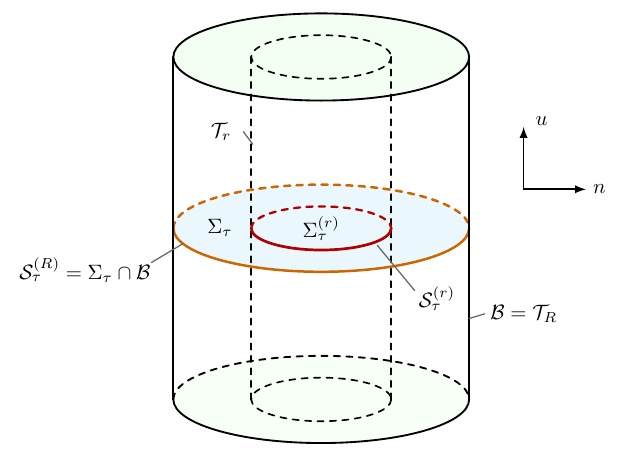}
    \caption{Schematic of the orthogonal temporal--radial foliation at a finite outer radius $R$. The outer solid cylinder denotes $\mathcal B=\mathcal T_R$ for a finite physical boundary, and the inner dashed cylinder is the auxiliary hypersurface $\mathcal T_r$. The blue disk represents the portion of $\Sigma_\tau$ enclosed by $\mathcal T_R$; its part inside the red curve is $\Sigma_\tau^{(r)}$. The orange and red curves are the outer cross-section $\mathcal S_\tau^{(R)}=\Sigma_\tau\cap\mathcal T_R$ and the interface cross-section $\mathcal S_\tau^{(r)}=\Sigma_\tau\cap\mathcal T_r$, respectively. The arrows indicate the unit normals $u$ and $n$, whose dual one-forms are $u=N\,\dd\tau$ and $n=P\,\dd\rho$, respectively. The upper and lower ends are understood to be joined by the prescribed thermal identification. For an asymptotic physical boundary, the outer cylinder represents a cutoff surface $\mathcal T_R$, and the limit $R\to\infty$ is taken according to the chosen asymptotic boundary prescription.}
    \label{fig:euclidean_geometry}
\end{figure}

Figure~\ref{fig:euclidean_geometry} illustrates this geometry with an outer radial surface $\mathcal T_R$ and an internal auxiliary interface $\mathcal T_r$. The outer surface represents either a finite physical boundary $\mathcal B=\mathcal T_R$ or a cutoff for an asymptotic boundary.

Let $e_A{}^\mu$ denote the coordinate tangent basis on $\mathcal S_\tau^{(r)}$. The evolution vectors associated with the temporal and radial coordinates are decomposed as
\begin{equation}
    (\partial_\tau)^\mu=Nu^\mu+\mathcal N^A e_A{}^\mu,
    \qquad
    (\partial_\rho)^\mu=Pn^\mu+\mathcal P^A e_A{}^\mu .
\end{equation}
Here $\mathcal N^A$ and $\mathcal P^A$ are the temporal and radial shifts, respectively. On a real Euclidean section, the unit normals satisfy $u^\mu u_\mu=n^\mu n_\mu=1$. Orthogonality and the adapted coordinates make both shifts tangent to $\mathcal S_\tau^{(r)}$, with $\mathcal N^\rho=\mathcal P^\tau=0$. The metric can then be written locally as
\begin{align}
    \dd s^2
    &=N^2\dd\tau^2+
      h_{ab}(\dd x^a+\mathcal N^a\dd\tau)
            (\dd x^b+\mathcal N^b\dd\tau)\notag\\
    &=N^2\dd\tau^2+P^2\dd\rho^2
      +\sigma_{AB}
       (\dd x^A+\mathcal N^A\dd\tau+\mathcal P^A\dd\rho)
       (\dd x^B+\mathcal N^B\dd\tau+\mathcal P^B\dd\rho).
    \label{eq:adm_metric}
\end{align}
In these coordinates, $\sqrt h=P\sqrt\sigma$. The extrinsic curvatures associated with the radial direction, for $\mathcal T_r$ and $\mathcal S_\tau^{(r)}$, are defined by
\begin{equation}
    \mathcal K_{ij}=\gamma_i{}^\mu\gamma_j{}^\nu
       \nabla_\mu n_\nu,
    \qquad
    k_{AB}=\sigma_A{}^a\sigma_B{}^bD_a n_b,
    \label{eq:radial_extrinsic_curvature_conventions}
\end{equation}
respectively. Their traces are $\mathcal K=\gamma^{ij}\mathcal K_{ij}$ and $k=\sigma^{AB}k_{AB}$. The Euclidean time slices have extrinsic curvature
\begin{equation}
    K_{ab}=h_a{}^\mu h_b{}^\nu\nabla_\mu u_\nu .
    \label{eq:temporal_extrinsic_curvature_definition}
\end{equation}

\begin{figure}[!htbp]
    \centering
    \includegraphics[width=0.6\textwidth]{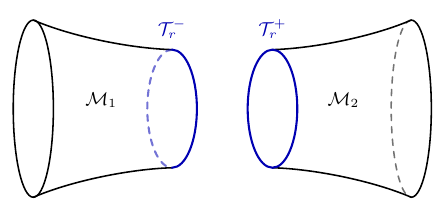}
    \caption{Schematic of the two subregions obtained by cutting the
    manifold along the auxiliary hypersurface $\mathcal T_r$.
    The displayed boundary components
    $\mathcal T_r^-\subset\partial\mathcal M_1$ and
    $\mathcal T_r^+\subset\partial\mathcal M_2$ are two copies of the
    same interface. For the initial smooth geometry, choosing $n$ to point from
    $\mathcal M_1$ toward $\mathcal M_2$ gives the outward normals
    $n_1^{\rm out}=n$ and $n_2^{\rm out}=-n$.
    The superscripts $-$ and $+$ label the two copies only. The gap is
    drawn solely to distinguish the copies after the cut; the Dirichlet
    or pN sewing conditions specified below determine which boundary
    data are identified.}
    \label{fig:manifold_split}
\end{figure}

Cutting the manifold along $\mathcal T_r$ yields the two subregions shown in Fig.~\ref{fig:manifold_split},
\begin{equation}
    \mathcal M=\mathcal M_1
       \cup_{\mathcal T_r^-\sim\mathcal T_r^+}\mathcal M_2,
    \qquad
    \mathcal T_r^-\subset\partial\mathcal M_1,
    \quad \mathcal T_r^+\subset\partial\mathcal M_2 .
    \label{eq:manifold_split}
\end{equation}
We label the two subregions such that the interior of $\mathcal M_1$ lies in $\rho<r$, whereas the interior of $\mathcal M_2$ lies in $\rho>r$. The previously defined truncated time slice is therefore $\Sigma_\tau^{(r)}=\Sigma_\tau\cap\mathcal M_1$. When present, the outer physical boundary $\mathcal B$ belongs to $\partial\mathcal M_2$. The geometric cut initially gives two restrictions of the same smooth configuration. In the variational problems considered below, however, the fields on the two subregions are allowed to vary independently, subject to the interface data specified by the chosen boundary representation.

We formulate the discussion primarily for $\mathcal M_1$; the construction for $\mathcal M_2$ is analogous. In the common-normal convention, $n$ denotes the radial unit normal pointing toward increasing $\rho$, normalized using the metric on each side. The outward normals are therefore $n_1^{\rm out}=n$ and $n_2^{\rm out}=-n$ on the respective interface copies. Normal-dependent interface quantities are expressed in this convention before taking their jumps.

To compare fields on the two interface copies, introduce an abstract interface $\mathcal T_r$ together with the identification maps $\iota_1:\mathcal T_r\to\mathcal T_r^-$ and $\iota_2:\mathcal T_r\to\mathcal T_r^+$ induced by the common $(\tau,x^A)$ coordinates. Normal contractions and projections are performed separately on each side using the convention above. The resulting intrinsic interface fields are then pulled back to the abstract interface and expressed in the same tangential frame. For any such field $X$, we define
\begin{equation}
    [X]=\iota_2^*X_2-\iota_1^*X_1
        = X_2-X_1 .
    \label{eq:common_normal_jump}
\end{equation}
Here the final expression is the shorthand used below for the difference defined using the pullbacks. The labels $-$ and $+$ correspond to the labels $1$ and $2$, respectively, so $X_2-X_1$ is the value on the $+$ side minus that on the $-$ side. The former notation is convenient in interface formulas, while the latter more clearly identifies the associated manifolds or physical regions; both sets of labels distinguish the same two sides. 

If the inner region $\mathcal M_1$ contains a codimension-two fixed-point set of the contractible thermal circle, we refer to this set as a \emph{bolt}. In the rotating case, this circle is defined by the combined thermal and angular identification. The region extending radially from a regular bolt to the auxiliary interface $\mathcal T_r$ is called the \emph{horizon cap}. A regular bolt is not a boundary component of the Euclidean manifold, but the ADM temporal foliation degenerates there. The associated contribution is defined by the small-tube prescription in Eq.~\eqref{eq:adm_bolt_compensation}. For complex rotating saddles, the relevant geometric quantities are defined by analytic continuation, and the notation introduced above is retained.

\subsection{Dirichlet Action, Boundary Variation, and Additivity}
\label{subsec:dirichlet_variational_principle}
The Euclidean Dirichlet action for Einstein gravity is
\begin{equation}
    I_{\rm D}[g]=-\frac{1}{16\pi}
       \int_{\mathcal M}\dd^{d+2}x\,\sqrt g\,(R-2\Lambda)
       -\frac{1}{8\pi}
       \int_{\partial\mathcal M}\dd^{d+1}x\,\sqrt\gamma\,
       \mathcal K.
    \label{eq:euclidean_lagrangian_action}
\end{equation}
The Gibbons--Hawking--York boundary term cancels the terms involving normal derivatives of $\delta g_{\mu\nu}$ in the variation of the bulk action, thereby yielding a well-posed variational principle with the induced metric held fixed at the boundary \cite{Gibbons:1976ue}. For the full manifold, $\partial\mathcal M$ ordinarily denotes the physical boundary $\mathcal B$. In the factorization construction below, however, the same Dirichlet variational principle will also be applied to subregions obtained by cutting along $\mathcal T_r$, in which case $\mathcal T_r$ becomes a boundary component of each subregion.

We now rewrite the Euclidean action in ADM form using the foliation of $\mathcal M_1$ by the constant-$\tau$ hypersurfaces $\Sigma_\tau^{(r)}$. With $\dot h_{ab}=\partial_\tau h_{ab}$, the extrinsic curvature of $\Sigma_\tau^{(r)}$ and its trace are
\begin{equation}
    K_{ab}=\frac{1}{2N}
       \bigl(\dot h_{ab}-D_a\mathcal N_b-D_b\mathcal N_a\bigr),
    \qquad K=h^{ab}K_{ab}.
    \label{eq:time_extrinsic_curvature}
\end{equation}
The Gauss--Codazzi relations express the tangential and mixed projections of the bulk curvature in terms of the intrinsic and extrinsic geometry of the slices. After the total divergence term in the Einstein--Hilbert Lagrangian is separated, the bulk ADM Lagrangian contains $\tau$ derivatives only through $K_{ab}$ and hence through $\dot h_{ab}$. The conjugate momentum is therefore
\begin{equation}
    p^{ab}
      =\frac{\delta I_{\rm D}}{\delta\dot h_{ab}}
      =\frac{\sqrt h}{16\pi}\,\pi^{ab},
    \qquad
    \pi^{ab}=K^{ab}-Kh^{ab}.
    \label{eq:adm_momentum}
\end{equation}
Thus $p^{ab}$ is a tensor density of weight one, whereas $\pi^{ab}$ is an ordinary tensor. We denote its trace by $\pi=h_{ab}\pi^{ab}$. The temporal Legendre transformation leaves the Dirichlet boundary data on $\mathcal T_r$ unchanged. In the resulting first-order formulation, $h_{ab}$ and $p^{ab}$ are treated as independent variables, with $\pi^{ab}=16\pi p^{ab}/\sqrt h$. The relation between the canonical momentum and extrinsic curvature in Eq.~\eqref{eq:adm_momentum} is then recovered by varying $p^{ab}$; eliminating the momentum through this equation reproduces the metric Dirichlet action. In the Hamiltonian formulation, $N$ and $\mathcal N^a$ act as Lagrange multipliers enforcing the Hamiltonian and momentum constraints.

The corresponding first-order Dirichlet action, also denoted by $I_{\rm D}$, is
\begin{equation}
\begin{aligned}
    I_{\rm D}[\mathcal M_1]
    &=\frac{1}{16\pi}\int_0^\beta\dd\tau
      \int_{\Sigma_\tau^{(r)}}\dd^{d+1}x\,\sqrt h\,
      \bigl(\pi^{ab}\dot h_{ab}
            -N\mathcal H-\mathcal N_a\mathcal H^a\bigr)
      \\
    &\quad+B_N[\mathcal T_r]+B_{\mathcal N}[\mathcal T_r]
          +I_{\rm bolt},
\end{aligned}
    \label{eq:adm_action}
\end{equation}
and
\begin{equation}
    \mathcal H
       =R[h]-2\Lambda+\pi^{ab}\pi_{ab}-\frac{1}{d}\pi^2,
    \qquad
    \mathcal H^a=-2D_b\pi^{ab}.
    \label{eq:adm_constraints}
\end{equation}
The terms $B_N$ and $B_{\mathcal N}$ are the radial boundary terms associated with the lapse and shift, respectively. The former results from combining the radial flux of the bulk total derivative with the GHY term, whereas the latter arises from integrating by parts the bulk term containing spatial derivatives of the shift. Explicitly,
\begin{align}
    B_N[\mathcal T_r]
       &=-\frac{1}{8\pi}\int_0^\beta\dd\tau
          \int_{\mathcal S_\tau^{(r)}}\dd^dx\,
          \sqrt\sigma\,Nk,
          \notag\\
    B_{\mathcal N}[\mathcal T_r]
       &=-\frac{1}{8\pi}\int_0^\beta\dd\tau
          \int_{\mathcal S_\tau^{(r)}}\dd^dx\,
          \sqrt\sigma\,\mathcal N_a n_b\pi^{ab}.
    \label{eq:adm_boundary_terms}
\end{align}
The term $I_{\rm bolt}$ accounts for a possible degeneration of the temporal foliation and vanishes when no bolt is present. If $\mathcal M_1$ contains a smooth bolt, we excise a tubular neighborhood of radius $\epsilon$ and denote its boundary by $\mathcal H_\epsilon$, with the outward normal of the punctured region pointing toward the excised bolt. The canonical decomposition is then performed on the punctured region. Restoring the original smooth cap and taking $\epsilon\to0$ gives
\begin{equation}
    I_{\rm bolt}
       =\lim_{\epsilon\to0}
        \left[
          B_N [\mathcal H_\epsilon] +B_{\mathcal N} [\mathcal H_\epsilon]
          +\frac{1}{8\pi}
           \int_{\mathcal H_\epsilon}\dd^{d+1}x\,
                 \sqrt\gamma\,\mathcal K
        \right].
    \label{eq:adm_bolt_compensation}
\end{equation}
The final term compensates for the GHY contribution introduced on the artificial tubular boundary, while the sum of the first two terms collects the corresponding ADM boundary terms on $\mathcal H_\epsilon$. Their combined limit gives the finite contribution associated with the degeneration of the temporal foliation.

To make the radial structure of the action manifest, we further decompose the intrinsic scalar curvature of $\Sigma_\tau^{(r)}$ near $\mathcal S_\tau^{(r)}$ as
\begin{equation}
    R[h]=R[\sigma]+k^2-k_{AB}k^{AB}
         -2D_a(n^ak)+2D_a a^a,
    \qquad
    a^a=n^bD_bn^a.
    \label{eq:gauss_codazzi_radial}
\end{equation}
Integrating the curvature divergences and the shift term $-\mathcal N_a\mathcal H^a$ by parts over the spatial slice yields radial surface contributions that cancel $B_N[\mathcal T_r]$ and $B_{\mathcal N}[\mathcal T_r]$, respectively. In the adapted orthogonal foliation, the acceleration term involves only tangential derivatives and contributes no radial boundary term. The resulting bulk representation contains no spatial derivatives of $\pi^{ab}$, and radial derivatives occur at most to first order, acting only on $N$, $\mathcal N^A$, and $\sigma_{AB}$. This identifies $(N,\mathcal N^A,\sigma_{AB})$ as the Dirichlet data on $\mathcal T_r$ and explains why the other canonical fields need not be fixed there.

The same construction applies to the outer region $\mathcal M_2$, with boundary terms evaluated using its outward normals, and yields the same set of Dirichlet data on its side of the interface.

We now specify the boundary data and their conjugate responses. In the adapted foliation, the metric induced on $\mathcal T_r$ takes the ADM form
\begin{equation}
    \gamma_{ij}\dd x^i\dd x^j=N^2\dd\tau^2
       +\sigma_{AB}(\dd x^A+\mathcal N^A\dd\tau)
                    (\dd x^B+\mathcal N^B\dd\tau).
    \label{eq:boundary_adm_metric}
\end{equation}
Once the foliation and coordinate identification are fixed, specifying $\gamma_{ij}$ is equivalent to specifying the boundary data

\begin{equation}
    \mathcal C_{\rm D}=(N,\mathcal N^A,\sigma_{AB})_{\mathcal T_r}.
    \label{eq:dirichlet_boundary_data}
\end{equation}
In the metric formulation, the responses conjugate to these data are encoded in the Brown--York tensor. With the Euclidean action and outward-normal convention adopted above, we define
\begin{equation}
    \tau^{ij}=\frac{2}{\sqrt\gamma}
       \left.\frac{\delta I_{\rm D}}{\delta\gamma_{ij}}
       \right|_{\rm EOM}
       =\frac{1}{8\pi}
         (\mathcal K^{ij}-\mathcal K\gamma^{ij}).
    \label{eq:brown_york_stress_tensor}
\end{equation}
The expression on the right defines $\tau^{ij}$ in terms of extrinsic curvature without requiring the metric to satisfy the equations of motion. The EOM restriction removes the bulk contribution to the action variation, allowing $\tau^{ij}$ to be interpreted as the response of the on-shell action to variations of the boundary metric. Its temporal, mixed, and spatial projections define the quasilocal energy density, momentum density, and spatial stress, respectively \cite{Brown:1992br}:
\begin{equation}
    \varepsilon=u_i u_j\tau^{ij},
    \qquad j_A=-\sigma_A{}^i u^j\tau_{ij},
    \qquad s^{AB}=\sigma^A{}_i\sigma^B{}_j\tau^{ij}.
    \label{eq:by_projections}
\end{equation}
For the orthogonal foliation considered here, using the relation between the canonical momentum and extrinsic curvature in Eq.~\eqref{eq:adm_momentum} gives
\begin{equation}
    \varepsilon=-\frac{k}{8\pi},
    \qquad
    j_A=\frac{1}{8\pi}\sigma_A{}^a n_b\pi^b{}_a.
    \label{eq:by_energy_momentum_density}
\end{equation}
Thus the quasilocal energy density is determined by the extrinsic curvature of $\mathcal S_\tau^{(r)}$ within the spatial slice, while the momentum density is given by the normal projection of the temporal ADM momentum. In the first-order formulation, we use Eq.~\eqref{eq:by_energy_momentum_density} to define $\varepsilon$ and $j_A$ on the independent canonical field space. The canonical momentum response agrees with its geometric Brown--York projection upon imposing the relation between the canonical momentum and extrinsic curvature. In terms of these responses, the radial boundary terms in Eq.~\eqref{eq:adm_boundary_terms} take the form
\begin{equation}
    B_N[\mathcal T_r]+B_{\mathcal N}[\mathcal T_r]
       =\int_{\mathcal T_r}\dd\tau\,\dd^dx\,\sqrt\sigma
          (N\varepsilon-\mathcal N^A j_A).
    \label{eq:adm_boundary_terms_by}
\end{equation}

Accordingly, the on-shell variation of the Dirichlet action at $\mathcal T_r$ takes the form
\begin{equation}
    \left.\delta I_{\rm D}\right|_{{\rm EOM},\mathcal T_r}
       =\int_{\mathcal T_r}\dd\tau\,\dd^dx\,\sqrt\sigma
       \left(\varepsilon\,\delta N
        -j_A\,\delta\mathcal N^A
        +\frac{N}{2}s^{AB}\delta\sigma_{AB}\right).
    \label{eq:dirichlet_boundary_variation}
\end{equation}
It is convenient to introduce the weight-one densities conjugate to $N$ and $\mathcal N^A$,
\begin{equation}
    \mathcal E=\sqrt\sigma\,\varepsilon,
    \qquad \mathcal J_A=\sqrt\sigma\,j_A.
    \label{eq:densitized_by_data}
\end{equation}
Since $\sigma_{AB}$ remains an independent boundary datum, variations of these densitized responses also involve variations of the measure,
\begin{equation}
\begin{aligned}
    \delta\mathcal E
       &=\sqrt\sigma\,\delta\varepsilon
          +\frac{1}{2}\mathcal E\sigma^{AB}\delta\sigma_{AB},\\
    \delta\mathcal J_A
       &=\sqrt\sigma\,\delta j_A
          +\frac{1}{2}\mathcal J_A\sigma^{BC}\delta\sigma_{BC}.
\end{aligned}
    \label{eq:densitized_by_variations}
\end{equation}
The Brown--York responses introduced here are defined with respect to the unsubtracted action. On the auxiliary interface $\mathcal T_r$, they will serve as local boundary data for the sewing construction. At a physical asymptotic boundary, by contrast, finite Hamiltonian charges generally require an appropriate reference subtraction or renormalization prescription, which will be specified when particular geometries are considered. The interface transformations introduced below act only on the auxiliary boundary data and leave the prescribed physical boundary conditions unchanged.

We can now establish the additivity of the Dirichlet action. Consider first a smooth configuration on $\mathcal M$ and restrict it to the two subregions. The induced data on the two copies of the interface obey
\begin{equation}
    [N]=0,\qquad[\mathcal N^A]=0,\qquad[\sigma_{AB}]=0.
    \label{eq:dirichlet_matching}
\end{equation}
For a smooth configuration, the bulk integration domain splits into the two subregions, while the artificial interface contributions cancel between the two sides because their outward normals have opposite orientations. The Dirichlet action is therefore additive,
\begin{equation}
    I[\mathcal M;\Phi_1,\Phi_2]
       =I_{\rm D}[\mathcal M_1;\Phi_1]
        +I_{\rm D}[\mathcal M_2;\Phi_2].
    \label{eq:adm_additivity}
\end{equation}
Here $\Phi_s$ denotes the canonical fields $(h_{ab},p^{ab},N,\mathcal N^a)$ on $\mathcal M_s$, with $s=1,2$. The subscript D specifies the Dirichlet data imposed on the auxiliary interface $\mathcal T_r$; any physical boundary components retain their prescribed physical boundary conditions and the associated action terms.

We extend the action to piecewise smooth canonical histories, regular up to the nondegenerate interface on each side, using the bulk representation described above, which contains radial derivatives only up to first order. The left-hand side of Eq.~\eqref{eq:adm_additivity} is understood in this sense. On this enlarged field space, $\Phi_1$ and $\Phi_2$ are chosen independently subject to the matching conditions in Eq.~\eqref{eq:dirichlet_matching}; their conjugate responses may have finite jumps. Although the explicit interface terms need not cancel, the bulk representation has an integrable action density with no delta functions at the interface, so its integration domain splits directly into $\mathcal M_1$ and $\mathcal M_2$. Reversing the integrations by parts separately in the two subregions restores the Dirichlet boundary terms and reproduces Eq.~\eqref{eq:adm_additivity}. In the metric formulation, for a continuous metric with a jump in extrinsic curvature, the noncancelling interface GHY terms reproduce the distributional curvature contribution at $\mathcal T_r$; they do not represent an additional physical interface action. This is the additivity mechanism discussed in Refs.~\cite{Draper:2022xzl,Draper:2023bhg}, expressed here in preparation for the boundary representations considered below.

Now keep the physical boundary data fixed and any bolts regular, and suppose that the fields in both subregions satisfy the full canonical bulk equations, including the relation between the canonical momentum and extrinsic curvature. We continue to identify their Dirichlet data across the interface, but allow the common values themselves to vary, so that their variations are identified on the two sides. In the absence of an independent interface action, the resulting variation of the sewn action is
\begin{equation}
    \left.\delta I\right|_{\rm EOM}
       =\int_{\mathcal T_r}\dd\tau\,\dd^dx
       \left(-[\mathcal E]\delta N
        +[\mathcal J_A]\delta\mathcal N^A
        -\frac{\sqrt\sigma}{2}[Ns^{AB}]\delta\sigma_{AB}\right).
    \label{eq:dirichlet_sewn_boundary_variation}
\end{equation}
Stationarity with respect to the common interface data requires $[\mathcal E]=0$, $[\mathcal J_A]=0$, and $[Ns^{AB}]=0$. With the matched Dirichlet data and a nonzero common lapse, these conditions are equivalent to continuity of the full Brown--York tensor in the common-normal convention.

\subsection{Partial Legendre Transform and Additivity of the pN Action}
\label{subsec:adm_sewing}

To fix the densitized energy and momentum responses on the auxiliary interface while retaining the cross-section metric as Dirichlet data, we perform a boundary Legendre transform with respect to $N$ and $\mathcal N^A$ \cite{Brown:1992bq}. On the copy of $\mathcal T_r^\pm$ belonging to each subregion, we add
\begin{equation}
    B_{\rm Leg}^{\mathcal T_r^\pm}
       =\pm\int_{\mathcal T_r^\pm}\dd\tau\,\dd^dx
             (N\mathcal E-\mathcal N^A\mathcal J_A),
    \label{eq:partial_neumann_legendre_term}
\end{equation}
where the responses are evaluated using the common-normal convention. Since $\mathcal T_r^-$ belongs to $\mathcal M_1$ and $\mathcal T_r^+$ to $\mathcal M_2$, the Legendre term carries a minus sign for $\mathcal M_1$ and a plus sign for $\mathcal M_2$. These signs follow from the opposite outward-normal orientations of the two subregions. On each subregion, with the corresponding interface copy understood, we define
\begin{equation}
    I_{\rm pN}=I_{\rm D}+B_{\rm Leg}^{\mathcal T_r}.
    \label{eq:partial_neumann_action_definition}
\end{equation}
For the Dirichlet action in Eq.~\eqref{eq:adm_action}, this Legendre term precisely cancels $B_N+B_{\mathcal N}$ on the corresponding copy of $\mathcal T_r$. Hence, for either subregion $\mathcal M_s$ ($s=1,2$), the resulting action takes the form
\begin{equation}
\begin{aligned}
    I_{\rm pN}[\mathcal M_s]
    &=\frac{1}{16\pi}\int_0^\beta\dd\tau
       \int_{\Sigma_\tau\cap\mathcal M_s}\dd^{d+1}x\,\sqrt h\,
       (\pi^{ab}\dot h_{ab}-N\mathcal H
             -\mathcal N_a\mathcal H^a)
        +I_{\rm rest}[\mathcal M_s].
\end{aligned}
    \label{eq:partial_neumann_action}
\end{equation}
Here $I_{\rm rest}[\mathcal M_s]$ collects any bolt contribution and the prescribed physical boundary terms, including the ADM pairings and required subtraction or counterterms. These terms are unaffected by the interface Legendre transform.

Keeping the physical boundary conditions fixed and any bolts regular, the on-shell interface variation for the inner region $\mathcal M_1$, whose outward normal agrees with the chosen common orientation, is
\begin{equation}
    \left.\delta I_{\rm pN}\right|_{{\rm EOM},\mathcal T_r}
       =\int_{\mathcal T_r}\dd\tau\,\dd^dx
       \left(-N\delta\mathcal E
        +\mathcal N^A\delta\mathcal J_A
        +\frac{\sqrt\sigma N}{2}s^{AB}\delta\sigma_{AB}\right).
    \label{eq:partial_neumann_boundary_variation}
\end{equation}
For $\mathcal M_2$, the right-hand side carries an overall minus sign when its responses are expressed in the common-normal convention. The same fixed pN data therefore define the variational problem on either side:
\begin{equation}
    \mathcal C_{\rm pN}
       =(\mathcal E,\mathcal J_A;\sigma_{AB})_{\mathcal T_r},
    \label{eq:partial_neumann_boundary_data}
\end{equation}
\begin{equation}
    \delta\mathcal E=0,\qquad
    \delta\mathcal J_A=0,\qquad
    \delta\sigma_{AB}=0.
    \label{eq:partial_neumann_variations}
\end{equation}
The term ``partial-Neumann'' refers to the fact that the Legendre transform exchanges the fixed variables only in the boundary canonical pairs associated with the lapse and tangential shift. The cross-section metric $\sigma_{AB}$ remains subject to Dirichlet boundary conditions, and its conjugate stress $s^{AB}$ is therefore not fixed.

Using the common-normal convention, we identify the pN data on the two interface copies:
\begin{equation}
    [\sigma_{AB}]=0,\qquad
    [\mathcal E]=0,\qquad
    [\mathcal J_A]=0.
    \label{eq:partial_neumann_sewing_conditions}
\end{equation}
Because Eq.~\eqref{eq:partial_neumann_boundary_variation} contains only variations of $(\mathcal E,\mathcal J_A;\sigma_{AB})$, the conjugate sources $N$ and $\mathcal N^A$ remain unfixed. The pN sewing conditions therefore allow
\begin{equation}
    [N]\quad\text{and}\quad[\mathcal N^A]
    \quad\text{need not vanish}.
    \label{eq:lapseshift_jump_allowed}
\end{equation}
With the common pN interface data held fixed, stationarity of the subregion actions does not require $N$ or $\mathcal N^A$ to be continuous across the interface.

The additivity mechanism can be seen by comparing the purely bulk Dirichlet and pN representations. For the source--response part of the radial derivative terms, evaluated on the constant-$\rho$ cross-sections, the relevant identity obtained by integration by parts is
\begin{equation}
    \mathcal E\,\partial_\rho N
       -\mathcal J_A\,\partial_\rho\mathcal N^A
       =-N\,\partial_\rho\mathcal E
       +\mathcal N^A\,\partial_\rho\mathcal J_A
       +\partial_\rho(N\mathcal E-\mathcal N^A\mathcal J_A).
    \label{eq:pn_radial_derivative_exchange}
\end{equation}
The left-hand side occurs in the purely bulk Dirichlet representation. In the pN representation, the corresponding terms are $-N\partial_\rho\mathcal E+\mathcal N^A\partial_\rho\mathcal J_A$, while the remaining radial derivatives act on $\sigma_{AB}$ and are at most first order. For piecewise smooth canonical fields regular up to the interface on each side, the matching conditions in Eq.~\eqref{eq:partial_neumann_sewing_conditions} therefore yield an integrable bulk action density without interface delta functions, even when $N$ and $\mathcal N^A$ have finite jumps. The original uncut action, evaluated in this representation under the pN matching conditions, consequently decomposes as
\begin{equation}
    I[\mathcal M;\Phi_1,\Phi_2]
       =I_{\rm pN}[\mathcal M_1;\Phi_1]
        +I_{\rm pN}[\mathcal M_2;\Phi_2].
    \label{eq:pn_original_action_additivity}
\end{equation}
When the Dirichlet data $(N,\mathcal N^A;\sigma_{AB})$ are identified instead, Eq.~\eqref{eq:adm_additivity} gives the decomposition into Dirichlet subregion actions. The relation between the two decompositions is encoded in Eq.~\eqref{eq:pn_radial_derivative_exchange}: its left-hand side contains the Dirichlet radial derivative terms, while the first two terms on the right are their pN counterparts. The remaining total derivative accounts for the change of interface boundary terms under the partial Legendre transform. The corresponding subregion boundary terms therefore implement two decompositions of the same uncut action, with the prescribed physical boundary conditions unchanged.

Once the subregion fields satisfy the full canonical bulk equations, stationarity under variations of the common pN data in the unrestricted sewing problem requires $[N]=0$, $[\mathcal N^A]=0$, and $[Ns^{AB}]=0$. Section~\ref{sec:quasilocal_constrained_instantons} relates the two classical decompositions to changes of the interface boundary representation in the gravitational path integral and distinguishes fixed-data saddles from stationary points of the subsequent integral over the common data.

\section{Sewing the Gravitational Path Integral and Constrained Instantons}
\label{sec:quasilocal_constrained_instantons}

The preceding section established Dirichlet and pN decompositions of the same uncut action. We now relate their path integral representations by boundary integral transforms and select sectors with fixed common pN data. Their constrained saddles may have lapse and tangential shift jumps. Stationarity under the subsequent integral over the common data imposes the additional matching conditions.

\subsection{Path Integral Sewing and Boundary Representations}
\label{subsec:path_integral_dirichlet_sewing}

We retain the geometry, orientations, and units of Sec.~\ref{sec:adm_factorization} and collect the independent ADM variables into
\begin{equation}
    \Phi=(h_{ab},p^{ab},N,\mathcal N^a).
\end{equation}
For prescribed physical boundary data, the Euclidean gravitational path integral is formally
\begin{equation}
    \mathcal Z[\mathcal M]
    =\int_{\Gamma_{\mathcal M}}
    \mathscr D\mu_{\mathcal M}(\Phi)\,
    e^{-I[\mathcal M;\Phi]}.
    \label{eq:gravitational_canonical_path_integral}
\end{equation}
Here $I[\mathcal M;\Phi]$ is the uncut canonical action, including the terms required by the prescribed physical boundary conditions and all bolt contributions. The labels D and pN below distinguish the representations on the auxiliary interface. All functional integrals are understood with a common regulator, compatible gauge fixing, and appropriate integration contours. The functional measures include the gauge-fixing determinants, zero-mode and residual gauge factors, and any required boundary and normalization factors. Their contributions to the exponent are assumed to be subleading in the semiclassical expansion considered here.

Cutting along $\mathcal T_r$ introduces the common Dirichlet data $\mathcal C_{\rm D}=(N,\mathcal N^A,\sigma_{AB})$. The subregion amplitudes integrate the weights $e^{-I_{\rm D}[\mathcal M_s]}$ at fixed interface data, with the respective physical boundary conditions and interior regularity conditions retained. Using the action additivity in Eq.~\eqref{eq:adm_additivity}, we obtain the formal Dirichlet sewing relation~\cite{Draper:2022xzl},
\begin{equation}
\begin{aligned}
    \mathcal Z[\mathcal M]
    ={}&\int \mathscr D\mu_{\rm D}
    (N,\mathcal N^A,\sigma_{AB})\,
    \mathcal Z_{\rm D}
    [\mathcal M_1\mid N,\mathcal N^A,\sigma_{AB}]
    \\
    &\hspace{16mm}\times
    \mathcal Z_{\rm D}
    [N,\mathcal N^A,\sigma_{AB}\mid\mathcal M_2].
    \label{eq:dirichlet_factorization}
\end{aligned}
\end{equation}
At fixed common data, the two interior integrals are performed independently; the common data are then integrated once. No matching of conjugate responses is imposed on individual off-shell configurations. In the semiclassical limit, response matching follows from stationarity with respect to the common Dirichlet data.

We next change the representation on the auxiliary interface at each fixed value of $\sigma_{AB}$. Write $q_s=(N_s,\mathcal N_s^A)$ and $\xi=(\mathcal E,\mathcal J_A)$, and define the source--response pairing
\begin{equation}
    \mathcal P_s[\mathcal E,\mathcal J_A]
    =
    \int_{\mathcal T_r}\dd\tau\,\dd^dx\,
    \bigl(N_s\mathcal E-\mathcal N_s^A\mathcal J_A\bigr).
    \label{eq:ci_source_response_pairing}
\end{equation}
The densities $\mathcal E$ and $\mathcal J_A$ initially denote independent transform labels conjugate to the boundary sources. Their signs follow the common-normal convention adopted for the Brown--York responses. The pN amplitudes are defined by
\begin{align}
    \mathcal Z_{\rm pN}
    [\mathcal M_1\mid\mathcal E,\mathcal J_A;\sigma_{AB}]
    &={}
    \int_{\Gamma_1}\mathscr D\mu_1(q_1;\xi,\sigma_{AB})\,
    e^{\mathcal P_1[\mathcal E,\mathcal J_A]}
    \mathcal Z_{\rm D}
    [\mathcal M_1\mid N_1,\mathcal N_1^A,\sigma_{AB}],
    \label{eq:pn_boundary_transform_1}
    \\
    \mathcal Z_{\rm pN}
    [\mathcal E,\mathcal J_A;\sigma_{AB}\mid\mathcal M_2]
    &={}
    \int_{\Gamma_2}\mathscr D\mu_2(q_2;\xi,\sigma_{AB})\,
    e^{-\mathcal P_2[\mathcal E,\mathcal J_A]}
    \mathcal Z_{\rm D}
    [N_2,\mathcal N_2^A,\sigma_{AB}\mid\mathcal M_2].
    \label{eq:pn_boundary_transform_2}
\end{align}
The induced source measures $\mathscr D\mu_s(q_s;\xi,\sigma_{AB})$ and their contours are chosen together with the pN sewing measure so that the two boundary transforms are dual with respect to the Dirichlet sewing pairing.

The exponential kernels supply the interface terms, changing the actions in the Dirichlet weights to $I_{\rm D}[\mathcal M_1]-\mathcal P_1$ and $I_{\rm D}[\mathcal M_2]+\mathcal P_2$. The relative signs follow from the opposite outward-normal orientations and agree with Eq.~\eqref{eq:partial_neumann_legendre_term}. At a saddle of the boundary source integrals, these terms reproduce the classical pN Legendre terms, as established below using Eq.~\eqref{eq:ci_response_label_matching}.

Under this duality prescription, the boundary transforms in Eqs.~\eqref{eq:pn_boundary_transform_1} and \eqref{eq:pn_boundary_transform_2} rewrite the Dirichlet sewing formula \eqref{eq:dirichlet_factorization} in the following equivalent pN representation of the same partition function:
\begin{equation}
\begin{aligned}
    \mathcal Z[\mathcal M]
    ={}&\int\mathscr D\mu_{\rm pN}
    (\mathcal E,\mathcal J_A,\sigma_{AB})\,
    \mathcal Z_{\rm pN}
    [\mathcal M_1\mid\mathcal E,\mathcal J_A;\sigma_{AB}]
    \\
    &\hspace{14mm}\times
    \mathcal Z_{\rm pN}
    [\mathcal E,\mathcal J_A;\sigma_{AB}\mid\mathcal M_2],
    \label{eq:partial_neumann_factorization}
\end{aligned}
\end{equation}
where $\mathscr D\mu_{\rm pN}$ is the induced measure on the common pN data. Equality of the lapse and tangential shift sources on the two sides is recovered after integration over the complete response data. The transformation acts only on the auxiliary interface, leaving the physical boundary conditions and their thermodynamic ensemble unchanged. Appendix~\ref{app:ci_boundary_transform} gives the formal reconstruction of the Dirichlet sewing relation from the pN representation.

For notational convenience, we henceforth denote functional measures by their integration variables, leaving the full measure factors and contour prescriptions implicit unless needed explicitly. Delta functionals are normalized with respect to the corresponding data measures, and their multiplier representations include the required normalization factors.

\subsection{Fixed pN Data and Constrained Sectors}
\label{subsec:ci_fixed_sectors}

Starting from the pN sewing formula \eqref{eq:partial_neumann_factorization}, we select a sector by fixing the common pN data, denoted by $\mathcal C_{\rm pN}$ in the preceding section. We abbreviate these integration variables as $\mathcal C$ and denote their prescribed values by $\mathcal C_{\mathcal T}$:
\begin{equation}
    \mathcal C=(\mathcal E,\mathcal J_A;\sigma_{AB}),
    \qquad
    \mathcal C_{\mathcal T}
    =(\overline{\mathcal E},\overline{\mathcal J}_A;
      \overline\sigma_{AB}).
    \label{eq:common_pn_interface_data}
\end{equation}
The barred quantities are held fixed within a sector, with $\overline\sigma=\det(\overline\sigma_{AB})$. Here $\mathscr D\mathcal C$ denotes the pN sewing measure in Eq.~\eqref{eq:partial_neumann_factorization}. At each value of $\mathcal C$, the subregion fields induce the common $\sigma_{AB}$ on their respective interface copies, while their lapse and tangential shift are integrated independently in the pN amplitudes.

Inserting a delta functional into the integral over the common data in Eq.~\eqref{eq:partial_neumann_factorization} defines the fixed-sector amplitude,
\begin{equation}
\begin{aligned}
    \mathcal Z_{\rm constr}[\mathcal C_{\mathcal T}]
    ={}&\int\mathscr D\mathcal C\,
    \delta[\mathcal C-\mathcal C_{\mathcal T}]\,
    \mathcal Z_{\rm pN}[\mathcal M_1\mid\mathcal C]
    \mathcal Z_{\rm pN}[\mathcal C\mid\mathcal M_2].
    \label{eq:ci_fixed_sector_integral}
\end{aligned}
\end{equation}
The normalized delta functional evaluates the single integral over the common data at $\mathcal C=\mathcal C_{\mathcal T}$. The two remaining subregion amplitudes therefore give
\begin{equation}
    \mathcal Z_{\rm constr}[\mathcal C_{\mathcal T}]
    =\mathcal Z_{\rm pN}[\mathcal M_1\mid\mathcal C_{\mathcal T}]
     \mathcal Z_{\rm pN}[\mathcal C_{\mathcal T}\mid\mathcal M_2].
    \label{eq:constrained_instanton_factorization}
\end{equation}
Integrating over the sector labels with the same data measure recovers the full partition function,
\begin{equation}
    \mathcal Z[\mathcal M]
    =\int\mathscr D\mathcal C_{\mathcal T}\,
    \mathcal Z_{\rm constr}[\mathcal C_{\mathcal T}].
    \label{eq:constraint_sector_decomposition}
\end{equation}
These identities hold before any saddle-point approximation. Each fixed-sector amplitude still integrates over off-shell subregion configurations; we next characterize its constrained stationary configurations.

\subsection{Constrained Instantons and Interface Jumps}
\label{subsec:ci_interface_variation}

Fix $\mathcal C_{\mathcal T}$ and consider a semiclassical saddle of that sector. Each subregion must satisfy the full canonical field equations, its physical boundary conditions, and interior regularity, including smooth closure of the Euclidean circle at any nonextremal black hole bolt. Throughout the variation, the interface embedding, foliation, common time normalization, and angular frame are fixed. Variations are taken before restricting to a stationary black hole family.

Represent the constraint delta functional by an integral over the multiplier contour,
\begin{equation}
    \delta[\mathcal C-\mathcal C_{\mathcal T}]
    =\int_{\Gamma_\eta}\mathscr D\eta\,
    e^{-\langle\eta,\mathcal C-\mathcal C_{\mathcal T}\rangle_{\mathcal T_r}}.
    \label{eq:constraint_multiplier_representation}
\end{equation}
The target values $\mathcal C_{\mathcal T}$ remain fixed, while the common integration data $\mathcal C$ and the multipliers are varied independently. Multiplier variation enforces the constraint, with pairing
\begin{equation}
\begin{aligned}
    \langle\eta,\mathcal C-\mathcal C_{\mathcal T}\rangle_{\mathcal T_r}
    =\int_{\mathcal T_r}\dd\tau\,\dd^dx\,\bigl[
    &\eta_{\mathcal E}(\mathcal E-\overline{\mathcal E})
    +\eta_{\mathcal J}^{A}(\mathcal J_A-\overline{\mathcal J}_A)
    \\
    &+\eta_\sigma^{AB}(\sigma_{AB}-\overline\sigma_{AB})\bigr].
    \label{eq:ci_multiplier_pairing}
\end{aligned}
\end{equation}
The energy and momentum labels are already densities, whereas $\eta_\sigma^{AB}$ carries the density weight required to pair with $\sigma_{AB}$.

Combining this multiplier representation with the boundary transforms in Eqs.~\eqref{eq:pn_boundary_transform_1} and \eqref{eq:pn_boundary_transform_2} gives the augmented action
\begin{equation}
\begin{aligned}
    \widetilde I_\eta
    ={}&I_{\rm D}[\mathcal M_1]+I_{\rm D}[\mathcal M_2]
    -\mathcal P_1[\mathcal E,\mathcal J_A]
    +\mathcal P_2[\mathcal E,\mathcal J_A]
    \\
    &+\langle\eta,\mathcal C-\mathcal C_{\mathcal T}\rangle_{\mathcal T_r}.
    \label{eq:ci_extended_augmented_action}
\end{aligned}
\end{equation}
Here $\mathcal E$ and $\mathcal J_A$ remain independent transform labels. Let $\mathcal E_{(s)}$, $\mathcal J_{(s)A}$, and $s_{(s)}^{AB}$ denote the actual Brown--York responses of each subregion, all expressed using the common-normal convention. Once the bulk field equations hold, independent source variations give
\begin{equation}
\begin{aligned}
    \left.\delta\widetilde I_\eta\right|_{\delta N_s,\delta\mathcal N_s}
    =\int_{\mathcal T_r}\dd\tau\,\dd^dx\,\bigl[
    &(\mathcal E_{(1)}-\mathcal E)\delta N_1
     -(\mathcal J_{(1)A}-\mathcal J_A)\delta\mathcal N_1^A
    \\
    &-(\mathcal E_{(2)}-\mathcal E)\delta N_2
     +(\mathcal J_{(2)A}-\mathcal J_A)\delta\mathcal N_2^A\bigr].
    \label{eq:ci_source_variation_matching}
\end{aligned}
\end{equation}
The source saddle equations are therefore
\begin{equation}
    \mathcal E_{(1)}=\mathcal E_{(2)}=\mathcal E,
    \qquad
    \mathcal J_{(1)A}=\mathcal J_{(2)A}=\mathcal J_A.
    \label{eq:ci_response_label_matching}
\end{equation}
They identify the actual responses with the common labels while leaving the two lapse and tangential shift values independent. Under this identification, the kernel terms in Eq.~\eqref{eq:ci_extended_augmented_action} reproduce the classical pN Legendre terms. The augmented action then takes the geometric pN form
\begin{equation}
    I_\eta
    =I_{\rm pN}[\mathcal M_1]+I_{\rm pN}[\mathcal M_2]
     +\langle\eta,\mathcal C-\mathcal C_{\mathcal T}\rangle_{\mathcal T_r}.
    \label{eq:ci_augmented_action}
\end{equation}

After taking the independent variations, we impose the bulk field equations and the source saddle equations~\eqref{eq:ci_response_label_matching}. The remaining variation of $\widetilde I_\eta$ is given below. Its restriction to variations that preserve the identification of the response labels with the actual responses, with $\delta \mathcal E_{(s)}=\delta\mathcal E$ and $\delta \mathcal J_{(s)A}=\delta\mathcal J_A$, agrees with the variation of the classical pN form obtained by adding the two interface contributions in Eq.~\eqref{eq:partial_neumann_boundary_variation}, with $[X]=X_2-X_1$ and the opposite outward-normal orientations:
\begin{equation}
\begin{aligned}
    \left.\delta\widetilde I_\eta\right|_{\rm bulk,\;source\;EOM}
    =\int_{\mathcal T_r}\dd\tau\,\dd^dx\,\Bigl\{
    &([N]+\eta_{\mathcal E})\delta\mathcal E
     +(-[\mathcal N^A]+\eta_{\mathcal J}^{A})\delta\mathcal J_A
    \\
    &+\left(\eta_\sigma^{AB}
        -\frac{\sqrt\sigma}{2}[Ns^{AB}]\right)\delta\sigma_{AB}\Bigr\}
    +\langle\delta\eta,\mathcal C-\mathcal C_{\mathcal T}\rangle_{\mathcal T_r}.
    \label{eq:ci_interface_variation}
\end{aligned}
\end{equation}
Multiplier variations impose $\mathcal C=\mathcal C_{\mathcal T}$, while variations of the independent common data fix
\begin{equation}
    \eta_{\mathcal E}=-[N],\qquad
    \eta_{\mathcal J}^{A}=[\mathcal N^A],\qquad
    \eta_\sigma^{AB}=\frac{\sqrt\sigma}{2}[Ns^{AB}].
    \label{eq:ci_multiplier_jump_relations}
\end{equation}
Appendix~\ref{app:ci_independent_variation} gives the full variation with independent response labels.

A pair of subregion solutions with the same prescribed pN data, the required physical boundary conditions, and regular interiors is stationary under variations preserving $\mathcal C=\mathcal C_{\mathcal T}$. Choosing the multipliers according to Eq.~\eqref{eq:ci_multiplier_jump_relations} makes the augmented action stationary as well. Conversely, a stationary point of the augmented problem obeys these subregion conditions. We call such a pair a \emph{constrained instanton}~\cite{Affleck:1980mp,Cotler:2020lxj}. This equivalence assumes a regular constraint surface and independent admissible data after accounting for gauge redundancies and boundary identities.

Equation~\eqref{eq:ci_multiplier_jump_relations} shows that the constraint variation balances the jumps conjugate to the fixed pN data. Nonzero lapse and tangential shift jumps are supported by the corresponding multipliers. Their thermal and rotational interpretations are developed in Sec.~\ref{sec:free_energy_landscape}.

\subsection{Semiclassical Weights and Extremization over Interface Data}
\label{subsec:ci_action_landscape}

At a constrained saddle, denoted by $\star$, the multiplier term vanishes because $\mathcal C=\mathcal C_{\mathcal T}$, without requiring $\eta_\star=0$. Using the source saddle equations \eqref{eq:ci_response_label_matching}, the stationary action at fixed $\mathcal C_{\mathcal T}$ is
\begin{equation}
    I_{\rm constr}[\mathcal C_{\mathcal T}]
    = \widetilde I_{\eta,\star}
    =I_{\eta,\star}
    =I_{{\rm pN},\star}[\mathcal M_1]
     +I_{{\rm pN},\star}[\mathcal M_2].
    \label{eq:ci_stationary_action}
\end{equation}
By Eq.~\eqref{eq:pn_original_action_additivity}, this is also the original uncut action evaluated on the constrained configuration with the pN sewing prescription.

The fixed-sector amplitude has the semiclassical expansion
\begin{equation}
    \mathcal Z_{\rm constr}[\mathcal C_{\mathcal T}]
    \simeq\sum_{b}
    \mathcal A_b[\mathcal C_{\mathcal T}]\,
    e^{-I_{{\rm constr},b}[\mathcal C_{\mathcal T}]}.
    \label{eq:ci_semiclassical_weight}
\end{equation}
Here $b$ labels saddle pairs compatible with the chosen contour, and $\mathcal A_b$ includes the fluctuation, measure, and contour factors within each fixed sector.

To compare different sectors, vary $\mathcal C_{\mathcal T}$ along a smooth branch of contributing saddles, suppressing $b$ and holding the physical boundary conditions and interface identification fixed. Stationarity with respect to the fields and multipliers leaves only the explicit dependence on the target values, giving the envelope relation
\begin{equation}
    \frac{\delta I_{\rm constr}}{\delta\mathcal C_{\mathcal T}}
    =-\eta_\star.
    \label{eq:ci_envelope_relation}
\end{equation}
Together with Eq.~\eqref{eq:ci_multiplier_jump_relations}, this yields
\begin{equation}
    \delta I_{\rm constr}
    =\int_{\mathcal T_r}\dd\tau\,\dd^dx\,
    \left(
    [N]\,\delta\overline{\mathcal E}
    -[\mathcal N^A]\,\delta\overline{\mathcal J}_A
    -\frac{\sqrt{\overline\sigma}}{2}[Ns^{AB}]
       \delta\overline\sigma_{AB}\right).
    \label{eq:ci_action_gradient_differential}
\end{equation}
All jumps are evaluated on this branch; the derivative with respect to $\overline\sigma_{AB}$ holds the densitized energy and momentum data fixed.

At leading classical order, an interior saddle of the outer integral in Eq.~\eqref{eq:constraint_sector_decomposition} requires stationarity with respect to all freely varying independent local data. By Eq.~\eqref{eq:ci_envelope_relation}, this sets the corresponding multipliers $\eta_\star$ to zero, giving
\begin{equation}
    \frac{\delta I_{\rm constr}}{\delta\mathcal C_{\mathcal T}}=0
    \quad\Longrightarrow\quad
    [N]=0,\qquad[\mathcal N^A]=0,\qquad[Ns^{AB}]=0.
    \label{eq:full_sewing_stationarity}
\end{equation}
Together with the common $\mathcal E$, $\mathcal J_A$, and $\sigma_{AB}$, these conditions match the induced metric and its full Brown--York response at the nondegenerate auxiliary interface.

If the interface data are parametrized by a finite-dimensional family $\mathcal C_{\mathcal T}(\lambda^m)$, stationarity of the constrained on-shell action along this family requires
\begin{equation}
    \frac{\partial I_{\rm constr}}{\partial\lambda^m}
    =-\left\langle\eta_\star,
       \frac{\partial\mathcal C_{\mathcal T}}{\partial\lambda^m}
      \right\rangle_{\mathcal T_r}=0.
    \label{eq:ci_reduced_stationarity}
\end{equation}
Vanishing multipliers, $\eta_\star=0$, therefore imply stationarity of the constrained on-shell action with respect to $\lambda^m$. Conversely, stationarity enforces $\eta_\star=0$, and hence the matching conditions in Eq.~\eqref{eq:full_sewing_stationarity}, provided that the parameter-induced variations of the interface data are sufficient to detect every nonzero mismatch allowed within the chosen family. The next section specifies the thermodynamic boundary data and evaluates the constrained action for stationary black holes to construct the generalized free energy landscape.

\section{Generalized Free Energy from Constrained Instantons}
\label{sec:free_energy_landscape}

We now apply the construction to stationary axisymmetric black holes in Einstein gravity. We first consider a physical asymptotic boundary $\mathcal B$ whose Dirichlet data specify a grand canonical ensemble. In a fixed asymptotic time normalization and angular frame, its thermal and rotational sources are $(\beta=T^{-1},\Omega_{\mathcal B}^I)$, where the real angular velocities $\Omega_{\mathcal B}^I$ correspond to independent rotations. The remaining boundary geometry and external parameters are held fixed. We display these source arguments explicitly in $\mathcal Z(\beta,\Omega_{\mathcal B}^I)$ and $\mathcal Z_{\rm constr}[\mathcal C_{\mathcal T}; \beta,\Omega_{\mathcal B}^I]$.

\begin{figure}[t]
    \centering
    \includegraphics[width=0.6\textwidth]{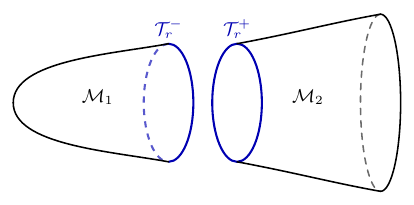}
    \caption{A constrained black hole sector obtained by cutting along the auxiliary surface $\mathcal T_r$. Its two copies bound the inner region $\mathcal M_1$ and the outer region $\mathcal M_2$. The pN sewing conditions identify $(\mathcal E,\mathcal J_A;\sigma_{AB})$, equivalently $(\varepsilon,j_A,\sigma_{AB})$, with responses expressed in the common-normal convention. The lapse and tangential shift on $\mathcal T_r^-$ and $\mathcal T_r^+$ need not agree. The outer region extends from $\mathcal T_r^+$ to the physical asymptotic boundary $\mathcal B$.}
    \label{fig:constrained_black_hole_sector}
\end{figure}

As illustrated in Fig.~\ref{fig:constrained_black_hole_sector}, the auxiliary surface $\mathcal T_r$ separates the inner region $\mathcal M_1$, containing a regular horizon cap, from the outer region $\mathcal M_2$, which extends to $\mathcal B$. The common pN data select a constrained sector, whereas the data at $\mathcal B$ specify the physical ensemble. The auxiliary cut introduces no additional physical boundary.

For a given nonextremal black hole solution, cap regularity determines the Hawking period $\beta_H=T_H^{-1}$ and the angular identification associated with the horizon angular velocity $\Omega_H^I$, defined in the same time normalization and angular frame. The prescribed sources at $\mathcal B$ need not equal these values. A conical construction represents the thermal mismatch by a singularity at the bolt~\cite{Li:2022oup,Wu:2025jes}. Here, independent lapse and shift sources on the two interface copies allow the cap to remain regular under the common coordinate identification. We construct the two solutions, verify their pN matching conditions, and evaluate Eq.~\eqref{eq:ci_stationary_action} to obtain the constrained free energy landscape.

\subsection{Smooth Subregion Saddles and Interface Matching}
\label{subsec:landscape_saddles}

The two subregion solutions must share the prescribed local data. In this section, bars on the prescribed interface values are suppressed, so that
\begin{equation}
    \mathcal C_{\mathcal T}
    =
    \left(
    \mathcal E,\,\mathcal J_A\,;\,\sigma_{AB}
    \right)_{\mathcal T_r}.
    \label{eq:constraint_data_rotating}
\end{equation}

Choose a nonextremal stationary axisymmetric black hole solution with fixed solution parameters and reference time and angular frame. The exterior region is a restriction of this solution, with Euclidean time $\tau$ and period $\beta$ prescribed by the data at $\mathcal B$. We use angular coordinates with the helical identification in Eq.~\eqref{eq:general_euclidean_identifications}. Unless stated otherwise, $N$, $\mathcal N^A$, $j_A$, and $\mathcal J_A$ denote Euclidean fields and responses, which may be complex on rotating saddles. In these coordinates, the exterior metric is
\begin{align}
    \dd s^2
    &=
    N^2\dd\tau^2
    +
    P^2\dd\rho^2
    +
    \sigma_{AB}
    \left(\dd x^A+\mathcal N^A\dd\tau+\mathcal P^A\dd\rho\right)
    \notag\\
    &\qquad\qquad\times
    \left(\dd x^B+\mathcal N^B\dd\tau+\mathcal P^B\dd\rho\right).
    \label{eq:outer_region_metric}
\end{align}

The inner region is obtained by pulling back the regular cap of this same solution, whose regularity data are $(\beta_H,\Omega_H^I)$, under the coordinate transformation below. We keep the radial coordinate and interface embedding fixed and use local angular coordinates adapted to the mutually commuting axial Killing fields $\varphi_I^A$; the remaining cross-section coordinates are unaffected by the twist. The cap coordinates $(\tau_H,x_H^A)$ are related to the exterior coordinates by a constant rescaling and a rigid axial twist:
\begin{equation}
    \alpha=\frac{\beta_H}{\beta}=\frac{T}{T_H},
    \qquad
    \tau_H=\alpha\tau,
    \qquad
    x_H^A=x^A+\nu^A\tau .
    \label{eq:cap_coordinate_map}
\end{equation}
Here $\nu^A=\nu^I\varphi_I^A$, with constant coefficients $\nu^I$. With $t=-i\tau$, the exterior and cap identifications are, respectively,
\begin{equation}
    (\tau,x^A)
    \sim
    \left(\tau+\beta,
    x^A-i\beta\Omega_{\mathcal B}^I\varphi_I^A\right),
    \qquad
    (\tau_H,x_H^A)
    \sim
    \left(\tau_H+\beta_H,
    x_H^A-i\beta_H\Omega_H^I\varphi_I^A\right).
    \label{eq:general_euclidean_identifications}
\end{equation}
In the fixed time normalization and angular frame, compatibility with Eq.~\eqref{eq:cap_coordinate_map} requires
\begin{equation}
    \alpha\beta=\beta_H,
    \qquad
    \nu^I=i\left(\Omega_{\mathcal B}^I-\alpha\Omega_H^I\right).
    \label{eq:thermal_rotational_source_map}
\end{equation}
Given the prescribed boundary sources $(\beta,\Omega_{\mathcal B}^I)$ and the regularity data $(\beta_H,\Omega_H^I)$ of the chosen nonextremal solution, these relations determine $(\alpha,\nu^I)$. For real angular velocities, $\nu^I$ is purely imaginary. When the two sets of sources agree, $\alpha=1$ and $\nu^I=0$.

Pulling back the regular cap to the common coordinates gives
\begin{align}
    \dd s^2
    &=
    \alpha^2N^2\dd\tau^2
    +
    P^2\dd\rho^2
    +
    \sigma_{AB}
    \left[\dd x^A+\left(\alpha\mathcal N^A+\nu^A\right)\dd\tau+\mathcal P^A\dd\rho\right]
    \notag\\
    &\qquad\qquad\times
    \left[\dd x^B+\left(\alpha\mathcal N^B+\nu^B\right)\dd\tau+\mathcal P^B\dd\rho\right].
    \label{eq:inner_cap_metric}
\end{align}
As restrictions or local pullbacks of the same solution, both metrics satisfy the bulk field equations. Since the transformation acts along Killing directions and leaves the radial coordinate unchanged, they share the same spatial ADM geometry, interface embedding, and radial normal in a neighborhood of the interface. Their lapse and tangential shift values satisfy
\begin{equation}
    N_{2}=N,
    \qquad
    N_{1}=\alpha N,
    \qquad
    \mathcal N_{2}^A=\mathcal N^A,
    \qquad
    \mathcal N_{1}^A=\alpha\mathcal N^A+\nu^A,
    \label{eq:lapse_shift_jump}
\end{equation}
Together, these jumps encode the thermal and rotational mismatches. We now verify that the lapse and shift relations in Eq.~\eqref{eq:lapse_shift_jump} preserve the common pN interface data.

With $[X]=X_2-X_1$ and the common-normal convention specified by $n$, Eq.~\eqref{eq:by_energy_momentum_density} gives
\begin{equation}
    [\varepsilon]
    =
    -\frac{[k]}{8\pi},
    \qquad
    [j_A]
    =
    \frac{1}{8\pi}
    \sigma_A{}^a n_b[\pi^b{}_a] .
    \label{eq:general_by_response_jumps}
\end{equation}
Because the two spatial ADM geometries coincide in a neighborhood of the interface and have the same embedding and radial normal, both $\sigma_{AB}$ and its normal derivative agree. Hence $[\sigma_{AB}]=[k]=0$; equality of the metric values at the interface alone would not imply the second condition. Equation~\eqref{eq:by_energy_momentum_density} then gives
\begin{equation}
    [\varepsilon]=0.
    \label{eq:spatial_energy_response_matching}
\end{equation}
The momentum response also depends on the extrinsic curvature $K_{ab}$ of the constant-$\tau$ hypersurfaces. We extend the rigid axial twist into a neighborhood of the interface as $\nu^a=\nu^I\varphi_I^a$, where the $\varphi_I^a$ are Killing fields of the full common spatial metric. Substituting Eq.~\eqref{eq:lapse_shift_jump} into Eqs.~\eqref{eq:time_extrinsic_curvature} and~\eqref{eq:adm_momentum}, with constant $\alpha$, gives the following jump. The trace term in the momentum drops out under the normal--tangential projection because
$\sigma_A{}^a n^b h_{ab}=0$:
\begin{equation}
    [j_A]
    =
    \frac{1}{16\pi\alpha N}
    \sigma_A{}^a n^b
    \left(
    \alpha\dot h^{(2)}_{ab}
    -\dot h^{(1)}_{ab}
    +(\mathcal L_\nu h)_{ab}
    \right).
    \label{eq:general_momentum_response_jump}
\end{equation}
The time derivatives are evaluated in the common interface coordinates. Stationarity gives $\dot h^{(1)}_{ab}=\dot h^{(2)}_{ab}=0$, while the Killing twist satisfies $\mathcal L_\nu h_{ab}=0$. Equation~\eqref{eq:general_momentum_response_jump} therefore gives $[j_A]=0$. Together with the matching of the spatial geometry and energy response, this yields
\begin{equation}
    [\sigma_{AB}]
    =
    [\varepsilon]
    =
    [j_A]
    =0.
    \label{eq:stationary_local_response_matching}
\end{equation}
Since $[\sigma_{AB}]=0$, the response conditions $[\varepsilon]=[j_A]=0$ in Eq.~\eqref{eq:stationary_local_response_matching} are equivalent to $[\mathcal E]=[\mathcal J_A]=0$, as required by Eq.~\eqref{eq:partial_neumann_sewing_conditions}. Thus the metric~\eqref{eq:inner_cap_metric} on $\mathcal M_1$ and the metric~\eqref{eq:outer_region_metric} on $\mathcal M_2$ satisfy the local pN sewing conditions at $\mathcal T_r$, while their lapse and tangential shift may jump across the interface.

\subsection{Evaluation of the Constrained Black Hole Action}
\label{subsec:landscape_weight}

We now evaluate the partial-Neumann actions of the inner metric~\eqref{eq:inner_cap_metric} and the exterior metric~\eqref{eq:outer_region_metric} constructed in Sec.~\ref{subsec:landscape_saddles}. The matching, regularity, and physical boundary conditions established there make this pair a constrained instanton, with multipliers given by Eq.~\eqref{eq:ci_multiplier_jump_relations}.

For the thermodynamic evaluation, we make explicit the added subtraction or counterterm contribution $I_{\mathcal B}^{\rm ren}$ already included in the prescribed physical boundary action. It is evaluated with the same radial cutoff as the unrenormalized action, and only their sum is required to have a finite asymptotic limit. These terms occur only at the asymptotic boundary $\mathcal B$ and, with its boundary data fixed, do not modify the pN sewing conditions or the multiplier equations at the auxiliary interface.

Write $I_{s,{\rm os}}= I_{{\rm pN},\star}[\mathcal M_s]$ for each subregion action and $I_{\rm bulk}^{(s)}$ for its canonical bulk term in Eq.~\eqref{eq:adm_action}, with the interface Legendre term defined in Eq.~\eqref{eq:partial_neumann_legendre_term}.

To evaluate these actions, we express the boundary terms in the common periodic angular coordinates and treat the bolt contribution using the small-tube prescription~\eqref{eq:adm_bolt_compensation}. The normal on $\mathcal H_\epsilon$ points outward from the punctured cap toward the bolt. The resulting subregion actions are
\begin{align}
    I_{1,{\rm os}}
    &=
    I_{\rm bulk}^{(1)}
    +(B_N+B_{\mathcal N})|_{\mathcal T_r^-}
    +B_{\rm Leg}^{\mathcal T_r^-}
    \notag\\
    &\quad+\lim_{\epsilon\to0}\Bigg[
    (B_N+B_{\mathcal N})|_{\mathcal H_\epsilon}
    +\frac{1}{8\pi}
    \int_{\mathcal H_\epsilon}\dd^{d+1}x\,
    \sqrt{\gamma}\,\mathcal K\Bigg]
    =
    -S_{\rm BH},
    \notag\\
    I_{2,{\rm os}}
    &=
    I_{\rm bulk}^{(2)}
    +(B_N+B_{\mathcal N})|_{\mathcal T_r^+}
    +B_{\rm Leg}^{\mathcal T_r^+}
    +(B_N+B_{\mathcal N})|_{\mathcal B}
    +I_{\mathcal B}^{\rm ren}
    =
    \beta H_{\mathcal B}.
    \label{eq:boundary_contribution_bookkeeping}
\end{align}
The Legendre term cancels the Dirichlet pairing on each interface copy, while stationarity and the bulk constraints make the canonical bulk terms vanish. The positive sign preceding the GHY integral compensates for the artificial boundary term introduced by excising the tube. For the inner metric~\eqref{eq:inner_cap_metric}, the coordinate map and identifications fixed in Sec.~\ref{subsec:landscape_saddles} preserve the regularity of the cap. Using angular coordinates $\phi^I$ adapted to the axial Killing fields $\varphi_I$, the cap's regular corotating angles coincide with the common periodic angles $\widetilde\phi^I=\phi^I+i\Omega_{\mathcal B}^I\tau$. In these coordinates, $(B_N+B_{\mathcal N})|_{\mathcal H_\epsilon}$ vanishes as the tube shrinks, and the compensating term gives $I_{1,{\rm os}}=-A_H/4=-S_{\rm BH}$~\cite{Gibbons:1976ue,Banados:1993qp,Draper:2022xzl}. The regularity relation $\alpha\beta=\beta_H$ in Eq.~\eqref{eq:thermal_rotational_source_map} ensures that the time rescaling introduces no additional factor.

For the stationary exterior metric~\eqref{eq:outer_region_metric}, the remaining physical boundary term integrates over the prescribed period $\beta$ to give $I_{2,{\rm os}}=\beta H_{\mathcal B}$. Here $H_{\mathcal B}$ is the renormalized boundary Hamiltonian generating the thermal evolution specified by the boundary identification in Sec.~\ref{subsec:landscape_saddles}. In the reference time normalization and angular frame fixed there,
\begin{equation}
    H_{\mathcal B}
    =
    E_{\mathcal B}-\Omega_{\mathcal B}^I J_{\mathcal B,I}.
    \label{eq:asymptotic_grand_canonical_generator}
\end{equation}
Here $E_{\mathcal B}$ is the Hamiltonian charge associated with the reference time translation, and $J_{\mathcal B,I}$ are the real Lorentzian angular momenta. These charges incorporate the same subtraction or counterterm prescription as the physical boundary action.\footnote{\label{fn:thermal_charge_conventions} With $t=-i\tau$, the real axial charge on a boundary cross-section $\mathcal S$ is obtained from the Euclidean momentum response as $J_I=-i\int_{\mathcal S}\dd^dx\,\sqrt{\sigma}\,j_A\varphi_I^A$. In the periodic angular coordinates, the shift becomes $\widetilde{\mathcal N}^A=\mathcal N^A-i\Omega_{\mathcal B}^I\varphi_I^A$; these relations give the term $-\Omega_{\mathcal B}^I J_{\mathcal B,I}$ in the ADM boundary pairing. The asymptotic charges are evaluated using renormalized responses at a finite radial cutoff before taking the limit, with the lapse weighting retained in the Hamiltonian energy, in particular for asymptotically AdS boundaries~\cite{Brown:1992br,Balasubramanian:1999re,Hohm:2010jc}.}

Adding the cap and exterior actions in Eq.~\eqref{eq:boundary_contribution_bookkeeping}, as prescribed by Eq.~\eqref{eq:ci_stationary_action}, gives the constrained black hole action:
\begin{equation}
    I_{\rm constr}=I_{\eta,\star}
    =\beta\bigl(E_{\mathcal B}-\Omega_{\mathcal B}^I J_{\mathcal B,I}\bigr)-S_{\rm BH}.
    \label{eq:black_hole_constrained_action}
\end{equation}
The corresponding contribution at leading exponential order is
\begin{equation}
    \mathcal Z_{\rm constr}[\mathcal C_{\mathcal T};\beta,\Omega_{\mathcal B}^I]
    \simeq
    \exp\left[
    -\beta
    \left(
    E_{\mathcal B}
    -\Omega_{\mathcal B}^I J_{\mathcal B,I}
    \right)
    +S_{\rm BH}
    \right].
    \label{eq:constrained_saddle_approx}
\end{equation}
Here $\simeq$ denotes the contribution of the chosen saddle, with fluctuation factors and other saddles suppressed. The entropy factor thus comes from the regular horizon cap. For a compact stationary geometry with two regular caps and no physical asymptotic boundary, Eq.~\eqref{eq:ci_stationary_action} gives $I_{\rm constr}=-(S_b+S_c)$, as discussed in Sec.~\ref{subsec:desitter_black_holes}. Section~\ref{sec:modified_gravity} extends the constrained variation to the full auxiliary field data of the local higher-curvature theories considered here, for which the regular cap contributes the Wald entropy.

\subsection{Reduced Landscape and Equilibrium Conditions}
\label{subsec:landscape_reduction}

Consider a retained branch of the nonextremal stationary axisymmetric saddle pairs constructed in Sec.~\ref{subsec:landscape_saddles}. Along this family, the physical Dirichlet data, interface embedding, and foliation are held fixed. We retain the charge definitions and normalization conventions of Sec.~\ref{subsec:landscape_weight}. Choose collective coordinates $\lambda^m$, $m=1,\ldots,N_{\rm coll}$, whose map to the complete local data $\mathcal C_{\mathcal T}(\lambda)$ is injective on this branch. Horizon and rotation parameters, or physical charges, can provide such coordinates. Each point labels a sector with its own fixed local data. The constrained action~\eqref{eq:black_hole_constrained_action} then defines the leading classical generalized free energy:
\begin{equation}
\begin{aligned}
    \mathcal F_{\rm gen}(\lambda^m;\,T,\Omega_{\mathcal B}^I)
    &=\frac{I_{\rm constr}[\mathcal C_{\mathcal T}(\lambda)]}{\beta}
    \\
    &=E_{\mathcal B}(\lambda^m)
    -\Omega_{\mathcal B}^I J_{\mathcal B,I}(\lambda^m)
    -T S_{\rm BH}(\lambda^m),
    \qquad T=\beta^{-1}.
\end{aligned}
    \label{eq:generalized_free_energy_rotating}
\end{equation}
The prescribed sources $(T,\Omega_{\mathcal B}^I)$ can differ from the horizon regularity values $(T_H,\Omega_H^I)$ of each member. Summing the corresponding sector weights along this branch gives the reduced integral
\begin{equation}
    \mathcal Z_{\rm red}(\beta,\Omega_{\mathcal B}^I)
    \simeq
    \int \dd^{N_{\rm coll}}\lambda\,\mu(\lambda)\,
    \exp\left[-\beta\,
    \mathcal F_{\rm gen}(\lambda^m;\,T = 1/\beta,\Omega_{\mathcal B}^I)
    \right],
    \label{eq:landscape_partition_lambda}
\end{equation}
Here $\mu(\lambda)$ includes the induced measure and semiclassical prefactors. Contributions from other admissible branches are summed separately. The full interface integral remains Eq.~\eqref{eq:constraint_sector_decomposition}. The landscape $\mathcal F_{\rm gen}(\lambda)$ determines the leading weight before this reduced integration. Note that the normalized ensemble average of $\mathcal F_{\rm gen}$ is distinct from the thermodynamic potential $-T\ln\mathcal Z_{\rm red}$~\cite{Liu:2025iei,Anand:2026oam}.

At leading classical order, candidate equilibrium configurations in the interior of the retained family satisfy the stationarity condition
\begin{equation}
    \left.
    \frac{\partial \mathcal F_{\rm gen}}{\partial\lambda^m}
    \right|_{T,\Omega_{\mathcal B}^I}
    =
    0.
    \label{eq:landscape_stationarity}
\end{equation}
The boundary temperature, angular velocities, asymptotic spatial geometry, and remaining external parameters are held fixed in this variation. As in Eq.~\eqref{eq:ci_reduced_stationarity}, variation along this family includes the induced changes of all local sewing data, including $\sigma_{AB}$.

Along the underlying family of stationary black hole solutions, variations of the solution parameters satisfy the first law
\begin{equation}
    \dd E_{\mathcal B}
    =
    T_H\,\dd S_{\rm BH}
    +\Omega_H^I\,\dd J_{\mathcal B,I}.
    \label{eq:first_law_rotating}
\end{equation}
Combining this first law with Eq.~\eqref{eq:generalized_free_energy_rotating} gives
\begin{equation}
    \left.\dd\mathcal F_{\rm gen}\right|_{T,\Omega_{\mathcal B}^I}
    =
    (T_H-T)\,\dd S_{\rm BH}
    +
    (\Omega_H^I-\Omega_{\mathcal B}^I)\,\dd J_{\mathcal B,I} .
    \label{eq:landscape_variation}
\end{equation}
Suppose that the Jacobian of $\lambda^m\mapsto(S_{\rm BH},J_{\mathcal B,I})$ has full row rank at the point under consideration, so that $\dd S_{\rm BH}$ and $\dd J_{\mathcal B,I}$ are linearly independent on the retained family. Equation~\eqref{eq:landscape_stationarity} is then locally equivalent to
\begin{equation}
    T=T_H,
    \qquad
    \Omega_{\mathcal B}^I=\Omega_H^I.
    \label{eq:equilibrium_conditions_rotating}
\end{equation}
If the rank condition is not satisfied, stationarity constrains only the projections of these mismatches onto the available variations. Within the paired family constructed in Sec.~\ref{subsec:landscape_saddles}, the equalities~\eqref{eq:equilibrium_conditions_rotating} imply $\alpha=1$ and $\nu^I=0$ by Eq.~\eqref{eq:thermal_rotational_source_map}, so the cap and exterior join into the same smooth black hole solution. At these points, the generalized free energy gives the classical grand potential of the corresponding black hole saddle:
\begin{equation}
    \mathcal F_{\rm eq}
    =
    E_{\mathcal B}
    -T_H S_{\rm BH}
    -\Omega_H^I J_{\mathcal B,I} .
    \label{eq:on_shell_grand_free_energy}
\end{equation}

\section{Black Hole Applications in Einstein Gravity}   \label{sec:bh_solutions}

We now apply the construction of Sec.~\ref{sec:free_energy_landscape} to Schwarzschild, Kerr, and Schwarzschild--de Sitter solutions, and discuss the proposed Kerr--de Sitter extension. The asymptotically flat examples pair a regular horizon cap with an exterior region, whose combined action is given by Eq.~\eqref{eq:black_hole_constrained_action}. Compact de Sitter configurations instead contain two regular horizon caps and no physical asymptotic boundary, so their constrained action is the negative sum of the black hole and cosmological horizon entropies, $I_{\rm constr}=-(S_b+S_c)$.

\subsection{Schwarzschild Black Hole}
    \label{subsec:schwarzschild_solution}
The Schwarzschild family provides the simplest spherically symmetric realization of the factorized construction. For a solution with horizon radius $r_h$, we place the spherical auxiliary interface $\mathcal T_r$ at $r=r_T>r_h$. This divides the Euclidean geometry into an inner region $\mathcal M_1$, covering $r_h\leq r\leq r_T$, and an exterior region $\mathcal M_2$, covering $r_T\leq r<\infty$ and extending to the physical asymptotic boundary $\mathcal B$. Specializing Eqs.~\eqref{eq:outer_region_metric} and~\eqref{eq:inner_cap_metric} to the static, nonrotating case gives the following metrics in common Euclidean coordinates $(\tau,r,\theta,\phi)$:
    \begin{equation}
    \begin{aligned}
        \left.\dd s^2\right|_{\mathcal M_2}
        &=
        f(r)\dd\tau^2
        +
        \frac{\dd r^2}{f(r)}
        +
        r^2\dd\Omega_2^2,
        \\
        \left.\dd s^2\right|_{\mathcal M_1}
        &=
        \alpha^2 f(r)\dd\tau^2
        +
        \frac{\dd r^2}{f(r)}
        +
        r^2\dd\Omega_2^2.
    \end{aligned}
        \label{eq:schwarzschild_metric}
    \end{equation}
Both metrics are built from the same Schwarzschild function,
    \begin{equation}
        f(r)=1-\frac{r_h}{r},
        \qquad
        M=\frac{r_h}{2},
        \label{eq:schwarzschild_metric_function}
    \end{equation}
where $r_h$ is the horizon radius. We use $\dd\Omega_2^2=\Omega_{AB}\dd x^A\dd x^B=\dd\theta^2+\sin^2\theta\,\dd\phi^2$ and impose $\tau\sim\tau+\beta$, with $\beta=T^{-1}$ fixed at $\mathcal B$. The inner cap is constructed by the relative lapse rescaling
    \begin{equation}
        \alpha=\frac{\beta_H}{\beta}=\frac{T}{T_H}.
        \label{eq:schwarzschild_temperature}
    \end{equation}
Here $\beta_H=T_H^{-1}=4\pi r_h$ is the Euclidean period required for regularity at the Schwarzschild bolt. The cap coordinate $\tau_H=\alpha\tau$ of Eq.~\eqref{eq:cap_coordinate_map} has period $\alpha\beta=\beta_H$, so the inner metric in Eq.~\eqref{eq:schwarzschild_metric} closes smoothly at the bolt. Both regions share the same spatial line element and vanishing tangential shift, with lapse functions $N_2=\sqrt f$ and $N_1=\alpha\sqrt f$. Thus, $T\neq T_H$ is encoded in a lapse jump at $\mathcal T_r$, while the spatial geometry agrees across the interface.

Since the Schwarzschild solution is static and nonrotating, its momentum response vanishes:
    \begin{equation}
        \left.\mathcal J_A\right|_{\mathcal T_r}
        =
        \left.\sqrt\sigma\,j_A\right|_{\mathcal T_r}
        =0.
        \label{eq:schwarzschild_vanishing_momentum}
    \end{equation}
Fixing the interface radius $r_T$ also fixes the round interface metric:
    \begin{equation}
        \left.\sigma_{AB}\dd x^A\dd x^B\right|_{\mathcal T_r^\pm}
        =
        r_T^2\left(\dd\theta^2+\sin^2\theta\,\dd\phi^2\right),
        \qquad
        x^A=(\theta,\phi).
        \label{eq:schwarzschild_fixed_interface_metric}
    \end{equation}
Consequently, $[\sigma_{AB}]=0$ and $\delta\sigma_{AB}=0$, so neither $\mathcal J_A$ nor $\sigma_{AB}$ supplies a collective coordinate within the retained Schwarzschild family. The remaining sector datum is the energy response. Combining Eqs.~\eqref{eq:densitized_by_data} and~\eqref{eq:by_energy_momentum_density} gives,
\begin{equation}
        \left.\mathcal E\right|_{\mathcal T_r}
        =
        -\left.\frac{\sqrt{\sigma}}{8\pi}\,k\right|_{\mathcal T_r},
        \label{eq:schwarzschild_interface_data}
\end{equation}
where $k=\sigma^{AB}k_{AB}$ is the extrinsic curvature trace of $\mathcal S_\tau^{(r)}$ within $\Sigma_\tau$, with $k_{AB}$ defined in Eq.~\eqref{eq:radial_extrinsic_curvature_conventions}. With the auxiliary surface $\mathcal T_r$ held fixed, fixing the energy response is therefore equivalent to fixing $k$.

The traces of the extrinsic curvature, evaluated using the common spatial metric and normal, agree on the two interface copies:
\begin{equation}
    k_{\mathcal T_r^-}=k_{\mathcal T_r^+}=k
    =\frac{2}{r_T}\sqrt{1-\frac{r_h}{r_T}},
    \qquad
    r_h=r_T\left[1-\left(\frac{kr_T}{2}\right)^2\right].
    \label{eq:schwarzschild_interface_k}
\end{equation}
At fixed $r_T$, this gives a one-to-one map between $0<r_h<r_T$ and $0<k<2/r_T$. Within this branch, fixing the local energy response therefore fixes $r_h$, or equivalently $M=r_h/2$. The lapse is not prescribed as part of the interface data; for a given sector and boundary period, its values on the constructed saddle follow from the exterior solution and cap regularity.

The area law $S_{\rm BH}=A_H/4$ and the standard asymptotically flat subtraction for the Hamiltonian energy defined in Sec.~\ref{subsec:landscape_weight} give
    \begin{equation}
        S_{\rm Schw}(r_h)=\pi r_h^2,
        \qquad
        E_{\mathcal B}=M=\frac{r_h}{2}.
        \label{eq:schwarzschild_entropy_energy}
    \end{equation}
Here $E_{\mathcal B}$ is the physical Hamiltonian energy, distinct from the unrenormalized interface response. Applying Eq.~\eqref{eq:black_hole_constrained_action} gives the action and leading weight of the fixed sector,
\begin{equation}
\begin{aligned}
    I_{\rm constr}^{\rm Schw}[k;\beta]
    &=\beta M(k)-S_{\rm Schw}(k),
    \\
    \mathcal Z_{\rm constr}^{\rm Schw}[k;\beta]
    &\simeq e^{-\beta M(k)+S_{\rm Schw}(k)}.
\end{aligned}
    \label{eq:schwarzschild_fixed_sector_weight}
\end{equation}
This evaluation does not impose $T=T_H$. Dividing the action by $\beta$ gives the generalized free energy of Eq.~\eqref{eq:generalized_free_energy_rotating} in the nonrotating sector,
    \begin{equation}
        \mathcal F_{\rm gen}^{\rm Schw}(r_h;T)
        =
        E_{\mathcal B}(r_h)-T S_{\rm Schw}(r_h)
        =
        \frac{r_h}{2}
        -
        \pi T r_h^2 .
        \label{eq:schwarzschild_generalized_free_energy}
    \end{equation}
The constrained action and the associated generalized free energy, expressed in terms of $r_h$, are independent of the auxiliary radius $r_T$. The cut can always be placed outside any finite horizon. At fixed $T$, applying the stationarity condition~\eqref{eq:landscape_stationarity} with respect to $r_h$ gives
    \begin{equation}
        \frac{\partial\mathcal F_{\rm gen}^{\rm Schw}}{\partial r_h}
        =
        \frac{1}{2}
        -
        2\pi T r_h
        =
        \frac{1}{2}
        \left(
        1-\frac{T}{T_H}
        \right)
        =
        0.
        \label{eq:schwarzschild_stationarity}
    \end{equation}
The stationary radius is $r_h^\star=1/(4\pi T)$ and satisfies $T=T_H$. At this point, the generalized free energy gives the classical Helmholtz free energy of the equilibrium Schwarzschild saddle,
    \begin{equation}
        \mathcal F_{\rm eq}^{\rm Schw}
        =
        M-T_H S_{\rm Schw}
        =
        \frac{r_h}{4}
        =
        \frac{M}{2}.
        \label{eq:schwarzschild_on_shell_free_energy}
    \end{equation}
At fixed $T$, $\partial_{r_h}^2\mathcal F_{\rm gen}^{\rm Schw}=-2\pi T<0$, so this stationary point is a local maximum of the reduced landscape. This reproduces the familiar canonical instability associated with the negative heat capacity of an asymptotically flat Schwarzschild black hole. A physical cavity at a finite radius~\cite{York:1986it,Gregory:2001bd} or asymptotically AdS boundary conditions~\cite{Hawking:1982dh,Prestidge:1999uq} modify the infrared thermodynamic problem and can therefore change this stability conclusion.

The integral over admissible fixed interface sectors within this spherical branch can be expressed equivalently in terms of the local datum $k$ or the collective coordinate $r_h$,
\begin{equation}
\begin{aligned}
    \mathcal Z_{\rm red}^{\rm Schw}(\beta)
    &\simeq
    \int_0^{2/r_T}\dd k\,\mu_k(k)\,
    e^{-\beta M(k)+S_{\rm Schw}(k)}
    \\
    &=
    \int_0^{r_T}\dd r_h\,\mu_r(r_h)\,
    e^{-\beta M(r_h)+S_{\rm Schw}(r_h)}.
\end{aligned}
    \label{eq:schwarzschild_integral_over_k}
\end{equation}
The measure $\mu_k$ includes the fluctuation determinants, and $\mu_r$ is induced by the change of variables, including the absolute Jacobian. The classical exponent agrees with the conical evaluation for the same Einstein black hole family~\cite{Li:2022oup}. It has the thermodynamic structure used in reduced black hole ensembles~\cite{Cheng:2024hxh,Cheng:2024efw,Ali:2024adt,Liu:2025iei}, with the present sectors defined by local interface data.

Formally extending the retained Schwarzschild family to arbitrarily large $r_h$, with the auxiliary cut chosen outside the horizon, exposes the unbounded growth of $-\beta\mathcal F_{\rm gen}^{\rm Schw}=\pi r_h^2-\beta r_h/2$ as $r_h\to\infty$. Unless the measure provides a compensating exponential suppression, this yields the usual asymptotically flat canonical divergence. Nevertheless, each fixed sector retains a finite constrained action, represented here by a smooth cap and a lapse jump.

    \subsection{Kerr Black Hole}
    \label{subsec:kerr_solution}
For a Kerr solution with outer horizon radius $r_+$, we use the cap and exterior metrics of Eqs.~\eqref{eq:inner_cap_metric} and~\eqref{eq:outer_region_metric} in common Euclidean Boyer--Lindquist coordinates $(\tau,r,\theta,\phi)$. The auxiliary surface $\mathcal T_r$ at fixed $r=r_T>r_+$ separates the inner cap $\mathcal M_1$ from the exterior $\mathcal M_2$, which extends to the physical asymptotic boundary $\mathcal B$.

We parametrize the Kerr family by $r_+$ and the rotation parameter $a$. The Boyer--Lindquist radial function is
    \begin{equation}
        \Delta(r)=r^2-2Mr+a^2,
        \qquad
        \Delta(r_+)=0.
        \label{eq:kerr_radial_function}
    \end{equation}
Solving $\Delta(r_+)=0$ fixes the mass, with the angular momentum expressed in the same parameters:
    \begin{equation}
        M=\frac{r_+^2+a^2}{2r_+},
        \qquad
        J=\frac{a\left(r_+^2+a^2\right)}{2r_+}.
        \label{eq:kerr_mass_angular_momentum}
    \end{equation}
We restrict attention to the nonextremal branch $r_+>|a|$. With the stationary Killing field normalized at infinity and the angular velocity measured relative to an asymptotically nonrotating frame, the horizon data are
    \begin{equation}
        S_{\rm Kerr}
        =
        \pi\left(r_+^2+a^2\right),
        \qquad
        T_H
        =
        \frac{r_+^2-a^2}
        {4\pi r_+\left(r_+^2+a^2\right)},
        \qquad
        \Omega_H
        =
        \frac{a}{r_+^2+a^2}.
        \label{eq:kerr_thermodynamic_data}
    \end{equation}

Rotation gives a nonzero axial momentum response, and, unlike Schwarzschild, the interface metric $\sigma_{AB}$ at fixed $r_T$ generally varies with $(r_+,a)$. Each sector fixes the complete local pN data
    \begin{equation}
        \mathcal C_{\mathcal T}^{\rm Kerr}
        =
        \left(
        \mathcal E,\,\mathcal J_A\,;\,\sigma_{AB}
        \right)_{\mathcal T_r}.
        \label{eq:kerr_interface_data}
    \end{equation}
These data match by Eq.~\eqref{eq:stationary_local_response_matching}, since the cap and exterior share the same spatial ADM geometry near $\mathcal T_r$ and are related by the rigid transformation~\eqref{eq:cap_coordinate_map}. In the Boyer--Lindquist frame, $\mathcal J_\theta=0$; with $\varphi^A=(\partial_\phi)^A$, the real angular momentum follows from the generally complex Euclidean response as
    \begin{equation}
        J_{\mathcal T}
        =
        -i\int_{\mathcal S_\tau^{(r)}}
        \dd^2x\,\mathcal J_A\varphi^A ,
        \label{eq:kerr_interface_angular_momentum}
    \end{equation}
with the Euclidean charge convention of footnote~\ref{fn:thermal_charge_conventions}. For stationary vacuum Kerr, $J_{\mathcal T}=J_{\mathcal B}=J$~\cite{Brown:1992br,Brown:1992bq}. The full local response $\mathcal J_A$ remains the sewing datum.

With the interface and reference conventions fixed as in Sec.~\ref{subsec:landscape_reduction}, we restrict to a branch on which $(r_+,a)\mapsto\mathcal C_{\mathcal T}^{\rm Kerr}$ is injective. Matching the local data then selects common solution parameters, while the lapse and axial shift obey
    \begin{equation}
    \begin{aligned}
        \left.N_1\right|_{\mathcal T_r^-}
        &=
        \alpha\left.N_2\right|_{\mathcal T_r^+},
        \\
        \left.\mathcal N_1^A\right|_{\mathcal T_r^-}
        &=
        \alpha\left.\mathcal N_2^A\right|_{\mathcal T_r^+}
        +\nu^\phi\varphi^A.
    \end{aligned}
        \label{eq:kerr_interface_sources}
    \end{equation}
For independently prescribed sources $(T,\Omega)$, where $\Omega=\Omega_{\mathcal B}$, the cap regularity conditions~\eqref{eq:thermal_rotational_source_map} give
\begin{equation}
    \nu^\phi=i\left(\Omega-\alpha\Omega_H\right),
    \qquad
    \alpha=\frac{\beta_H}{\beta}=\frac{T}{T_H}.
    \label{eq:kerr_angular_twist}
\end{equation}

With the standard asymptotically flat charge prescription of Sec.~\ref{subsec:landscape_weight}, $E_{\mathcal B}=M$ and $J_{\mathcal B}=J$. Equation~\eqref{eq:generalized_free_energy_rotating} then gives
    \begin{equation}
    \begin{split}
        \mathcal F_{\rm gen}^{\rm Kerr}(r_+,a;T,\Omega)
        &=
        M(r_+,a)
        -
        \Omega J(r_+,a)
        -
        T S_{\rm Kerr}(r_+,a)
        \\
        &=
        \frac{r_+^2+a^2}{2r_+}
        -
        \Omega\,
        \frac{a\left(r_+^2+a^2\right)}{2r_+}
        -
        \pi T\left(r_+^2+a^2\right).
    \end{split}
    \label{eq:kerr_generalized_free_energy}
    \end{equation}
The corresponding leading contribution of this fixed sector is
\begin{equation}
        \mathcal Z_{\rm constr}^{\rm Kerr}[r_+,a;\beta,\Omega]
        \simeq
        \exp\!\left[-\beta {\mathcal F_{\rm gen}^{\rm Kerr}}(r_+,a;T,\Omega)\right].
        \label{eq:kerr_fixed_sector_weight}
\end{equation}
To apply the stationarity criterion of Sec.~\ref{subsec:landscape_reduction} at fixed $(T,\Omega)$, note that
\begin{equation}
    \det\frac{\partial(S_{\rm Kerr},J)}{\partial(r_+,a)}
    =\frac{\pi(r_+^2+a^2)^2}{r_+^2}\neq0.
    \label{eq:kerr_thermodynamic_jacobian}
\end{equation}
Thus an interior stationary point satisfies $T=T_H$ and $\Omega=\Omega_H$. At this smooth Kerr saddle, the generalized free energy reduces to
\begin{equation}
    \mathcal F_{\rm eq}^{\rm Kerr}
    =
    M-\Omega_H J-T_H S_{\rm Kerr}
    =
    \frac{r_+^2+a^2}{4r_+}
    =
    \frac{M}{2}.
    \label{eq:kerr_on_shell_free_energy}
\end{equation}

Equation~\eqref{eq:kerr_generalized_free_energy} provides a formal generalized free energy landscape for the asymptotically flat rotating ensemble. For $\Omega\neq0$, the Lorentzian thermal generator $\chi=\partial_t+\Omega\partial_\phi$ has asymptotic norm $g(\chi,\chi)\sim-1+\Omega^2r^2\sin^2\theta$ and becomes spacelike sufficiently far from the rotation axis. A rigidly rotating thermal bath therefore cannot extend to infinity. This limitation concerns the global thermal interpretation; the fixed-sector constrained action remains well defined.

\subsection{de Sitter Black Holes}
\label{subsec:desitter_black_holes}
For de Sitter black holes, the relevant Euclidean region lies between the black hole and cosmological bolts. The auxiliary interface therefore separates two horizon caps, with no physical asymptotic boundary. We briefly recover the Schwarzschild--de Sitter result of Refs.~\cite{Draper:2022xzl,Draper:2023bhg} within the present partial-Neumann framework, before turning to the rotating case.

\paragraph{Schwarzschild--de Sitter.}
For a generic nondegenerate Schwarzschild--de Sitter solution, the black hole and cosmological horizons have different Hawking temperatures in the same static time normalization, so a single Euclidean period cannot make the standard metric regular at both horizons. Following Ref.~\cite{Draper:2022xzl}, the constrained construction accommodates this mismatch through a relative lapse rescaling, allowing both caps to remain regular while the lapse jumps at the auxiliary interface.

For a nondegenerate solution with $0<M<\ell/(3\sqrt{3})$, let $r_b$ and $r_c$ denote the black hole and cosmological horizon radii. Choose $r_b<r_T<r_c$ and let $\mathcal M_1$ and $\mathcal M_2$ be the corresponding caps. The metric function is
    \begin{equation}
        f(r)=1-\frac{2M}{r}-\frac{r^2}{\ell^2},
        \qquad
        \Lambda=\frac{3}{\ell^2}.
        \label{eq:sds_metric_function}
    \end{equation}
In the static time normalization, define $T_i=|f'(r_i)|/(4\pi)$, $\beta_i=T_i^{-1}$, and $S_i=\pi r_i^2$ for $i\in\{b,c\}$. Taking the cosmological cap as reference, choose the common period $\beta_c$ and set $\tau_b=\alpha\tau$ on the black hole cap, with $\alpha=\beta_b/\beta_c=T_c/T_b<1$. The metrics have the form of Eq.~\eqref{eq:schwarzschild_metric}: both caps are regular, and their different regularity periods produce a lapse jump at $\mathcal T_r$.

The pN data on this static, spherically symmetric family consist of $\mathcal E$, $\mathcal J_A=0$, and $\sigma_{AB}=r_T^2\Omega_{AB}$. At fixed $(\ell,r_T)$, fixing $\mathcal E$ is equivalent, by Eq.~\eqref{eq:by_energy_momentum_density}, to fixing the common-normal curvature trace $k_T=2\sqrt{f(r_T)}/r_T$, which uniquely determines $M$. Thus $M$ serves as a collective label for these full local data.

As in the small-tube evaluation of Sec.~\ref{subsec:landscape_weight}, the canonical bulk terms vanish and the pN terms cancel the auxiliary interface pairings. Each regular bolt contributes $-S_i$ without an extra time-rescaling factor. The constrained action and leading weight are therefore
\begin{equation}
    \begin{aligned}
        I_{\rm constr}^{\rm SdS}[M]
        &=
        -S_b(M)-S_c(M),
        \\
        \mathcal Z_{\rm constr}^{\rm SdS}[M]
        &\simeq
        \exp\!\left[S_b(M)+S_c(M)\right].
    \end{aligned}
        \label{eq:sds_two_cap_weight}
    \end{equation}
This classical result is independent of the auxiliary radius $r_T$. With no physical boundary, there is no $\beta E_{\mathcal B}$ term. Normalizing relative to empty de Sitter, whose entropy is $S_{\rm dS}=\pi\ell^2$, gives the entropy-deficit factor $\exp[S_b(M)+S_c(M)-S_{\rm dS}]$, in agreement with Refs.~\cite{Draper:2022xzl,Draper:2023bhg}.

\paragraph{Kerr--de Sitter.}
We now apply the same construction to Kerr--de Sitter, developing the rotating extension suggested in Ref.~\cite{Draper:2023bhg}. In Boyer--Lindquist coordinates, the radial function is
    \begin{equation}
        \Delta_r
        =
        \left(r^2+a^2\right)
        \left(1-\frac{r^2}{\ell^2}\right)
        -2mr.
        \label{eq:kds_delta}
    \end{equation}
On a nondegenerate branch with black hole and cosmological roots $r_b<r_c$, place $\mathcal T_r$ between them. Both caps are constructed from the same stationary Kerr--de Sitter solution with parameters $(m,a,\ell)$, using the same interface embedding and common-normal convention. Let $(T_i,\Omega_i,S_i)$ denote the positive temperature, angular velocity, and entropy of each horizon in a fixed time normalization and angular frame, and set $\beta_i=T_i^{-1}$. We use the cosmological cap as reference, with the common identification $(\tau,\phi)\sim(\tau+\beta_c,\phi-i\beta_c\Omega_c)$. Applying Eqs.~\eqref{eq:cap_coordinate_map} and~\eqref{eq:thermal_rotational_source_map} to the black hole cap gives
    \begin{equation}
        \alpha=\frac{\beta_b}{\beta_c}=\frac{T_c}{T_b},
        \qquad
        \nu^\phi=i\left(\Omega_c-\alpha\Omega_b\right).
        \label{eq:kds_relative_sources}
    \end{equation}
The reference identification $(\beta_c,\Omega_c)$ is determined by the cosmological horizon of each solution. In general, both a time rescaling and an axial twist are needed to make the two caps individually regular. Their generally complex metrics share the same spatial ADM geometry in a neighborhood of the interface, and the rigid twist is generated by an axial Killing field of this common spatial metric. Equation~\eqref{eq:stationary_local_response_matching} therefore ensures matching of the complete local pN data,
    \begin{equation}
        \mathcal C_{\mathcal T}^{\rm KdS}
        =
        \left(\mathcal E,\,\mathcal J_A\,;\,\sigma_{AB}\right)_{\mathcal T_r}
        \label{eq:kds_interface_data}
    \end{equation}
including the generally nonzero axial momentum response and the generally nonround cross-section geometry.

With the AdS radius $\ell$, the interface embedding, foliation, and reference conventions fixed, evaluating the local interface data for each Kerr--de Sitter solution defines the map $(m,a)\mapsto\mathcal C_{\mathcal T}^{\rm KdS}$. This gives a family of constrained sectors parametrized by $(m,a)$, with the full local fields in Eq.~\eqref{eq:kds_interface_data} supplying the boundary data in the factorized path integral.

The same two-cap evaluation gives the classical action and associated leading saddle weight:
\begin{equation}
    \begin{aligned}
        I_{\rm constr}^{\rm KdS}[\mathcal C_{\mathcal T}^{\rm KdS}]
        &=
        -S_b-S_c,
        \\
        \mathcal Z_{\rm constr}^{\rm KdS}[\mathcal C_{\mathcal T}^{\rm KdS}]
        &\simeq
        \exp\!\left[S_b+S_c\right].
    \end{aligned}
        \label{eq:kds_two_cap_weight}
    \end{equation}
This result agrees with the Kerr--de Sitter constrained instanton action obtained in the angular momentum representation~\cite{Wu:2004db}.

In both de Sitter examples, both horizon caps remain regular, with their thermal and rotational mismatches encoded by the interface lapse and shift jumps. The constrained weights can be expressed in terms of a generalized free energy landscape by choosing a common reference Euclidean period held fixed across sectors. This period sets the normalization of the landscape without representing an externally imposed inverse temperature.

    \section{Extension to \texorpdfstring{$f(R_{\mu\nu\rho\sigma})$}{f(Riemann)} Gravity}
    \label{sec:modified_gravity}
We now extend the factorized partition function and free energy landscape beyond Einstein gravity. Higher-curvature interactions modify both the boundary variation of the action and its radial derivative structure. The boundary variation, or equivalently the pullback of the presymplectic potential including the chosen boundary terms, determines the conjugate data and admissible boundary conditions~\cite{Iyer:1994ys,Harlow:2019yfa}. Additivity imposes a further requirement: the radial derivative structure must be examined under the proposed sewing conditions. We first review the temporal ADM decomposition and subsequent radial decomposition presented in Ref.~\cite{Deruelle:2009zk,Draper:2023bhg}. We then use this framework to construct the generalized free energy of black holes in higher-curvature gravity.

We consider local, parity-even $f(\mathrm{Riemann})$ gravity, where $f(\mathrm{Riemann})$ denotes a scalar $f(R_{\mu\nu\rho\sigma})$ constructed algebraically from the Riemann tensor and the metric, with no explicit covariant derivatives of the curvature. The corresponding Euclidean action in metric variables is
    \begin{equation}
        I_{{\rm E},{\rm met}}^f[g]
        =
        \frac{1}{2}
        \int_{\mathcal M}\dd^D x\,\sqrt g\,
        f(R_{\mu\nu\rho\sigma})
        +
        I_{\partial\mathcal M}^{f},
        \label{eq:fr_action_metric}
    \end{equation}
With our sign convention, the bulk term reduces to the Einstein--Hilbert form when $f=-(R-2\Lambda)/(8\pi)$. The functional $I_{\partial\mathcal M}^{f}$ denotes the boundary terms required by the chosen variational principle; at a physical asymptotic boundary, it also includes the subtraction or counterterm contributions needed to render the action and associated charges finite. This class includes general parity-even quadratic-curvature theories and, more broadly, the algebraic curvature operators appearing in gravitational effective field theories~\cite{Endlich:2017tqa,Draper:2023bhg}.

\subsection{Auxiliary Formulation and Dirichlet Boundary Data}
\label{subsec:metric_auxiliary_factorization}

To obtain a first-order temporal ADM description, we use an auxiliary field formulation, introducing tensors $\varrho_{\mu\nu\rho\sigma}$ and $\phi^{\mu\nu\rho\sigma}$ with the algebraic symmetries of the Riemann tensor; their indices are raised and lowered with $g_{\mu\nu}$. We adopt the geometric and orientation conventions of Sec.~\ref{sec:adm_factorization}. We illustrate the local boundary analysis on $\mathcal M_1$, denoting its interface copy $\mathcal T_r^-$ simply by $\mathcal T_r$, with outward normal $n$. Contributions from other boundary components are left implicit. The analysis on the interface copy $\mathcal T_r^+$ of $\mathcal M_2$ is analogous, with outward normal $-n$. With this convention, the covariant Euclidean action on $\mathcal M_1$ in the auxiliary field formulation is
    \begin{equation}
        I^f
        =
        \frac{1}{2}
        \int_{\mathcal M_1}\dd^D x\,\sqrt g\,
        \left[
        f(\varrho_{\mu\nu\rho\sigma})
        +
        \phi^{\mu\nu\rho\sigma}
        \left(
        R_{\mu\nu\rho\sigma}-\varrho_{\mu\nu\rho\sigma}
        \right)
        \right]
        +B_{\rm GHY}^{f}.
        \label{eq:auxiliary_fr_action}
    \end{equation}
The generalized GHY contribution displayed above is
\begin{equation}
    B_{\rm GHY}^{f}
    =
    2\int_{\mathcal T_r}\dd^{D-1}x\,
    \sqrt\gamma\,\mathcal K_{ij}\Phi_{(n)}^{ij},
    \label{eq:generalized_ghy_auxiliary}
\end{equation}
where the double-normal auxiliary projection is defined by
\begin{equation}
    \Phi_{(n)}^{ij}
    =
    \gamma^i{}_{\mu}\gamma^j{}_{\nu}
    \phi^{\mu\alpha\nu\beta}n_{\alpha}n_{\beta}.
    \label{eq:double_normal_auxiliary_projection}
\end{equation}
The boundary term $B_{\rm GHY}^{f}$ cancels the boundary contributions at $\mathcal T_r$ involving normal derivatives of $\delta g_{\mu\nu}$. The resulting Dirichlet problem for the metric and auxiliary fields fixes
\begin{equation}
    \delta\gamma_{ij}=0,
    \qquad
    \delta\Phi_{(n)}^{ij}=0
    \quad\text{on }\mathcal T_r.
    \label{eq:fr_covariant_dirichlet_data}
\end{equation}
In the Einstein--Hilbert limit, $B_{\rm GHY}^{f}$ reduces to the contribution of the standard GHY term in the action~\eqref{eq:euclidean_lagrangian_action} at $\mathcal T_r$. Varying with respect to $\phi^{\mu\nu\rho\sigma}$ and $\varrho_{\mu\nu\rho\sigma}$, respectively, gives the algebraic equations
    \begin{equation}
        \varrho_{\mu\nu\rho\sigma}=R_{\mu\nu\rho\sigma},
        \qquad
        \phi^{\mu\nu\rho\sigma}
        =
        \frac{\partial f}{\partial\varrho_{\mu\nu\rho\sigma}}.
        \label{eq:auxiliary_field_equations}
    \end{equation}
Eliminating the auxiliary fields through these algebraic equations recovers the metric theory defined by the action~\eqref{eq:fr_action_metric}. Before their elimination, the auxiliary formulation makes the Hamiltonian variables, constraints, and boundary data explicit.

Performing the temporal Legendre transform of $I^f$ in Eq.~\eqref{eq:auxiliary_fr_action}, including its boundary terms, gives the canonical Dirichlet action
    \begin{equation}
        I_{\rm D}^{f}
        =
        \int_0^\beta\dd\tau
        \int_{\Sigma_\tau\cap\mathcal M_1}\dd^{D-1}x\,
        \left(
        p^{ab}\dot h_{ab}
        +
        \Pi_{ab}\dot\Psi^{ab}
        -
        N\mathcal H^f
        -
        \mathcal N^a\mathcal H_a^f
        \right)
        +
        B_N^{f}
        +
        B_{\mathcal N}^{f},
        \label{eq:fr_dirichlet_action_schematic}
    \end{equation}
Compared with Eq.~\eqref{eq:adm_action}, the auxiliary formulation introduces an additional canonical pair $(\Psi^{ab},\Pi_{ab})$ alongside $(h_{ab},p^{ab})$. Here $\Psi_{ab}$ is the symmetric double-normal projection of $\phi^{\mu\nu\rho\sigma}$ along the temporal unit normal $u$. The remaining auxiliary projections enter algebraically and are retained as independent variables in the variational principle. The lapse and shift enforce $\mathcal H^f=\mathcal H_a^f=0$. Here $\mathcal H^f$ and $\mathcal H_a^f$ are densities of weight one, whereas the density factor in Eq.~\eqref{eq:adm_constraints} is displayed separately. The boundary terms, including the generalized GHY contribution, are expressed in the same canonical variables. The technical details are collected in Appendix~\ref{app:fr_canonical_details}.

We vary Eq.~\eqref{eq:fr_dirichlet_action_schematic} with the canonical fields treated as independent, without imposing Hamilton's relations on their variations. After integration by parts, the variation has the following contribution at $\mathcal T_r$, with the bulk field equation terms left implicit:
\begin{equation}
    \delta I_{\rm D}^{f}\big|_{\mathcal T_r}
    =\int_{\mathcal T_r}\dd\tau\,\dd^d x\,
    \left[
    \mathcal E^f\,\delta N
    -\mathcal J_A^f\,\delta\mathcal N^A
    +\mathcal P_f^{\mathsf a}\delta q_{\mathsf a}
    -4\sqrt\sigma\,\chi_2^{\parallel(AB)}\delta Z_{AB}
    \right].
    \label{eq:fr_quasilocal_density_definition}
\end{equation}
The generalized Brown--York responses $\mathcal E^f$ and $\mathcal J_A^f$, together with $\mathcal P_f^{\mathsf a}$, include the cross-section factor $\sqrt\sigma$ and carry density weight one. The collective label $\mathsf a$ indexes the remaining configuration data
\begin{equation}
    q_{\mathsf a}
    =
    \left\{
    \sigma_{AB},\,
    \theta_{AB}^{\parallel},\,
    \chi_A^{\parallel},\,
    \Psi_\perp
    \right\}.
    \label{eq:fr_remaining_configuration_data}
\end{equation}
with the individual entries defined by
\begin{equation}
\begin{aligned}
    \sigma_{AB}
    &=
    e_A{}^\mu e_B{}^\nu g_{\mu\nu},
    &
    \theta_{AB}^{\parallel}
    &=
    \sigma_A{}^a\sigma_B{}^b
    \Phi_{acbd}n^c n^d,
    \\
    \chi_A^{\parallel}
    &=
    \sigma_A{}^a\Phi_{abc}n^b n^c,
    &
    \Psi_\perp
    &=
    \Psi_{ab}n^a n^b.
\end{aligned}
    \label{eq:auxiliary_corner_projections}
\end{equation}
The temporal projections $\Phi_{abcd}$, $\Phi_{abc}$, and $\Psi_{ab}$ are defined in Eq.~\eqref{eq:auxiliary_time_projections}. The label $\parallel$ indicates that the free indices are tangent to $\mathcal S_\tau^{(r)}$, while $\perp$ denotes contraction with two radial normals. We also use the mixed projection $\chi_2^{\parallel AB}=\sigma^A{}_a\sigma^B{}_c\Phi^{abc}n_b$ and the projected momentum $\Pi_{AB}^{\parallel}=\sigma_A{}^a\sigma_B{}^b\Pi_{ab}$. The additional boundary variable is
\begin{equation}
    Z_{AB}
    =
    N\frac{\Pi_{AB}^{\parallel}}{\sqrt h}
    +\mathscr D_{(A}\mathcal N_{B)}.
    \label{eq:fr_additional_sewing_datum}
\end{equation}
Equation~\eqref{eq:fr_quasilocal_density_definition} vanishes for fixed Dirichlet data $\mathcal C_{\rm D}^{f}=(N,\mathcal N^A;q_{\mathsf a};Z_{AB})_{\mathcal T_r}$. Here $\mathcal E^f$ and $-\mathcal J_A^f$ are the source responses at fixed $(q_{\mathsf a},Z_{AB})$, while $\mathcal P_f^{\mathsf a}$ is conjugate to $q_{\mathsf a}$ at fixed $(N,\mathcal N^A,Z_{AB})$. Appendix~\ref{app:fr_boundary_responses} gives the derivation, the explicit energy and momentum responses, and their relation to the covariant boundary variation.

The same response functionals organize the radial boundary contribution to the Dirichlet ADM action as
\begin{equation}
    \left.(B_N^f+B_{\mathcal N}^f)\right|_{\mathcal T_r}
    =
    \int_{\mathcal T_r}\dd\tau\,\dd^d x\,
    \left(N\mathcal E^f-\mathcal N^A\mathcal J_A^f\right)
    +B_{\rm aux}.
    \label{eq:fr_dirichlet_boundary_by_decomposition}
\end{equation}
After extracting the Brown--York pairing in Eq.~\eqref{eq:fr_dirichlet_boundary_by_decomposition}, tangential integration by parts on the closed cross-sections gives
\begin{equation}
    B_{\rm aux}
    =-4\int_0^\beta\dd\tau
    \int_{\mathcal S_\tau^{(r)}}\dd^d x\,\sqrt\sigma\,
    \chi_{2}^{\parallel(AB)}Z_{AB}.
    \label{eq:fr_residual_auxiliary_boundary_term}
\end{equation}
On solutions, Hamilton's first equation $\Pi_{ab}/\sqrt h=K_{ab}$ and the orthogonal foliation imply
\begin{equation}
    Z_{AB}=\frac12\,\partial_\tau\sigma_{AB}.
    \label{eq:fr_Z_legendre_relation}
\end{equation}
Thus, when $\sigma_{AB}$ is independent of $\tau$, $Z_{AB}=0$ and $B_{\rm aux}$ makes no contribution to the on-shell action. The term must still be retained in the action for independent canonical variations. After imposing the Einstein--Hilbert auxiliary constraints, $\Phi^{abc}=0$, so $B_{\rm aux}=0$ and Eq.~\eqref{eq:fr_dirichlet_boundary_by_decomposition} reduces to the usual Einstein Brown--York boundary term.

\subsection{Partial-Neumann Sewing and Constrained Instantons}
\label{subsec:fr_pn_sewing}

On the same subregion $\mathcal M_1$, we perform a Legendre transform in the lapse and tangential shift sectors on $\mathcal T_r$, keeping $q_{\mathsf a}$ and $Z_{AB}$ fixed. The resulting action has the ADM form
\begin{equation}
\begin{aligned}
    I_{\rm pN}^{f}
    &=
    I_{\rm D}^{f}
    -
    \int_{\mathcal T_r}\dd\tau\,\dd^d x\,
    \left(
    N\mathcal E^f
    -\mathcal N^A\mathcal J_A^f
    \right)
    \\
    &=
    \int_0^\beta\dd\tau
    \int_{\Sigma_\tau\cap\mathcal M_1}\dd^{D-1}x\,
    \left(
    p^{ab}\dot h_{ab}
    +\Pi_{ab}\dot\Psi^{ab}
    -N\mathcal H^f
    -\mathcal N^a\mathcal H_a^f
    \right)
    +B_{\rm aux}.
\end{aligned}
    \label{eq:fr_partial_neumann_action}
\end{equation}
Varying Eq.~\eqref{eq:fr_partial_neumann_action} and using Eq.~\eqref{eq:fr_quasilocal_density_definition} directly gives the partial-Neumann interface variation
\begin{equation}
    \left.\delta I_{\rm pN}^{f}\right|_{\mathcal T_r}
    =\int_{\mathcal T_r}\dd\tau\,\dd^d x\,
    \left[
    -N\,\delta\mathcal E^f
    +\mathcal N^A\,\delta\mathcal J_A^f
    +\mathcal P_f^{\mathsf a}\,\delta q_{\mathsf a}
    -4\sqrt\sigma\,
    \chi_{2}^{\parallel(AB)}\delta Z_{AB}
    \right].
    \label{eq:fr_partial_neumann_variation}
\end{equation}
The boundary Legendre transform replaces the fixed interface sources $(N,\mathcal N^A)$ with their responses $(\mathcal E^f,\mathcal J_A^f)$, while retaining $(q_{\mathsf a},Z_{AB})$ as fixed data. Accordingly, the partial-Neumann variational problem with independent canonical fields fixes the local data
\begin{equation}
    \mathcal C_{\rm pN}^{f}
    =\left(\mathcal E^f,\mathcal J_A^f;q_{\mathsf a};Z_{AB}\right)_{\mathcal T_r}.
    \label{eq:fr_partial_neumann_data}
\end{equation}
Holding these data fixed makes the interface contribution to the variation vanish. They are specified pointwise on each cross-section $\mathcal S_\tau^{(r)}$, while the lapse and tangential shift are not separately fixed.

The same analysis applies to the interface copy $\mathcal T_r^+$ of $\mathcal M_2$. We then sew $\mathcal M_1$ and $\mathcal M_2$ across $\mathcal T_r$ by identifying the complete data in Eq.~\eqref{eq:fr_partial_neumann_data}, using the common-normal convention of Sec.~\ref{subsec:adm_geometry}:
\begin{equation}
    [\mathcal E^f]=0,
    \qquad
    [\mathcal J_A^f]=0,
    \qquad
    [q_{\mathsf a}]=0,
    \qquad
    [Z_{AB}]=0.
    \label{eq:fr_partial_neumann_sewing_conditions}
\end{equation}
Combining the canonical expressions in Appendix~\ref{app:fr_canonical_details} with the boundary response relations in Appendix~\ref{app:fr_boundary_responses}, we can write the full action~\eqref{eq:fr_partial_neumann_action}, including $B_{\rm aux}$, as a bulk integral involving at most first radial derivatives of the complete data in Eq.~\eqref{eq:fr_partial_neumann_data}. For piecewise smooth canonical fields regular up to $\mathcal T_r$ on each side, matching these data excludes interface delta functions in this representation, so splitting the integral and restoring the boundary terms yields the sum of the complete subregion actions. Thus, Eq.~\eqref{eq:fr_partial_neumann_sewing_conditions} ensures additivity off shell, while lapse function $N$ and the tangential shift $\mathcal N^A$ may remain discontinuous~\cite{Draper:2023bhg}. Applying the construction of Sec.~\ref{sec:quasilocal_constrained_instantons} with compatible sewing measures gives the formal factorization
\begin{equation}
    \mathcal Z^{f}[\mathcal M]
    =
    \int\mathscr D\mathcal C_{\rm pN}^{f}\,
    \mathcal Z_{\rm pN}^{f}
    \left[\mathcal M_1\mid\mathcal C_{\rm pN}^{f}\right]
    \mathcal Z_{\rm pN}^{f}
    \left[\mathcal C_{\rm pN}^{f}\mid\mathcal M_2\right].
    \label{eq:fr_partial_neumann_factorization}
\end{equation}
The source saddle equations identify the independent energy and momentum labels with the actual subregion responses. Each factor carries the prescribed physical boundary conditions. A delta functional fixing the complete common data then defines the fixed-sector amplitude:
\begin{equation}
    \mathcal Z_{\rm constr}^{f}
    \!\left[\mathcal C_{\rm pN}^{f}\right]
    =
    \mathcal Z_{\rm pN}^{f}
    \left[\mathcal M_1\mid\mathcal C_{\rm pN}^{f}\right]
    \mathcal Z_{\rm pN}^{f}
    \left[\mathcal C_{\rm pN}^{f}\mid\mathcal M_2\right].
    \label{eq:fr_constrained_factorization}
\end{equation}
We now construct black hole saddles in these sectors using the metrics in Eqs.~\eqref{eq:outer_region_metric} and~\eqref{eq:inner_cap_metric}. Their lapse and shift satisfy $N_1=\alpha N_2$ and $\mathcal N_1^A=\alpha\mathcal N_2^A+\nu^A$, with the constant rescaling and axial twist fixed by Eq.~\eqref{eq:thermal_rotational_source_map}. Assume that the two complete field configurations are local pullbacks of the same stationary axisymmetric solution of the metric and auxiliary field equations. The cap is regular, and the exterior satisfies the prescribed physical boundary conditions. In the common-normal convention, Appendix~\ref{app:fr_stationary_sewing} then gives $[q_{\mathsf a}]=[\mathcal E^f]=[\mathcal J_A^f]=0$. Using Eq.~\eqref{eq:fr_stationary_Z_map}, stationarity and axial symmetry give $Z_{AB}^{(1)}=Z_{AB}^{(2)}=0$. These solutions therefore satisfy all the sewing conditions in Eq.~\eqref{eq:fr_partial_neumann_sewing_conditions}, while their lapse and shift may differ to accommodate thermal and rotational mismatches. At fixed $\mathcal C_{\rm pN}^{f}$, this pair is a stationary point of the action and hence a constrained instanton, as verified by the full variation in Appendix~\ref{app:ci_auxiliary_variation}.

\subsection{Constrained Actions and Generalized Free Energy}
\label{subsec:fr_constrained_free_energy}

We now evaluate the action on the constrained geometry discussed in Sec.~\ref{subsec:fr_pn_sewing}, with the exterior $\mathcal M_2$ and regular inner cap $\mathcal M_1$ described by Eqs.~\eqref{eq:outer_region_metric} and~\eqref{eq:inner_cap_metric}, respectively. The canonical bulk integrand in Eq.~\eqref{eq:fr_partial_neumann_action} vanishes in each region because the stationary solutions satisfy
\begin{equation}
    \dot h_{ab}=0,
    \qquad
    \dot\Psi^{ab}=0,
    \qquad
    \mathcal H^f=\mathcal H_a^f=0.
    \label{eq:fr_stationary_bulk_vanishing}
\end{equation}
The auxiliary interface term $B_{\rm aux}$ also vanishes on these saddles by Eqs.~\eqref{eq:fr_residual_auxiliary_boundary_term} and~\eqref{eq:fr_Z_legendre_relation}. With the Brown--York pairing removed by the partial-Neumann transform, the remaining action contributions come from the physical boundary of $\mathcal M_2$ and the regular bolt in $\mathcal M_1$ corresponding to the black hole horizon.

For the exterior region $\mathcal M_2$, the physical boundary $\mathcal B$ carries fixed Dirichlet sources, including $(T,\Omega_{\mathcal B}^I)$ and the independent auxiliary sources. Integrating the stationary Hamiltonian boundary pairing over the period $\beta$ gives
\begin{equation}
\begin{aligned}
    I_{2,{\rm os}}^{f}
    &=
    \left.\left(B_N^f+B_{\mathcal N}^f\right)\right|_{\mathcal B}
    +I_{\mathcal B}^{\rm ren}
    \\
    &=
    \beta\left(
    E_{\mathcal B}^{f}
    -\Omega_{\mathcal B}^I J_{\mathcal B,I}^{f}
    \right).
\end{aligned}
    \label{eq:fr_asymptotic_hamiltonian_boundary}
\end{equation}
Here $I_{\mathcal B}^{\rm ren}$ supplies the prescribed subtraction or counterterms. The resulting charges are finite and integrable: $E_{\mathcal B}^{f}$ is the Hamiltonian energy of the reference time translation, and $J_{\mathcal B,I}^{f}$ are the real physical angular momenta, with the conventions of footnote~\ref{fn:thermal_charge_conventions}.

For regular inner cap $\mathcal M_1$, the temporal foliation degenerates at the regular bolt $\mathcal H$ corresponding to the black hole horizon. The bolt is not a physical boundary; we evaluate its contribution using the small-tube prescription of Sec.~\ref{subsec:landscape_weight}. In a regular corotating frame, the ADM boundary term on the regulating tube vanishes as the tube shrinks. Restoring the excised neighborhood then supplies the compensating generalized GHY contribution,
\begin{equation}
    I_{1,{\rm os}}^{f}
    =I_{\rm cap}^{f}
    =-S_f,
    \label{eq:fr_regular_cap_action}
\end{equation}
where $S_f$ corresponds to the Wald entropy~\cite{Wald:1993nt,Iyer:1994ys,Iyer:1995kg}. With the Euclidean action normalization in Eq.~\eqref{eq:fr_action_metric}, it is given by
\begin{equation}
    S_f
    =
    -\pi
    \int_{\mathcal H}
    \dd^d x\,\sqrt\sigma\,
    \frac{\partial f}
    {\partial R_{\mu\nu\rho\sigma}}\,
    \epsilon_{\mu\nu}\epsilon_{\rho\sigma}.
    \label{eq:wald_entropy_general}
\end{equation}
Here $\epsilon_{\mu\nu}$ is the Euclidean binormal to $\mathcal H$, normalized by $\epsilon_{\mu\nu}\epsilon^{\mu\nu}=2$.

As in Eq.~\eqref{eq:ci_stationary_action}, adding the two subregion on-shell actions gives the constrained stationary action:
\begin{equation}
    I_{\rm constr}^{f}
    =I_{1,{\rm os}}^{f}+I_{2,{\rm os}}^{f}
    =
    \beta\left(
    E_{\mathcal B}^{f}
    -\Omega_{\mathcal B}^I J_{\mathcal B,I}^{f}
    \right)
    -S_f.
    \label{eq:fr_stationary_on_shell_action}
\end{equation}
The leading exponential weight of this saddle in the expansion~\eqref{eq:ci_semiclassical_weight} is
\begin{equation}
    \exp\left[-I_{\rm constr}^{f}\right]
    =
    \exp\left[
    -\beta\left(
    E_{\mathcal B}^{f}
    -\Omega_{\mathcal B}^I J_{\mathcal B,I}^{f}
    \right)
    +S_f
    \right].
    \label{eq:fr_constrained_saddle_weight}
\end{equation}
On the chosen branch of stationary field solutions, let $\lambda^m$, with $m=1,\ldots,N_{\rm coll}$, be collective coordinates whose map to the complete data $\mathcal C_{\rm pN}^{f}$ is injective. Restricting the action~\eqref{eq:fr_stationary_on_shell_action} to this family and dividing by the prescribed Euclidean period defines the generalized free energy:
\begin{equation}
\begin{aligned}
    \mathcal F_{\rm gen}^{f}(\lambda^m;T,\Omega_{\mathcal B}^I)
    &=\frac{I_{\rm constr}^{f}[\mathcal C_{\rm pN}^{f}(\lambda)]}{\beta}
    \\
    &=E_{\mathcal B}^f(\lambda^m)
    -\Omega_{\mathcal B}^I J_{\mathcal B,I}^f(\lambda^m)
    -T S_f(\lambda^m),
    \qquad T=\beta^{-1}.
\end{aligned}
    \label{eq:modified_generalized_free_energy}
\end{equation}
The prescribed sources $T$ and $\Omega_{\mathcal B}^I$ need not coincide with the horizon potentials $T_H$ and $\Omega_H^I$, and the result follows from regular subregion geometries without a small-deficit expansion.

For variations within the underlying stationary family, with the couplings, spatial geometry and independent auxiliary sources at $\mathcal B$ held fixed, the first law~\eqref{eq:first_law_rotating} holds with physical charges and Wald entropy~\cite{Wald:1993nt,Iyer:1994ys,Iyer:1995kg}. Following Sec.~\ref{subsec:landscape_reduction}, an interior stationary point of $\mathcal F_{\rm gen}^{f}$ at fixed $(T,\Omega_{\mathcal B}^I)$ satisfies $T=T_H$ and $\Omega_{\mathcal B}^I=\Omega_H^I$ at leading classical order, provided $S_f$ and $J_{\mathcal B,I}^{f}$ can be varied independently within this family. The two regions then join smoothly, and $\mathcal F_{\rm gen}^{f}$ equals the classical grand potential.

\subsection{Static AdS Black Holes in Einstein--Gauss--Bonnet Gravity}
\label{subsec:egb_factorization}

We first consider static, spherically symmetric AdS black holes in five-dimensional EGB gravity~\cite{Boulware:1985wk,Wheeler:1985nh,Cai:2001dz}. The Gauss--Bonnet invariant is the quadratic Lovelock density and yields second-order metric field equations~\cite{Lanczos:1938sf,Lovelock:1971yv}. In the convention of Eq.~\eqref{eq:fr_action_metric}, the theory is defined by
\begin{equation}
    f_{\rm GB}
    =-\frac{1}{8\pi}
      \left[R+\frac{12}{L^2}
      +\alpha_{\rm GB}\mathcal G\right],
    \qquad
    \mathcal G
    =R_{\mu\nu\rho\sigma}R^{\mu\nu\rho\sigma}
      -4R_{\mu\nu}R^{\mu\nu}+R^2,
    \label{eq:egb_lagrangian}
\end{equation}
Here $\Lambda=-6/L^2$ and $\alpha_{\rm GB}$ has dimensions of length squared. In the auxiliary formulation of Sec.~\ref{subsec:metric_auxiliary_factorization}, $f_{\rm GB}(\varrho)$ denotes the same curvature polynomial with the curvature tensor replaced by $\varrho_{\mu\nu\rho\sigma}$. The corresponding covariant subregion action, including the generalized GHY term, is
\begin{equation}
\begin{aligned}
    I^{\rm GB}[\mathcal M_s]
    &=\frac12\int_{\mathcal M_s}\dd^5x\,\sqrt g\,
      \left[f_{\rm GB}(\varrho)
      +\phi_{\rm GB}^{\mu\nu\rho\sigma}
       (R_{\mu\nu\rho\sigma}-\varrho_{\mu\nu\rho\sigma})\right]
      +B_{\rm GHY}^{\rm GB},
    \\
    B_{\rm GHY}^{\rm GB}
    &=2\int_{\partial\mathcal M_s}\dd^4x\,\sqrt\gamma\,
      \mathcal K_{ij}\Phi_{(n)}^{ij}.
\end{aligned}
    \label{eq:egb_auxiliary_action}
\end{equation}
Here $\Phi_{(n)}^{ij}$ is the double-normal projection of $\phi_{\rm GB}^{\mu\nu\rho\sigma}$ defined in Eq.~\eqref{eq:double_normal_auxiliary_projection}, with the outward normal on each boundary component. Eliminating the auxiliary fields recovers the EGB bulk action specified by Eq.~\eqref{eq:egb_lagrangian}, with the boundary prescription inherited from Eq.~\eqref{eq:egb_auxiliary_action}.

We choose the branch continuously connected to Einstein gravity, with $0<8\alpha_{\rm GB}/L^2<1$. Its metric function is~\cite{Cai:2001dz}
\begin{equation}
    F(r)=1+\frac{r^2}{4\alpha_{\rm GB}}
      \left(1-\sqrt{1-\frac{8\alpha_{\rm GB}}{L^2}
                         +\frac{8\alpha_{\rm GB}\mu}{r^4}}\right).
    \label{eq:egb_metric_function}
\end{equation}
Here $\mu$ is the mass parameter. The effective AdS radius $\ell$ satisfies $\ell^{-2}=(1-\Delta)/(4\alpha_{\rm GB})$, with $\Delta=\sqrt{1-8\alpha_{\rm GB}/L^2}>0$. The Einstein limit is understood continuously as $\alpha_{\rm GB}\to0$.

An auxiliary surface $\mathcal T_r$ at $r=r_T>r_h$ separates the inner cap $\mathcal M_1$ from the exterior $\mathcal M_2$, which extends to the physical asymptotic boundary $\mathcal B$. Let $\dd\Omega_3^2=\Omega_{AB}\dd x^A\dd x^B$ denote the metric of the unit three-sphere, of volume $2\pi^2$. Specializing Eqs.~\eqref{eq:outer_region_metric} and~\eqref{eq:inner_cap_metric} to this static family gives
\begin{equation}
\begin{aligned}
    \dd s_2^2
    &=F(r)\dd\tau^2+\frac{\dd r^2}{F(r)}+r^2\dd\Omega_3^2,
    && r_T\leq r<\infty,
    \\
    \dd s_1^2
    &=\alpha^2F(r)\dd\tau^2+\frac{\dd r^2}{F(r)}+r^2\dd\Omega_3^2,
    && r_h\leq r\leq r_T.
\end{aligned}
    \label{eq:egb_paired_metrics}
\end{equation}
Both metrics have $\tau\sim\tau+\beta$, with the period fixed at $\mathcal B$. The shifts and imposed angular velocities both vanish. At fixed couplings, varying $\mu$ preserves the prescribed asymptotic auxiliary sources, and the net auxiliary contribution to the boundary variation vanishes at infinity. The relative lapse normalization $\alpha=\beta_H/\beta$ is fixed by cap regularity, as in Eq.~\eqref{eq:cap_coordinate_map}. The horizon condition $F(r_h)=0$ and the regularity condition give
\begin{equation}
    \mu=r_h^2+2\alpha_{\rm GB}+\frac{r_h^4}{L^2},
    \qquad
    T_H=\frac{F'(r_h)}{4\pi}
       =\frac{r_h(1+2r_h^2/L^2)}{2\pi(r_h^2+4\alpha_{\rm GB})},
    \qquad
    \alpha\beta=\beta_H=T_H^{-1}.
    \label{eq:egb_horizon_data}
\end{equation}
Cap regularity thus fixes the relative lapse normalization, allowing the exterior period $\beta$ to be prescribed independently of $T_H$.

The paired configurations must satisfy the sewing conditions in Eq.~\eqref{eq:fr_partial_neumann_sewing_conditions}. On solutions, the auxiliary tensor associated with Eq.~\eqref{eq:egb_lagrangian} is
\begin{equation}
\begin{aligned}
    \phi_{\rm GB}^{\mu\nu\rho\sigma}
    =-\frac{1}{8\pi}\Big[&g^{\mu[\rho}g^{\sigma]\nu}
    +2\alpha_{\rm GB}\big(
      R^{\mu\nu\rho\sigma}
      -2g^{\mu[\rho}R^{\sigma]\nu}
      +2g^{\nu[\rho}R^{\sigma]\mu}
      +R\,g^{\mu[\rho}g^{\sigma]\nu}\big)\Big].
\end{aligned}
    \label{eq:egb_auxiliary_tensor}
\end{equation}
In the common-normal convention, $n=\sqrt F\,\partial_r$ on both copies, whereas $u_1=(\alpha\sqrt F)^{-1}\partial_\tau$ and $u_2=F^{-1/2}\partial_\tau$. Contracting the auxiliary tensor with these unit normals cancels the constant lapse rescaling. Explicit calculation shows that the configuration data $q_{\mathsf a}$ in Eq.~\eqref{eq:fr_remaining_configuration_data} take the following common values at $r=r_T$, with $s=1,2$:
\begin{equation}
\begin{aligned}
    \sigma_{AB}^{(s)}&=r_T^2\Omega_{AB},
    &\theta_{AB}^{\parallel(s)}&=-\frac{V(r_T)}{16\pi}\sigma_{AB},
    \\
    \Psi_\perp^{(s)}&=-\frac{U(r_T)}{8\pi},
    &\chi_A^{\parallel(s)}&=0.
\end{aligned}
    \label{eq:egb_auxiliary_projections}
\end{equation}
The radial functions $U(r)$ and $V(r)$ are given by
\begin{equation}
\begin{aligned}
    U(r)
    &=3\sqrt{1-\frac{8\alpha_{\rm GB}}{L^2}+\frac{8\alpha_{\rm GB}\mu}{r^4}}-2,
    \\
    V(r)
    &=\frac{3-\dfrac{24\alpha_{\rm GB}}{L^2}
                +\dfrac{8\alpha_{\rm GB}\mu}{r^4}}
            {\sqrt{1-\dfrac{8\alpha_{\rm GB}}{L^2}
                      +\dfrac{8\alpha_{\rm GB}\mu}{r^4}}}-2.
\end{aligned}
    \label{eq:egb_projection_functions}
\end{equation}
For these static solutions with vanishing shifts, $\mathcal J_{(s)A}^{\rm GB}=Z_{AB}^{(s)}=0$. Direct evaluation of the energy response in Eq.~\eqref{eq:fr_brown_york_energy_density} for the paired configurations, with the normal derivative taken before setting $r=r_T$, gives
\begin{equation}
\begin{aligned}
    \mathcal E_{(s)}^{\rm GB}
    &=\sqrt\sigma \left.(k\Psi_\perp+D_n\Psi_\perp)\right|_{r=r_T}=\sqrt\sigma \frac{-3\sqrt{F(r_T)}}{8\pi r_T}V(r_T),
    \qquad s=1,2.
\end{aligned}
    \label{eq:egb_interface_energy_derivation}
\end{equation}
Equations~\eqref{eq:egb_auxiliary_projections} and~\eqref{eq:egb_interface_energy_derivation}, together with $\mathcal J_{(s)A}^{\rm GB}=Z_{AB}^{(s)}=0$, establish all the sewing conditions in Eq.~\eqref{eq:fr_partial_neumann_sewing_conditions}. The lapse may still jump, with $[N]=(1-\alpha)\sqrt{F(r_T)}$. By the variational argument of Sec.~\ref{subsec:fr_pn_sewing}, the matched pair therefore defines a constrained instanton.

At fixed couplings and interface radius, the complete pN data in Eq.~\eqref{eq:fr_partial_neumann_data} are in one-to-one correspondence with $\mu$ on this branch, as follows from Eqs.~\eqref{eq:egb_auxiliary_projections}--\eqref{eq:egb_interface_energy_derivation}. Through the horizon relation~\eqref{eq:egb_horizon_data}, $r_h$ therefore serves as a collective label for these fixed-data sectors.

For the exterior $\mathcal M_2$, we use AdS background subtraction with zero energy assigned to the $\mu=0$ solution. The renormalized Hamiltonian at the physical boundary is
\begin{equation}
    H_{\mathcal B}
    =E_{\mathcal B}^{\rm GB}
    =\frac{3\pi\mu}{8},
    \label{eq:egb_renormalized_energy}
\end{equation}
in agreement with the EGB mass~\cite{Cai:2001dz}. For the inner cap $\mathcal M_1$, the ADM boundary pairing vanishes on a shrinking tube around the regular bolt. Restoring the excised neighborhood gives the generalized GHY contribution
\begin{equation}
    \lim_{\epsilon\to0^+}B_{\rm GHY}^{\rm GB}(r_h+\epsilon)
    =-\frac{A_H}{4}\left(1+\frac{12\alpha_{\rm GB}}{r_h^2}\right)
      =-S_{\rm GB},
    \qquad A_H=2\pi^2r_h^3,
    \label{eq:egb_bolt_contribution}
\end{equation}
Here $S_{\rm GB}$ is the Wald entropy, and the regularity condition in Eq.~\eqref{eq:egb_horizon_data} removes the lapse rescaling from the bolt contribution. Appendix~\ref{app:egb_boundary_contributions} evaluates both physical boundary and bolt contributions explicitly. Since $Z_{AB}^{(s)}=0$ on these static saddles, $B_{\rm aux}^{(s)}$ vanishes on both interface copies by Eq.~\eqref{eq:fr_residual_auxiliary_boundary_term}.

Applying Eq.~\eqref{eq:fr_stationary_on_shell_action} to this matched static pair gives the fixed-sector action
\begin{equation}
    I_{\rm constr}^{\rm GB}
    =I_{2,{\rm os}}^{\rm GB}+I_{1,{\rm os}}^{\rm GB}
    =\beta E_{\mathcal B}^{\rm GB}-S_{\rm GB}.
    \label{eq:egb_constrained_action}
\end{equation}
Dividing by the prescribed Euclidean period gives the generalized free energy landscape at fixed temperature $T$:
\begin{equation}
\begin{aligned}
    \mathcal F_{\rm gen}^{\rm GB}(r_h;T)
    &=\frac{I_{\rm constr}^{\rm GB}}{\beta}
      =E_{\mathcal B}^{\rm GB}(r_h)-T S_{\rm GB}(r_h)
    \\
    &=\frac{3\pi}{8}\left(r_h^2+2\alpha_{\rm GB}+\frac{r_h^4}{L^2}\right)
      -\frac{\pi^2T}{2}\left(r_h^3+12\alpha_{\rm GB}r_h\right).
\end{aligned}
    \label{eq:egb_generalized_free_energy}
\end{equation}
The landscape is independent of the auxiliary radius $r_T$. The interface can therefore be placed anywhere outside the horizon for each member of the family. This expression agrees with the generalized free energy obtained from the regularized conical Euclidean action~\cite{Li:2023men}. The energy and entropy obey the first law, giving
\begin{equation}
    \left.\frac{\partial\mathcal F_{\rm gen}^{\rm GB}}{\partial r_h}\right|_{T,L,\alpha_{\rm GB}}
    =(T_H-T)\frac{\dd S_{\rm GB}}{\dd r_h},
    \qquad
    \frac{\dd S_{\rm GB}}{\dd r_h}
    =\frac{3\pi^2}{2}(r_h^2+4\alpha_{\rm GB})>0.
    \label{eq:egb_landscape_extremum}
\end{equation}
The positive entropy derivative implies that interior stationary points satisfy $T=T_H$. At these points, $\alpha=1$, the lapse jump disappears, and the two regions join into a smooth Euclidean EGB black hole.

\subsection{BTZ Black Holes in New Massive Gravity}
\label{subsec:nmg_factorization}

We next consider the rotating, locally AdS$_3$ BTZ branch of three-dimensional NMG, a parity-even quadratic-curvature theory~\cite{Bergshoeff:2009hq}. We verify its complete auxiliary sewing data, charge normalization, and rotating generalized free energy. In the convention of Eq.~\eqref{eq:fr_action_metric}, the theory is defined by
\begin{equation}
    f_{\rm NMG}
    =
    -\frac{1}{8\pi}
    \left[
    R-2\Lambda
    +\alpha_{\rm NMG}
    \left(
    R_{\mu\nu}R^{\mu\nu}
    -\frac{3}{8}R^2
    \right)
    \right].
    \label{eq:nmg_lagrangian}
\end{equation}
Here $\alpha_{\rm NMG}$ is the quadratic curvature coupling, and the Euclidean bulk action is $I_{{\rm E},{\rm bulk}}^{\rm NMG}[g]=\tfrac12\int_{\mathcal M}\dd^3x\,\sqrt g\,f_{\rm NMG}$. The quadratic curvature terms make the metric field equations fourth order. Nevertheless, NMG admits a locally AdS$_3$ BTZ black hole branch satisfying
\begin{equation}
    R_{\mu\nu}=-\frac{2}{\ell^2}g_{\mu\nu},
    \qquad
    R=-\frac{6}{\ell^2},
    \qquad
    \Lambda=-\frac{1}{\ell^2}+\frac{\alpha_{\rm NMG}}{4\ell^4}.
    \label{eq:nmg_ads_branch}
\end{equation}
The corresponding Lorentzian black hole metric is
\begin{align}
    \dd s_{\rm L}^2
    &=-F(r)\dd t^2+\frac{\dd r^2}{F(r)}
    +r^2\left(\dd\phi-\frac{j}{2r^2}\dd t\right)^2,
    \notag\\
    F(r)
    &=-m+\frac{r^2}{\ell^2}+\frac{j^2}{4r^2}
    =\frac{(r^2-r_+^2)(r^2-r_-^2)}{\ell^2r^2},
    \label{eq:nmg_btz_metric}
\end{align}
Here $m$ and $j$ are the real Lorentzian mass and rotation parameters, respectively, and $\phi$ has period $2\pi$. The parameter $j$ is distinct from the local Euclidean response $\mathcal J_A^{\rm NMG}$ and remains real under the continuation $t=-i\tau$. We consider the nonextremal branch $r_+>|r_-|$, where $r_+>0$ is the outer horizon radius and $r_-$ is signed, with $|r_-|$ the inner horizon radius and ${\rm sgn}(r_-)={\rm sgn}(j)$. These quantities satisfy
\begin{equation}
    m=\frac{r_+^2+r_-^2}{\ell^2},
    \qquad
    j=\frac{2r_+r_-}{\ell}.
    \label{eq:nmg_btz_parameters}
\end{equation}
This continuation gives the Euclidean shift $\mathcal N^\phi=ij/(2r^2)$, so the saddle is generally complex for real $j$. Its thermodynamic horizon data are
\begin{equation}
    T_H
    =
    \frac{r_+^2-r_-^2}{2\pi\ell^2r_+},
    \qquad
    \Omega_H
    =
    \frac{r_-}{\ell r_+}
    =
    \frac{j}{2r_+^2}.
    \label{eq:nmg_btz_horizon_data}
\end{equation}

Differentiating Eq.~\eqref{eq:nmg_lagrangian} with respect to the Riemann tensor gives the auxiliary tensor $\phi_{\rm NMG}^{\mu\nu\rho\sigma}$. On the locally AdS$_3$ branch, it reduces to the particularly simple form
\begin{equation}
    \phi_{\rm NMG}^{\mu\nu\rho\sigma}
    =
    -\frac{\Xi_{\rm NMG}}{8\pi}
    g^{\mu[\rho}g^{\sigma]\nu},
    \qquad
    \Xi_{\rm NMG}
    =
    1+\frac{\alpha_{\rm NMG}}{2\ell^2}.
    \label{eq:nmg_auxiliary_ads_reduction}
\end{equation}
On this branch, the auxiliary tensor is covariantly constant and equals $\Xi_{\rm NMG}$ times its Einstein value. The relevant radial and temporal projections, together with the radial decomposition of the latter, are given in Appendix~\ref{app:nmg_auxiliary_reduction}. In particular, $(\theta_{\phi\phi}^{\parallel},\chi_\phi^{\parallel},\Psi_\perp)$ are fixed by $\sigma_{\phi\phi}$ and the couplings, while $\chi_{2}^{\parallel\phi\phi}=0$. Hence
\begin{equation}
    B_{\rm aux}^{\rm NMG}=0
    \qquad\text{on the locally AdS$_3$ branch}.
    \label{eq:nmg_residual_auxiliary_vanishing}
\end{equation}
Independently, Hamilton's first equation and stationarity give $Z_{\phi\phi}=0$, as in Sec.~\ref{subsec:fr_pn_sewing}. Either $Z_{\phi\phi}=0$ or the branch identity $\chi_2^{\parallel\phi\phi}=0$ suffices to make $B_{\rm aux}^{\rm NMG}$ vanish; both hold on these saddles and are imposed only after the independent canonical variation.

Following the paired construction of Sec.~\ref{subsec:fr_pn_sewing}, we place the auxiliary interface at $r=r_T>r_+$, separating a smooth inner horizon cap $\mathcal M_1$ from the exterior region $\mathcal M_2$. The two complete field configurations are local pullbacks of the same stationary BTZ solution under the constant Killing reparametrization in Eq.~\eqref{eq:cap_coordinate_map}, with $(\alpha,\nu^\phi)$ determined by Eq.~\eqref{eq:thermal_rotational_source_map}. We compare all normal-dependent data using the common-normal convention; the interface cross-section is a circle with $\sigma_{\phi\phi}=r_T^2$.

We now verify the matching of the NMG boundary data explicitly. Before restriction to the stationary locally AdS$_3$ branch, the complete canonical partial-Neumann data inherited from Eq.~\eqref{eq:fr_partial_neumann_data} are
\begin{equation}
    \mathcal C_{\rm pN}^{\rm NMG}
    =
    \left\{
    \mathcal E^{\rm NMG},\,
    \mathcal J_\phi^{\rm NMG}\,;\,
    \sigma_{\phi\phi},\,
    \theta_{\phi\phi}^{\parallel},\,
    \chi_\phi^{\parallel},\,
    \Psi_\perp\,;\,
    Z_{\phi\phi}
    \right\}_{\mathcal T_r}.
    \label{eq:nmg_complete_interface_data}
\end{equation}
Restricting these data to the stationary locally AdS$_3$ saddle constructed above, the common BTZ spatial geometry gives $[\sigma_{\phi\phi}]=0$. Equation~\eqref{eq:nmg_corner_auxiliary_data} then gives $[\theta_{\phi\phi}^{\parallel}]=[\chi_\phi^{\parallel}]=[\Psi_\perp]=0$. The unrenormalized interface energy and momentum responses evaluate to
\begin{equation}
\begin{aligned}
    \mathcal E_{(1)}^{\rm NMG}
    =\mathcal E_{(2)}^{\rm NMG}
    &=-\frac{\Xi_{\rm NMG}}{8\pi}\sqrt{F(r_T)},
    \\
    \mathcal J_{(1)\phi}^{\rm NMG}
    =\mathcal J_{(2)\phi}^{\rm NMG}
    &=\frac{i\Xi_{\rm NMG}}{16\pi}j .
\end{aligned}
    \label{eq:nmg_interface_response_evaluation}
\end{equation}
Equation~\eqref{eq:nmg_interface_response_evaluation} gives $[\mathcal E^{\rm NMG}]=[\mathcal J_\phi^{\rm NMG}]=0$. The densitized Euclidean momentum response $\mathcal J_\phi^{\rm NMG}$ is purely imaginary for real $j$. Hamilton's first equation also gives, on each side,
\begin{equation}
    Z_{\phi\phi}^{(s)}
    =
    \frac12\partial_\tau\sigma_{\phi\phi}
    =0,
    \qquad s=1,2.
    \label{eq:nmg_Z_matching}
\end{equation}
The data in Eq.~\eqref{eq:nmg_complete_interface_data} therefore satisfy all the sewing conditions in Eq.~\eqref{eq:fr_partial_neumann_sewing_conditions}. At fixed couplings and interface radius, with $\Xi_{\rm NMG}\neq0$, Eq.~\eqref{eq:nmg_interface_response_evaluation} determines  $m$ and $j$ from the energy response and momentum response. The map $(m,j)\mapsto\mathcal C_{\rm pN}^{\rm NMG}$ is thus injective on this branch, so these parameters can serve as collective labels for fixed-data sectors.

For the BTZ branch, we fix $(\ell,\alpha_{\rm NMG})$ and adopt the standard asymptotically AdS$_3$ boundary conditions and counterterm prescription, keeping the sources for additional modes fixed. The Hamiltonian charges conjugate to the standard time translation and axial rotation equal the BTZ values in Einstein gravity multiplied by $\Xi_{\rm NMG}$~\cite{Hohm:2010jc}. The Wald entropy has the same rescaling. Hence
\begin{equation}
    E_{\mathcal B}^{\rm NMG}
    =\frac{\Xi_{\rm NMG}}{8}m,
    \qquad
    J_{\mathcal B}^{\rm NMG}
    =\frac{\Xi_{\rm NMG}}{8}j,
    \qquad
    S_{\rm NMG}
    =\frac{\pi\Xi_{\rm NMG}}{2}r_+.
    \label{eq:nmg_charges_entropy}
\end{equation}
These physical charges and the entropy are real. The angular momentum uses the $-i$ continuation of the Euclidean response, and the Hamiltonian energy retains the lapse weighting in the asymptotic cutoff limit. Appendix~\ref{app:nmg_auxiliary_reduction} verifies both normalizations explicitly for this branch.

Applying Eq.~\eqref{eq:fr_stationary_on_shell_action} to this matched BTZ pair gives the fixed-sector action
\begin{equation}
    I_{\rm constr}^{\rm NMG}
    =
    \beta
    \left(
    E_{\mathcal B}^{\rm NMG}
    -\Omega_{\mathcal B}J_{\mathcal B}^{\rm NMG}
    \right)
    -S_{\rm NMG},
    \label{eq:nmg_constrained_action}
\end{equation}
Dividing by the prescribed Euclidean period gives
\begin{equation}
\begin{aligned}
    \mathcal F_{\rm gen}^{\rm NMG}(m,j;T,\Omega_{\mathcal B})
    &=\frac{I_{\rm constr}^{\rm NMG}}{\beta}
    \\
    &=E_{\mathcal B}^{\rm NMG}
    -\Omega_{\mathcal B}J_{\mathcal B}^{\rm NMG}
    -T S_{\rm NMG}
    \\
    &=\frac{\Xi_{\rm NMG}}8
    \left[m-\Omega_{\mathcal B}j-4\pi T r_+(m,j)\right].
\end{aligned}
    \label{eq:nmg_generalized_free_energy}
\end{equation}
At fixed $(\ell,\alpha_{\rm NMG})$, the factor $\Xi_{\rm NMG}$ rescales the BTZ landscape in Einstein gravity. The auxiliary radius $r_T$ does not appear in the result: the cut can be chosen outside the horizon for each finite $r_+$. The rescaled charges and entropy obey the BTZ first law at these fixed couplings. On the nonextremal BTZ branch with $r_+>|r_-|$ and $\Xi_{\rm NMG}\neq0$, the independence of the entropy and angular momentum variations follows explicitly from
\begin{equation}
    \det\frac{\partial(S_{\rm NMG},J_{\mathcal B}^{\rm NMG})}
    {\partial(r_+,r_-)}
    =\frac{\pi\Xi_{\rm NMG}^2r_+}{8\ell}\neq0.
    \label{eq:nmg_thermodynamic_jacobian}
\end{equation}
Away from extremality, the map from $(r_+,r_-)$ to $(m,j)$ is also invertible. At leading classical order, an interior stationary point of $\mathcal F_{\rm gen}^{\rm NMG}$ therefore satisfies $T=T_H$ and $\Omega_{\mathcal B}=\Omega_H$ by the criterion in Sec.~\ref{subsec:fr_constrained_free_energy}. These equalities give $\alpha=1$ and $\nu^\phi=0$, so the metric and auxiliary fields on both sides join into the same smooth Euclidean BTZ saddle. For $\Xi_{\rm NMG}=0$, the classical landscape vanishes identically on this BTZ family, so extremization imposes no thermal or rotational matching conditions.

When the boundary temperature differs from the Hawking temperature, the conical approach also yields a generalized free energy similar in form to Eq.~\eqref{eq:nmg_generalized_free_energy} at linear order in the conical deficit~\cite{Wu:2025jes}. For a direct comparison, we fix a BTZ background and vary only $\beta$ at $\Omega_{\mathcal B}=\Omega_H$, retaining the same physical boundary prescription. For each fixed smoothing scale $a_{\rm reg}>0$, we choose a regulated family that depends smoothly on $\delta=1-\beta/\beta_H$ and coincides with the original regular BTZ geometry at $\delta=0$. The conical action then has the expansion
\begin{equation}
    \begin{aligned}
        I_{\rm cone}^{\rm NMG}(a_{\rm reg})
        &={}
        \beta\left(
        E_{\mathcal B}^{\rm NMG}
        -\Omega_HJ_{\mathcal B}^{\rm NMG}
        \right)
        -S_{\rm NMG}
        +\mathcal O_{a_{\rm reg}}(\delta^2).
    \end{aligned}
    \label{eq:nmg_conical_comparison}
\end{equation}
In this NMG comparison, agreement at linear order does not determine the contributions at quadratic and higher orders in the conical deficit. These terms can depend on the smoothing prescription and need not remain finite as $a_{\rm reg}\to0$. The constrained construction obtains the thermodynamic form directly from regular subregions at fixed canonical interface data, without evaluating a conical defect at the bolt.

\section{Conclusions and Discussion}
\label{sec:conclusions}
We have applied gravitational path integral factorization to construct a semiclassical interpretation of generalized black hole free energy landscapes at fixed external temperature and angular velocities, without imposing their equilibrium relations with the horizon quantities. To implement this construction, boundary integral transforms relate the Dirichlet and partial-Neumann representations of the same formal partition function, with subregion actions that decompose the same uncut action under their respective sewing conditions. Fixing the complete local pN data on the auxiliary interface then selects constrained sectors, whose amplitudes reconstruct this partition function upon integration over those data. At the semiclassical level, independent variation of the constrained action, performed before restriction to stationary families, determines the multipliers supporting lapse and tangential shift jumps and establishes the regular subregion pairs constructed here as constrained instantons. The resulting constrained saddle actions determine the generalized free energy, whose dependence on the black hole parameters at fixed external sources defines the free energy landscape.

In Einstein gravity, a regular horizon cap contributes $-S_{\rm BH}$ to the action, while its paired exterior extending to a physical asymptotic boundary contributes $\beta(E_{\mathcal B}-\Omega_{\mathcal B}^I J_{\mathcal B,I})$. On a retained branch whose parameters map injectively to the complete interface data, dividing the sum of these actions by the prescribed period gives the generalized free energy,
\begin{equation}
    \mathcal F_{\rm gen}
    =
    E_{\mathcal B}
    -\Omega_{\mathcal B}^I J_{\mathcal B,I}
    -T S_{\rm BH}.
    \label{eq:conclusion_generalized_free_energy}
\end{equation}
At fixed external sources and leading classical order, the first law implies that an interior stationary point satisfies $T=T_H$ and $\Omega_{\mathcal B}^I=\Omega_H^I$, provided the retained variations independently span the entropy and angular momentum directions, as specified in Sec.~\ref{subsec:landscape_reduction}. Within the paired family constructed from the same underlying black hole solution, these conditions remove the jumps and restore a smooth equilibrium saddle. The Schwarzschild and Kerr examples provide explicit nonrotating and rotating realizations of these results, with classical free energy landscapes independent of the auxiliary interface radius.

The factorization underlying these landscapes was developed for Schwarzschild--de Sitter black holes~\cite{Draper:2022xzl}. We revisit this compact setting to interpret its constrained weights as a free energy landscape and develop the Kerr--de Sitter extension suggested in Ref.~\cite{Draper:2023bhg}. Relative lapse normalizations and, in the rotating case, axial twists keep both caps regular at fixed complete local pN data, giving $I_{\rm constr}=-(S_b+S_c)$ and the leading semiclassical weight $\exp(S_b+S_c)$. The rotating action agrees with Ref.~\cite{Wu:2004db}. A common reference Euclidean period held fixed across sectors normalizes the associated free energy landscape without representing an externally imposed inverse temperature.

We have also discussed the extension of this landscape construction to local, parity-even $f(\mathrm{Riemann})$ theories without curvature derivatives. For the stationary saddles constructed here with complete pN sewing and prescribed physical boundary conditions, the generalized free energy retains the form of Eq.~\eqref{eq:conclusion_generalized_free_energy}, with generalized Hamiltonian charges and Wald entropy. The first law again relates interior extrema to smooth equilibrium saddles under the same independence condition on the retained variations. The examples of five-dimensional static AdS black holes in EGB gravity and rotating BTZ black holes explicitly verify the complete pN sewing conditions and recover this thermodynamic form with the corresponding Hamiltonian charges and Wald entropy.

The Schwarzschild generalized free energy agrees with the asymptotically flat limit of the conical result~\cite{Li:2022oup}. The EGB landscape likewise agrees with the generalized free energy obtained from the regularized conical Euclidean action at arbitrary ensemble temperature~\cite{Li:2023men}. In more general $f(\mathrm{Riemann})$ theories, corrections beyond linear order in the conical deficit can depend on the smoothing prescription~\cite{Fursaev:1995ef,Wu:2025jes}. For a fixed BTZ background in NMG, with $\Omega_{\mathcal B}=\Omega_H$ and the smoothing regulator held fixed, the constrained and conical actions agree through linear order, as shown in Eq.~\eqref{eq:nmg_conical_comparison}. Our construction obtains the thermodynamic form from regular subregions without requiring a small-deficit expansion.

Constrained path integrals with codimension-two defects or corners use local geometric data and variational conditions to define fixed-area sectors and their Lovelock generalization at fixed Jacobson--Myers functional~\cite{Dong:2018seb,Dong:2019piw,Kastikainen:2023omj}. Related Lorentzian and Euclidean constructions yield black hole thermal weights, including cases with thermal and rotational mismatches encoded in horizon singularities~\cite{Marolf:2022ybi,Chen:2025leq}. For the corresponding stationary black holes, both the classical thermal weights~\cite{Marolf:2022ybi,Chen:2025leq} and entropy contributions in higher-curvature gravity~\cite{Marolf:2022ybi,Kastikainen:2023omj} agree with our results. These formulations complement conical evaluations of the gravitational action~\cite{Fursaev:1995ef,Grumiller:2012rt,Eune:2013qs,Li:2022oup,Wu:2025jes} by characterizing the relevant defect geometries as saddles under specified constraints. In the present construction, thermal and rotational mismatches are carried by lapse and tangential shift jumps across a finite codimension-one auxiliary interface separated from the horizons, with complete pN data held fixed and the horizon caps kept regular. Independent variation of the constrained action shows that nonzero lapse and shift jumps are supported by nonzero Lagrange multipliers, establishing the regular subregion pairs constructed here as constrained instantons.

For black holes with an asymptotic boundary, the auxiliary hypersurface $\mathcal T_r$ separates a regular horizon cap from the exterior region extending to infinity. This decomposition suggests a thermodynamic picture in which, at fixed complete pN data, the two regions may be viewed as constrained thermodynamic subsystems with matched quasilocal responses but generally different conjugate thermal and rotational potentials. In this picture, the Lagrange multipliers enforce the interface constraints and support these mismatches, allowing the two regions to be paired without requiring mutual thermodynamic equilibrium.

The present construction determines the leading semiclassical weights on selected nonextremal saddle branches. A more complete evaluation of the reduced ensemble requires the fluctuation prefactors and induced measure, together with an assessment of stability beyond the retained collective variables. For the rotating configurations considered here, admissible complex integration contours also remain to be determined. With these ingredients, the reduced integral would yield the thermodynamic potential $-T\ln\mathcal Z_{\rm red}$ and normalized averages of the generalized free energy~\cite{Cheng:2024hxh,Cheng:2024efw,Ali:2024adt,Liu:2025iei,Anand:2026oam}. Beyond these ensemble calculations, the thermodynamic picture above motivates connecting the present factorization framework to existing stochastic descriptions of relaxation and transitions on black hole free energy landscapes~\cite{Li:2020nsy,Liu:2021lmr,Liu:2023sbf,Yue:2024qms}. In particular, it would be useful to clarify the possible dynamical role of the thermal and rotational potential mismatches encoded by the interface multipliers.

\acknowledgments
This work was supported by the National Natural Science Foundation of China (Grants No. 12475055, No. 12475056, No.12247101, No. 12625504), the Fundamental Research Funds for the Central Universities (Grants No. lzujbky-2025-jdzx07), the Natural Science Foundation of Gansu Province (No. 22JR5RA389 and No. 25JRRA799).

\appendix

\section{Boundary Representations and Constrained Variations}
\label{app:ci_proof}

This appendix verifies the formal reconstruction of the original path integral from the boundary transforms introduced in the main text, gives the full constrained variation relating paired subregion saddles to constrained instantons, and extends that variation to auxiliary fields. We keep the artificial interface fixed and assume closed cross-sections with prescribed thermal identifications, so tangential integration by parts produces no endpoint terms. The physical boundary data and smooth-cap conditions remain fixed. We use $[X]=X_2-X_1$ and the common-normal convention of Sec.~\ref{subsec:adm_geometry}: on the $\mathcal M_1$ copy the outward normal is $+n$, whereas on the $\mathcal M_2$ copy it is $-n$.

\subsection{Boundary Transforms and Reconstruction of the Original Integral}
\label{app:ci_boundary_transform}

We retain the notation $q_s=(N_s,\mathcal N_s^A)$, $\xi=(\mathcal E,\mathcal J_A)$, and $\mathcal P_s[\mathcal E,\mathcal J_A]$ of Sec.~\ref{subsec:path_integral_dirichlet_sewing}, and write $q=(N,\mathcal N^A)$ for the common source data. The duality prescription adopted there is expressed by the following integrated completeness relation for any suitable test functional $\mathscr F(q_1,q_2;\sigma_{AB})$,
\begin{equation}
    \begin{aligned}
        &\int\mathscr D\mu_{\rm pN}(\mathcal E,\mathcal J_A,\sigma_{AB})
        \int_{\Gamma_1}\mathscr D\mu_1(q_1;\xi,\sigma_{AB})
        \int_{\Gamma_2}\mathscr D\mu_2(q_2;\xi,\sigma_{AB})
        \\
        &\qquad\times e^{\mathcal P_1[\mathcal E,\mathcal J_A]-\mathcal P_2[\mathcal E,\mathcal J_A]}
        \mathscr F(q_1,q_2;\sigma_{AB})
        \\
        &\quad=\int\mathscr D\mu_{\rm D}(N,\mathcal N^A,\sigma_{AB})\,
        \mathscr F(q,q;\sigma_{AB}).
    \end{aligned}
    \label{eq:ci_boundary_completeness}
\end{equation}
The source measures retain their $\xi$ dependence inside the response integral, while the common $\sigma_{AB}$ is held fixed in the source transforms and integrated once on each side of the relation. The full measures and contour prescription are those adopted in Sec.~\ref{subsec:path_integral_dirichlet_sewing}.

To reconstruct the Dirichlet sewing relation, take
\[
    \mathscr F(q_1,q_2;\sigma_{AB})
    =\mathcal Z_{\rm D}
    [\mathcal M_1\mid N_1,\mathcal N_1^A,\sigma_{AB}]\,
    \mathcal Z_{\rm D}
    [N_2,\mathcal N_2^A,\sigma_{AB}\mid\mathcal M_2].
\]
With this choice, Eqs.~\eqref{eq:pn_boundary_transform_1} and \eqref{eq:pn_boundary_transform_2} identify the left-hand side of Eq.~\eqref{eq:ci_boundary_completeness} with the pN sewing integral in Eq.~\eqref{eq:partial_neumann_factorization}, whereas its right-hand side directly reproduces the Dirichlet sewing integral in Eq.~\eqref{eq:dirichlet_factorization}. This establishes their formal equivalence under the duality prescription, before any saddle-point approximation. 

In the fixed-sector product of Eq.~\eqref{eq:constrained_instanton_factorization}, the common pN data are not integrated over, so the two source values remain independent. Appendix~\ref{app:ci_independent_variation} derives the corresponding saddle equations and identifies the response labels with the geometric subregion responses.

\subsection{Independent Response Labels and the Full Interface Variation}
\label{app:ci_independent_variation}

We now turn to the saddle equations in a fixed sector. To derive the boundary conditions from the transformed subregion path integrals, we treat $\mathcal E$ and $\mathcal J_A$ as independent transform labels, without yet identifying them with the actual responses $\mathcal E_{(s)}$ and $\mathcal J_{(s)A}$ on either side. The cross-section metric $\sigma_{AB}$ is shared configuration data, whereas $N_s$ and $\mathcal N_s^A$ are independent on the two sides. The target values $\mathcal C_{\mathcal T}=(\overline{\mathcal E},\overline{\mathcal J}_A;\overline\sigma_{AB})$ are held fixed throughout the variation.

The normalized multiplier representation in Eq.~\eqref{eq:constraint_multiplier_representation}, combined with the boundary transforms, gives the augmented action $\widetilde I_\eta$ with independent response labels in Eq.~\eqref{eq:ci_extended_augmented_action}.

After imposing the bulk equations, define $\mathscr S_{(s)}^{AB}=\sqrt\sigma N_s s_{(s)}^{AB}/2$. The two outward-normal Dirichlet variations are
\begin{align}
    \left.\delta I_{\rm D}[\mathcal M_1]\right|_{\mathcal T_r}
    &=\int_{\mathcal T_r}\dd\tau\,\dd^dx\,
    \left(\mathcal E_{(1)}\delta N_1
    -\mathcal J_{(1)A}\delta\mathcal N_1^A
    +\mathscr S_{(1)}^{AB}\delta\sigma_{AB}\right),
    \notag\\
    \left.\delta I_{\rm D}[\mathcal M_2]\right|_{\mathcal T_r}
    &=-\int_{\mathcal T_r}\dd\tau\,\dd^dx\,
    \left(\mathcal E_{(2)}\delta N_2
    -\mathcal J_{(2)A}\delta\mathcal N_2^A
    +\mathscr S_{(2)}^{AB}\delta\sigma_{AB}\right).
    \label{eq:ci_oriented_dirichlet_variations}
\end{align}
The resulting interface variation of $\widetilde I_\eta$ is
\begin{equation}
\begin{aligned}
    \left.\delta\widetilde I_\eta\right|_{\rm bulk\;EOM}
    =\int_{\mathcal T_r}\dd\tau\,\dd^dx\,\Bigl\{
    &(\mathcal E_{(1)}-\mathcal E)\delta N_1
     -(\mathcal J_{(1)A}-\mathcal J_A)\delta\mathcal N_1^A
    \\
    &-(\mathcal E_{(2)}-\mathcal E)\delta N_2
     +(\mathcal J_{(2)A}-\mathcal J_A)\delta\mathcal N_2^A
    \\
    &+([N]+\eta_{\mathcal E})\delta\mathcal E
     +(-[\mathcal N^A]+\eta_{\mathcal J}^{A})\delta\mathcal J_A
    \\
    &+(\eta_\sigma^{AB}-[\mathscr S^{AB}])\delta\sigma_{AB}\Bigr\}
     +\langle\delta\eta,\mathcal C-\mathcal C_{\mathcal T}\rangle_{\mathcal T_r}.
\end{aligned}
    \label{eq:ci_full_independent_variation}
\end{equation}
Together with the shared cross-section metric, the independent source and multiplier variations give
\begin{equation}
    \mathcal E_{(1)}=\mathcal E_{(2)}=\overline{\mathcal E},\qquad
    \mathcal J_{(1)A}=\mathcal J_{(2)A}=\overline{\mathcal J}_A,\qquad
    \sigma_{AB}^{(1)}=\sigma_{AB}^{(2)}=\overline\sigma_{AB}.
    \label{eq:ci_prescribed_response_equations}
\end{equation}
Since $[\mathscr S^{AB}]=\sqrt\sigma[Ns^{AB}]/2$, the remaining variations of the common interface data give Eq.~\eqref{eq:ci_multiplier_jump_relations}. Together with Eq.~\eqref{eq:ci_prescribed_response_equations}, this proves the equivalence of the saddle equations discussed in Sec.~\ref{subsec:ci_interface_variation}.

All variations here are performed before restricting to a stationary black hole ansatz. If the data are redundant or satisfy boundary identities after gauge reduction, one should first select independent data. Alternatively, the equations can be understood as identities on the admissible tangent space, in which case the multiplier components along redundant directions need not be unique. For a regular constraint surface, this is the usual Lagrange multiplier characterization of a constrained stationary point. Once Eq.~\eqref{eq:ci_prescribed_response_equations} holds, the multiplier term evaluates to zero, and Eq.~\eqref{eq:pn_original_action_additivity} yields Eq.~\eqref{eq:ci_stationary_action}. Setting the constraint term to zero before variation would discard the equations that support the source jumps.

If the target values are also integrated over, their variation gives $\delta_{\mathcal C_{\mathcal T}}\widetilde I_\eta=-\langle\eta,\delta\mathcal C_{\mathcal T}\rangle_{\mathcal T_r}$. At leading classical order, an interior stationary point of the integral over all independent target data therefore requires the corresponding multipliers to vanish, restoring the conjugate matching conditions. If $\mathcal C_{\mathcal T}$ is instead restricted to a collective family, only the projections along its tangent directions are imposed.

\subsection{Constrained Variation with Auxiliary Fields}
\label{app:ci_auxiliary_variation}

The constrained path integral in the auxiliary field formulation, Eq.~\eqref{eq:fr_constrained_factorization}, is defined at fixed complete data $\mathcal C_{\rm pN}^{f}=(\mathcal E^f,\mathcal J_A^f;q_{\mathsf a};Z_{AB})$. To characterize its saddles, we follow the construction with independent response labels in Appendix~\ref{app:ci_independent_variation}, applying the boundary transforms at fixed $q_{\mathsf a}$ and $Z_{AB}$ as specified in Sec.~\ref{subsec:fr_pn_sewing}. We exponentiate the delta functional selecting the complete data and vary the canonical fields, source values, transform labels, and multipliers independently. The bulk variations give the full auxiliary field equations, while the independent lapse and tangential shift variations identify the energy and momentum labels with the actual responses on both sides.

After deriving these equations, the transformed action can be written as the sum of the two complete $I_{\rm pN}^{f}$ actions and the multiplier pairing. As in Sec.~\ref{subsec:ci_interface_variation}, the remaining interface variation, restricted to variations that preserve the identification of the response labels with the actual responses, agrees with that of the geometric partial-Neumann form. Using the independent canonical boundary variation in Eq.~\eqref{eq:fr_partial_neumann_variation}, derived in Appendix~\ref{app:fr_boundary_responses}, the two geometric interface contributions sum in the common-normal convention to
\begin{equation}
\begin{split}
    \left.\delta\bigl(I_{\rm pN}^{f}[\mathcal M_1]
    +I_{\rm pN}^{f}[\mathcal M_2]\bigr)\right|_{\mathcal T_r}
    =\int_{\mathcal T_r}\dd\tau\,\dd^dx\,\Bigl[
    &[N]\delta\mathcal E^f-[\mathcal N^A]\delta\mathcal J_A^f
    \\
    &-[\mathcal P_f^{\mathsf a}]\delta q_{\mathsf a}
    +4\sqrt\sigma[\chi_2^{\parallel(AB)}]\delta Z_{AB}\Bigr].
\end{split}
    \label{eq:ci_auxiliary_summed_variation}
\end{equation}
Here the configuration data and $Z_{AB}$ are even under reversal of the radial normal at fixed temporal orientation; the oriented responses, including $\chi_2^{\parallel(AB)}$, account for the relative sign between the two sides. With the same positive sign for the multiplier pairing as in Eq.~\eqref{eq:ci_multiplier_pairing}, the remaining stationarity equations are
\begin{equation}
\begin{gathered}
    \eta_{\mathcal E}^{f}=-[N],\qquad
    \eta_{\mathcal J}^{f,A}=[\mathcal N^A],
    \\
    \eta_q^{\mathsf a}=[\mathcal P_f^{\mathsf a}],\qquad
    \eta_Z^{AB}=-4\sqrt\sigma[\chi_2^{\parallel(AB)}],
\end{gathered}
    \label{eq:ci_auxiliary_multiplier_relations}
\end{equation}
together with the prescribed values of all components of $\mathcal C_{\rm pN}^{f}$.

On the stationary saddles verified in Appendix~\ref{app:fr_stationary_sewing}, $B_{\rm aux}$ vanishes on the closed interfaces. The multiplier pairing vanishes after imposing the prescribed data. The resulting stationary action is Eq.~\eqref{eq:fr_stationary_on_shell_action}, including the generalized physical charges and the Wald entropy contribution from the cap.

\section{Auxiliary Field Formulation of \texorpdfstring{$f(R_{\mu\nu\rho\sigma})$}{f(Riemann)} Gravity}
\label{app:fr_technical_details}

This appendix collects the technical formulas used in Sec.~\ref{sec:modified_gravity} for local $f(\mathrm{Riemann})$ gravity, including the temporal canonical decomposition, radial boundary responses, stationary sewing relations, and the EGB and NMG examples. The geometry and common-normal convention follow Sec.~\ref{sec:adm_factorization}; the auxiliary field notation and complete boundary data are given in Secs.~\ref{subsec:metric_auxiliary_factorization} and~\ref{subsec:fr_pn_sewing}, respectively. These formulas provide the complete auxiliary field data used in the constrained variational proof of Appendix~\ref{app:ci_auxiliary_variation}.

\subsection{Temporal Canonical Decomposition}
\label{app:fr_canonical_details}

The temporal projections of $\phi^{\mu\nu\rho\sigma}$ are
\begin{equation}
    \Phi_{abcd}
    =h_a{}^\mu h_b{}^\nu h_c{}^\rho h_d{}^\sigma
      \phi_{\mu\nu\rho\sigma},
    \qquad
    \Phi_{abc}
    =h_a{}^\mu h_b{}^\nu h_c{}^\rho u^\sigma
      \phi_{\mu\nu\rho\sigma},
    \qquad
    \Psi_{ab}
    =2h_a{}^\mu h_b{}^\nu u^\rho u^\sigma
      \phi_{\mu\rho\nu\sigma}.
    \label{eq:auxiliary_time_projections}
\end{equation}
The corresponding projections of $\varrho_{\mu\nu\rho\sigma}$ are
\begin{equation}
\begin{aligned}
    \widehat{\varrho}_{abcd}
    &=
    h_a{}^\mu h_b{}^\nu h_c{}^\rho h_d{}^\sigma
    \varrho_{\mu\nu\rho\sigma},
    &
    \widehat{\varrho}_{abc}
    &=
    h_a{}^\mu h_b{}^\nu h_c{}^\rho u^\sigma
    \varrho_{\mu\nu\rho\sigma},
    \\
    \widehat{\varrho}_{ab}
    &=
    h_a{}^\mu h_b{}^\nu u^\rho u^\sigma
    \varrho_{\mu\rho\nu\sigma}.
\end{aligned}
    \label{eq:auxiliary_varrho_time_projections}
\end{equation}
Hats denote the temporal projections of $\varrho_{\mu\nu\rho\sigma}$; in particular, $\widehat{\varrho}_{ab}$ is its double-$u$ projection. The notation $[X]$ continues to denote the interface jump $X_2-X_1$. In Euclidean signature, the momenta conjugate to $h_{ab}$ and $\Psi^{ab}$ are~\cite{Deruelle:2009zk,Draper:2023bhg}
\begin{equation}
\begin{aligned}
    p^{ab}
    =\sqrt h\Bigg[
    &\Phi^{acdb}K_{cd}
    -\frac{2}{N}D_c\!\left(N\Phi^{c(ab)}\right)
    +\Psi^{c(a}K^{b)}{}_c
    +\frac12 K^{cd}\Psi_{cd}h^{ab}
    \\
    &+\frac12K\Psi^{ab}
    +\frac{1}{2N}\mathcal L_{Nu}\Psi^{ab}
    \Bigg],
    \qquad
    \Pi_{ab}=\sqrt h\,K_{ab}.
\end{aligned}
    \label{eq:fr_canonical_momenta}
\end{equation}
For independent canonical variables, $\Pi_{ab}/\sqrt h=K_{ab}$ follows from Hamilton's first equation, obtained by varying $p^{ab}$.

Writing $\Pi= h^{ab}\Pi_{ab}$ and abbreviating $f(\varrho_{\mu\nu\rho\sigma})$ as $f(\varrho)$, the generalized constraints in Eq.~\eqref{eq:fr_dirichlet_action_schematic} are
\begin{align}
    \mathcal H^f
    ={}&
    \sqrt h\left(
    D_aD_b\Psi^{ab}
    -\frac12 f(\varrho)
    +\Psi^{ab}\widehat{\varrho}_{ab}
    \right)
    \notag\\
    &+\frac{1}{\sqrt h}
    \left(
    2p^{ab}\Pi_{ab}
    -\Psi^{ab}\Pi_{ac}\Pi_b{}^c
    -\Pi\,\Psi^{ab}\Pi_{ab}
    \right)
    \notag\\
    &-\frac{\sqrt h}{2}\Phi^{abcd}
    \left[
    \frac{2}{h}\Pi_{ad}\Pi_{bc}
    +R_{abcd}[h]-\widehat{\varrho}_{abcd}
    \right]
    \notag\\
    &+2\sqrt h\,\Phi^{abc}
    \left[
    \widehat{\varrho}_{abc}
    -2D_a\!\left(\frac{\Pi_{bc}}{\sqrt h}\right)
    \right],
    \label{eq:fr_hamiltonian_constraint}\\
    \mathcal H_a^f
    ={}&
    2\sqrt h\,D_b
    \left(
    \frac{\Pi_{ac}}{\sqrt h}\Psi^{cb}
    \right)
    -2\sqrt h\,h_{ac}D_b
    \left(\frac{p^{cb}}{\sqrt h}\right)
    +\Pi_{bc}D_a\Psi^{bc}.
    \label{eq:fr_momentum_constraint}
\end{align}

For the orthogonal double foliation used in the main text, the radial boundary contributions on $\mathcal M_1$, with $n$ outward on $\mathcal T_r$, are
\begin{align}
    \left.B_{\mathcal N}^{f}\right|_{\mathcal T_r}
    ={}&
    \int_0^\beta\dd\tau
    \int_{\mathcal S_\tau^{(r)}}\dd^d x\,\sqrt\sigma\,
    \mathcal N^a
    \left(
    2\frac{\Pi_{ac}}{\sqrt h}\Psi^{cb}
    -2h_{ac}\frac{p^{cb}}{\sqrt h}
    \right)n_b,
    \label{eq:fr_shift_boundary_term}\\
    \left.B_N^{f}\right|_{\mathcal T_r}
    ={}&
    \int_0^\beta\dd\tau
    \int_{\mathcal S_\tau^{(r)}}\dd^d x\,\sqrt\sigma\,N
    \left[
    n_a\left(
    D_b\Psi^{ba}
    -\Psi^{ab}D_b\ln N
    \right)
    +2\mathcal K_{ij}\Phi_{(n)}^{ij}
    \right].
    \label{eq:fr_lapse_boundary_term}
\end{align}

In canonical variables, the mixed component of the generalized GHY term in Eq.~\eqref{eq:fr_lapse_boundary_term} contributes $4N\sqrt\sigma\,\chi_A^{\parallel}\Pi_{\parallel}^{A}/\sqrt h$ to the boundary integrand. Here $\chi_A^{\parallel}$ and $\Pi_{\parallel}^{A}$ are defined in Eqs.~\eqref{eq:auxiliary_corner_projections} and~\eqref{eq:fr_additional_corner_projections}, respectively. The momenta remain independent variables in the resulting canonical variational problem.

\paragraph{Einstein Limit of the Canonical Momentum.}
The metric momentum $p^{ab}$ in the auxiliary field formulation does not coincide directly with the Einstein momentum of Sec.~\ref{sec:adm_factorization}. In the Einstein limit, $\Psi^{ab}=-h^{ab}/(8\pi)$ and $\Phi_{abc}=0$. Pulling back the canonical one-form to this auxiliary constraint surface gives
\begin{equation}
    p^{ab}\delta h_{ab}+\Pi_{ab}\delta\Psi^{ab}
    =\left(p^{ab}+\frac{\Pi^{ab}}{8\pi}\right)\delta h_{ab},
    \qquad
    p_{\rm EH}^{ab}= p^{ab}+\frac{\Pi^{ab}}{8\pi}.
    \label{eq:einstein_reduced_canonical_momentum}
\end{equation}
Here the indices of $\Pi_{ab}$ are raised with $h^{ab}$. Using the momentum relations in Eq.~\eqref{eq:fr_canonical_momenta}, including $\mathcal L_{Nu}h^{ab}=-2NK^{ab}$, gives
\begin{equation}
    p^{ab}=-\frac{\sqrt h}{16\pi}(K^{ab}+Kh^{ab}),
    \qquad
    p_{\rm EH}^{ab}=\frac{\sqrt h}{16\pi}(K^{ab}-Kh^{ab}).
    \label{eq:einstein_momentum_limit_check}
\end{equation}
The reduced momentum $p_{\rm EH}^{ab}$ agrees with Eq.~\eqref{eq:adm_momentum}.

\subsection{Radial Boundary Variation and Generalized Brown--York Responses}
\label{app:fr_boundary_responses}

The additional radial decomposition components needed for the explicit response functions are
\begin{equation}
\begin{aligned}
    \Psi_{\parallel}^{AB}
    &=
    \sigma^A{}_a\sigma^B{}_b\Psi^{ab},
    &
    \Psi_{\parallel}^{A}
    &=
    \sigma^A{}_a\Psi^{ab}n_b,
    \\
    \Pi_A^{\parallel}
    &=
    \sigma_A{}^a\Pi_{ab}n^b,
    &
    \chi_{2}^{\parallel AB}
    &=
    \sigma^A{}_a\sigma^B{}_c\Phi^{abc}n_b.
\end{aligned}
    \label{eq:fr_additional_corner_projections}
\end{equation}
Indices on these cross-section tensors are raised and lowered with $\sigma_{AB}$; in particular, $\Pi_{\parallel}^{A}=\sigma^{AB}\Pi_B^{\parallel}$. We further define
\begin{equation}
    D_n= n^aD_a,
    \qquad
    a_A=\sigma_A{}^a n^bD_b n_a
    =-\mathscr D_A\ln P.
    \label{eq:fr_radial_derivative_acceleration}
\end{equation}
We begin with the independent canonical variation of Eq.~\eqref{eq:fr_dirichlet_action_schematic}. After integration by parts, its interface contribution has the form given in Eq.~\eqref{eq:fr_quasilocal_density_definition},
\begin{equation*}
    \left.\delta I_{\rm D}^{f}\right|_{\mathcal T_r}
    =\int_{\mathcal T_r}\dd\tau\,\dd^d x\,
    \left[
    \mathcal E^f\,\delta N
    -\mathcal J_A^f\,\delta\mathcal N^A
    +\mathcal P_f^{\mathsf a}\delta q_{\mathsf a}
    -4\sqrt\sigma\,\chi_2^{\parallel(AB)}\delta Z_{AB}
    \right].
\end{equation*}
Here $q_{\mathsf a}=\{\sigma_{AB},\theta_{AB}^{\parallel}, \chi_A^{\parallel},\Psi_\perp\}$ collects the remaining configuration data defined in Eq.~\eqref{eq:fr_remaining_configuration_data}. The bulk field equation terms are left implicit, and no Hamilton relation is imposed on the independent variations. The energy and momentum responses are~\cite{Draper:2023bhg}
\begin{align}
    \frac{\mathcal E^f}{\sqrt\sigma}
    ={}&
    2\mathscr D_A\Psi_{\parallel}^{A}
    -k_{AB}\Psi_{\parallel}^{AB}
    +k\Psi_\perp
    -2a_A\Psi_{\parallel}^{A}
    +D_n\Psi_\perp
    \notag\\
    &+2k^{AB}\theta_{AB}^{\parallel}
    +4\chi_A^{\parallel}\frac{\Pi_{\parallel}^{A}}{\sqrt h}
    +4\chi_{2}^{\parallel(AB)}
      \frac{\Pi_{AB}^{\parallel}}{\sqrt h},
    \label{eq:fr_brown_york_energy_density}\\
    \frac{\mathcal J_A^f}{\sqrt\sigma}
    ={}&
    -2\frac{\Pi_{Ac}}{\sqrt h}\Psi^{cb}n_b
    +2\sigma_{Aa}\frac{p^{ab}}{\sqrt h}n_b
    +4\sigma_{AC}\mathscr D_B
      \chi_{2}^{\parallel(BC)},
    \qquad
    \Pi_{Ac}=\sigma_A{}^a\Pi_{ac}.
    \label{eq:fr_brown_york_momentum_density}
\end{align}
The additional boundary variable is the combination introduced in Eq.~\eqref{eq:fr_additional_sewing_datum},
\begin{equation*}
    Z_{AB}
    =N\frac{\Pi_{AB}^{\parallel}}{\sqrt h}
    +\mathscr D_{(A}\mathcal N_{B)}.
\end{equation*}
For $N\neq0$, this definition is invertible with respect to $\Pi_{AB}^{\parallel}/\sqrt h$ at fixed $(N,\mathcal N^A,q_{\mathsf a})$.

To explain how the independent momentum variation leads to the source and $\delta Z_{AB}$ terms in Eq.~\eqref{eq:fr_quasilocal_density_definition}, first hold $q_{\mathsf a}$ fixed. Combining the radial boundary contributions from the bulk and generalized GHY variations gives
\begin{equation*}
\begin{aligned}
    \left.\delta I_{\rm D}^{f}\right|_{\mathcal T_r,\,\delta q=0}
    =\int_{\mathcal T_r}\dd\tau\,\dd^d x\,
    \Biggl[&
    \left(\mathcal E^f-4\sqrt\sigma\,
    \chi_2^{\parallel(AB)}\frac{\Pi_{AB}^{\parallel}}{\sqrt h}\right)\delta N
    \\
    &-\left(\mathcal J_A^f-4\sqrt\sigma\,\sigma_{AB}
    \mathscr D_C\chi_2^{\parallel(BC)}\right)\delta\mathcal N^A
    \\
    &-4N\sqrt\sigma\,\chi_2^{\parallel(AB)}
    \delta\!\left(\frac{\Pi_{AB}^{\parallel}}{\sqrt h}\right)\Biggr].
\end{aligned}
\end{equation*}
Rewriting the momentum variation in terms of $\delta Z_{AB}$ and integrating tangential derivatives by parts gives the source and $\delta Z_{AB}$ terms in Eq.~\eqref{eq:fr_quasilocal_density_definition}. Retaining the general variations $\delta q_{\mathsf a}$ in the full canonical calculation supplies the remaining coefficients $\mathcal P_f^{\mathsf a}$ at fixed $(N,\mathcal N^A,Z_{AB})$.

To pass to partial-Neumann boundary conditions, we subtract the variation of the Brown--York pairing and obtain Eq.~\eqref{eq:fr_partial_neumann_variation}, leaving these $q_{\mathsf a}$ coefficients unchanged in the independent canonical variation. The terms proportional to $Z_{AB}\delta\chi_2^{\parallel(AB)}$ are included in $-N\delta\mathcal E^f+\mathcal N^A\delta\mathcal J_A^f$, so $\chi_2^{\parallel(AB)}$ need not be fixed separately.

In the Einstein limit, $\Psi^{ab}=-h^{ab}/(8\pi)$ and $\theta_{AB}^{\parallel}=-\sigma_{AB}/(16\pi)$, while the mixed auxiliary projections vanish. Substituting these values and the reduced momentum~\eqref{eq:einstein_reduced_canonical_momentum} gives
\begin{equation}
    \mathcal E^{\rm EH}=-\frac{\sqrt\sigma\,k}{8\pi},
    \qquad
    \mathcal J_A^{\rm EH}
    =\frac{2\sqrt\sigma}{\sqrt h}\sigma_A{}^a n_b p_{{\rm EH}\,a}{}^b
    =\frac{\sqrt\sigma}{8\pi}\sigma_A{}^a n^bK_{ab}.
    \label{eq:einstein_boundary_response_check}
\end{equation}
The last equality holds on a classical solution. The orthogonality condition $\sigma_A{}^a n^bh_{ab}=0$ removes the trace term, recovering the Einstein responses in Eqs.~\eqref{eq:by_energy_momentum_density} and~\eqref{eq:densitized_by_data}. The convention for angular momentum in footnote~\ref{fn:thermal_charge_conventions} applies to this Euclidean response.

For the general auxiliary theory, the responses in Eqs.~\eqref{eq:fr_brown_york_energy_density} and~\eqref{eq:fr_brown_york_momentum_density} organize the ADM boundary terms into the Brown--York pairing and the auxiliary remainder in Eq.~\eqref{eq:fr_dirichlet_boundary_by_decomposition},
\begin{equation*}
\begin{aligned}
    B_{\rm aux}
    =-4\int_0^\beta\dd\tau
    \int_{\mathcal S_\tau^{(r)}}\dd^d x\,\sqrt\sigma
    \left[
    N\chi_{2}^{\parallel(AB)}
      \frac{\Pi_{AB}^{\parallel}}{\sqrt h}
    -\mathcal N_A \mathscr D_B
      \chi_{2}^{\parallel(AB)}
    \right].
\end{aligned}
\end{equation*}
On a closed cross-section, tangential integration by parts combines the two terms into $\chi_2^{\parallel(AB)}Z_{AB}$, giving Eq.~\eqref{eq:fr_residual_auxiliary_boundary_term} without imposing the field equations.

We now recover the on-shell boundary variation of the covariant auxiliary action $I^f$ in Eq.~\eqref{eq:auxiliary_fr_action} from the canonical result. Hamilton's first equation $\Pi_{ab}/\sqrt h=K_{ab}$ gives
\begin{equation}
    Z_{AB}
    =N\sigma_A{}^a\sigma_B{}^b K_{ab}
    +\mathscr D_{(A}\mathcal N_{B)}
    =\frac12\partial_\tau\sigma_{AB},
    \label{eq:fr_projected_hamilton_relation}
\end{equation}
as stated in Eq.~\eqref{eq:fr_Z_legendre_relation}. For variations among solutions, with the interface and thermal identifications held fixed, its linearization gives $\delta Z_{AB}=\tfrac12\partial_\tau\delta\sigma_{AB}$. Substituting this relation into the full canonical boundary variation and integrating by parts in Euclidean time gives the contribution $2\partial_\tau(\sqrt\sigma\,\chi_2^{\parallel(AB)})\delta\sigma_{AB}$; the endpoint term vanishes by periodicity. This contribution combines with the canonical coefficient of $\delta\sigma_{AB}$; the coefficients of $\delta N$, $\delta\mathcal N^A$, and the other components of $\delta q_{\mathsf a}$ are unchanged by this step.

Using the canonical momentum relations in Eq.~\eqref{eq:fr_canonical_momenta} on these solutions and retaining all $\delta q_{\mathsf a}$ terms gives the following boundary variation, which agrees with the direct variation of $I^f$:
\begin{equation}
\begin{aligned}
    \delta I^{f}\big|_{\mathcal T_r,{\rm os}}
    =\int_{\mathcal T_r}\dd\tau\,\dd^d x\,
    \Bigg[
    &\mathcal E^f\,\delta N
    -\mathcal J_A^f\,\delta\mathcal N^A
    +\frac{N\sqrt\sigma}{2}s_f^{AB}\delta\sigma_{AB}
    \\
    &+2N\sqrt\sigma\,k^{AB}\delta\theta_{AB}^{\parallel}
    +4N\sqrt\sigma\,\frac{\Pi_{\parallel}^{A}}{\sqrt h}\,
       \delta\chi_A^{\parallel}
    +\sqrt\sigma\,(D_nN)\delta\Psi_\perp
    \Bigg].
\end{aligned}
    \label{eq:fr_full_quasilocal_boundary_variation}
\end{equation}
The coefficients of $\delta q_{\mathsf a}$ in this expression define the densitized covariant responses $\mathcal P_{f,\mathrm{cov}}^{\mathsf a}$. On these solutions, with the momenta evaluated using Eq.~\eqref{eq:fr_canonical_momenta}, the coefficients of $\delta N$, $\delta\mathcal N^A$, $\delta\theta_{AB}^{\parallel}$, $\delta\chi_A^{\parallel}$, and $\delta\Psi_\perp$ coincide with their counterparts in the canonical variation~\eqref{eq:fr_quasilocal_density_definition}. The metric response $N\sqrt\sigma\,s_f^{AB}/2$ combines the canonical coefficient with the time-derivative contribution identified above; its explicit form is not needed here. The $\delta Z_{AB}$ contribution is absorbed into the metric response in Eq.~\eqref{eq:fr_full_quasilocal_boundary_variation} by using the projected Hamilton relation~\eqref{eq:fr_projected_hamilton_relation} and integrating by parts in Euclidean time.

\subsection{Stationary Sewing under Constant Killing Reparametrizations}
\label{app:fr_stationary_sewing}

We now verify that the stationary solution pairs constructed in Sec.~\ref{subsec:fr_pn_sewing} satisfy the sewing conditions in Eq.~\eqref{eq:fr_partial_neumann_sewing_conditions}. All sewing data are compared componentwise at the same $(\tau,\rho,x^A)$ in the exterior coordinates of Eq.~\eqref{eq:outer_region_metric}. The inner metric~\eqref{eq:inner_cap_metric} and its complete auxiliary configuration are obtained from the same stationary axisymmetric solution by the local map $F$ in Eq.~\eqref{eq:cap_coordinate_map}, with a constant time rescaling and a rigid axial twist. The angular coordinates are adapted to the axial Killing fields, and the complete auxiliary configuration shares the Killing symmetries of the metric.

The common coordinate evolution vector is $\partial_\tau$, and $\mathcal N_s=\mathcal N_s^A\partial_A$ denotes the tangential shift on side $s$. We write the exterior lapse and shift as $N$ and $\mathcal N$, as in Eq.~\eqref{eq:outer_region_metric}. The inner lapse and shift then satisfy
\begin{equation}
    N_1=\alpha N,
    \qquad
    \mathcal N_1=\alpha\mathcal N+\nu,
    \qquad
    \nu=\nu^I\varphi_I,
    \label{eq:fr_stationary_evolution_map}
\end{equation}
where $\alpha$ and $\nu^I$ are constants, and $\varphi_I$ are the mutually commuting axial Killing vector fields. The map is complexified when required by the rotating Euclidean saddle. The temporal unit normals obey
\begin{equation}
\begin{gathered}
    u_1=\frac{\partial_\tau-\alpha\mathcal N-\nu}{\alpha N},
    \qquad
    u_2=\frac{\partial_\tau-\mathcal N}{N},
    \\
    F_*u_1=u_2.
\end{gathered}
    \label{eq:fr_stationary_unit_normal_matching}
\end{equation}
The last equality compares vectors at points related by $F$; the untransported normals $u_1$ and $u_2$ generally have different components in the common coordinate basis.

On each fixed-time spatial slice, the spatial Jacobian of $F$ is the identity matrix in the adapted coordinates. Together with the pullback relations for the complete auxiliary tensors and $F_*u_1=u_2$, this implies that the two spatial ADM metrics and auxiliary projections have equal components at points related by $F$. Stationarity and axial symmetry then give equality at the same $(\tau,\rho,x^A)$. The common spatial metric and unchanged radial coordinate determine the same radial geometry, and we use the same orientation of $n$ on both sides. Since the spatial fields and metric agree throughout an interface neighborhood, their spatial covariant derivatives, including normal derivatives, agree as well. In particular, all four entries of $q_{\mathsf a}=\{\sigma_{AB},\theta_{AB}^{\parallel},\chi_A^{\parallel},\Psi_\perp\}$ match, establishing $[q_{\mathsf a}]=0$.

To establish matching of the energy and momentum responses, we next compare the canonical momenta. Using $\partial_\tau h_{ab}=0$ and $\mathcal L_\nu h_{ab}=0$, the ADM relation~\eqref{eq:time_extrinsic_curvature} gives
\begin{equation*}
    K_{ab}^{(1)}
    =-\frac{(\mathcal L_{\alpha\mathcal N+\nu}h)_{ab}}{2\alpha N}
    =-\frac{(\mathcal L_{\mathcal N}h)_{ab}}{2N}
    =K_{ab}^{(2)}.
\end{equation*}
This is the same geometric matching underlying Eq.~\eqref{eq:general_momentum_response_jump}. Hamilton's first equation then gives $\Pi_{ab}^{(1)}/\sqrt h=\Pi_{ab}^{(2)}/\sqrt h$. Together with the matched spatial fields and their derivatives, Eq.~\eqref{eq:fr_brown_york_energy_density} then gives $[\mathcal E^f]=0$.

To establish matching of the momentum response, we also compare $p^{ab}/\sqrt h$ using Eq.~\eqref{eq:fr_canonical_momenta}. In the spatial derivative term $N_s^{-1}D_c(N_s\Phi^{c(ab)})$, the constant $\alpha$ cancels between the numerator and denominator. For the other derivative term, $N_su_s=\partial_\tau-\mathcal N_s$ gives
\begin{equation*}
    \frac{1}{N_s}\mathcal L_{N_su_s}\Psi^{ab}
    =\frac{1}{N_s}\left(\partial_\tau\Psi^{ab}
    -\mathcal L_{\mathcal N_s}\Psi^{ab}\right).
\end{equation*}
Using $\partial_\tau\Psi^{ab}=0$ and $\mathcal L_\nu\Psi^{ab}=0$, this expression has the same value on both sides. The remaining terms in the momentum relation depend on fields that already match. The quantity $p^{ab}/\sqrt h$ therefore also matches. Substituting these results into Eq.~\eqref{eq:fr_brown_york_momentum_density} gives $[\mathcal J_A^f]=0$ pointwise on the interface.

Finally, substituting the matched $\Pi_{AB}^{\parallel}/\sqrt h$ and the source relations~\eqref{eq:fr_stationary_evolution_map} into the definition~\eqref{eq:fr_additional_sewing_datum} gives
\begin{equation}
    Z_{AB}^{(1)}
    =\alpha Z_{AB}^{(2)}
    +\frac12(\mathcal L_\nu\sigma)_{AB}.
    \label{eq:fr_stationary_Z_map}
\end{equation}
Equation~\eqref{eq:fr_projected_hamilton_relation} and stationarity give $Z_{AB}^{(2)}=\tfrac12\partial_\tau\sigma_{AB}=0$, while axial symmetry gives $\mathcal L_\nu\sigma_{AB}=0$. Hence $Z_{AB}^{(1)}=Z_{AB}^{(2)}=0$, completing the verification of the sewing conditions. By Eq.~\eqref{eq:fr_residual_auxiliary_boundary_term}, this also implies that $B_{\rm aux}$ vanishes on each interface copy.

This automatic matching applies to the stationary solution pairs described above. For independent canonical fields, or geometries not related by this construction, the sewing conditions must be imposed separately.

\subsection{Boundary Contributions in Einstein--Gauss--Bonnet Gravity}
\label{app:egb_boundary_contributions}

We evaluate the physical boundary and bolt contributions for the paired EGB geometries of Sec.~\ref{subsec:egb_factorization}, using the auxiliary action in Eq.~\eqref{eq:egb_auxiliary_action}. At an outer cutoff $r=R$, the unrenormalized energy response follows from Eq.~\eqref{eq:egb_interface_energy_derivation} with $r_T$ replaced by $R$. Its integral over the three-sphere is
\begin{equation}
    E_{\rm BY}^{\rm GB}(R)
    =\int_{\mathcal S_\tau^{(R)}}\dd^3x\,\mathcal E^{\rm GB}
    =-\frac{3\pi}{4}R^2\sqrt{F(R)}\,V(R).
    \label{eq:egb_cutoff_energy}
\end{equation}
For AdS background subtraction, we take the $\mu=0$ geometry, with $F_0(r)=1+r^2/\ell^2$ and $V_0=1-12\alpha_{\rm GB}/\ell^2$. At the same cutoff radius $R$, the induced spatial metrics already agree, so matching the reference lapse to $N_R=\sqrt{F(R)}$ and using the same Euclidean period $\beta$ ensure identical induced boundary metrics and proper thermal periods. The integrated reference response is $E_{{\rm BY},0}^{\rm GB}(R)=-3\pi R^2\sqrt{F_0(R)}V_0/4$. The prescribed contribution $I_{\mathcal B}^{\rm ren}$ in Eq.~\eqref{eq:fr_asymptotic_hamiltonian_boundary} is supplied by reference subtraction, yielding
\begin{equation}
\begin{aligned}
    I_{\mathcal B}^{\rm ren}(R)
    &=-\int_0^\beta\dd\tau\,N_R E_{{\rm BY},0}^{\rm GB}(R),
    \\
    H_R^{\rm ren}
    &=N_R\left[E_{\rm BY}^{\rm GB}(R)-E_{{\rm BY},0}^{\rm GB}(R)\right],
    \qquad N_R=\sqrt{F(R)}.
\end{aligned}
    \label{eq:egb_reference_hamiltonian}
\end{equation}
This prescription assigns zero energy to the $\mu=0$ solution. Taking the asymptotic limit of the subtracted Hamiltonian gives
\begin{equation}
    H_{\mathcal B}=\lim_{R\to\infty}H_R^{\rm ren}
    =\frac{3\pi\mu}{8},
    \label{eq:egb_mass_normalization_check}
\end{equation}
This is the Hamiltonian energy of the EGB black hole and reproduces Eq.~\eqref{eq:egb_renormalized_energy}.

For the inner cap, using the common radial normal $n=\sqrt F\,\partial_r$, we obtain the lapse-weighted energy response on the tube $r=r_h+\epsilon$ from Eq.~\eqref{eq:egb_interface_energy_derivation} and $N_1=\alpha\sqrt{F}$,
\begin{equation*}
\begin{aligned}
    \left.N_1 \mathcal E_{(1)}^{\rm GB}\right|_{r=r_h+\epsilon}
    &=-\left.\frac{3\alpha\sqrt\sigma}{8\pi r}F(r)V(r)\right|_{r=r_h+\epsilon}
    \\
    &=-\frac{3\alpha F'(r_h)V(r_h)}{8\pi r_h}
      \left.\sqrt\sigma\right|_{r=r_h}\epsilon+O(\epsilon^2),
\end{aligned}
\end{equation*}
so the ADM boundary pairing on the shrinking tube vanishes as $\epsilon\to0^+$. Restoring the excised neighborhood contributes the generalized GHY term in Eq.~\eqref{eq:egb_auxiliary_action}, with the normal pointing outward from that neighborhood. Taking the bolt limit gives
\begin{equation}
\begin{aligned}
    \lim_{\epsilon\to0^+}B_{\rm GHY}^{\rm GB}(r_h+\epsilon)
    &=-\frac{\alpha\beta\pi}{4}
      \left[\frac{r_h^3F'(r_h)}{2}U(r_h)+3r_h^2F(r_h)V(r_h)\right]
    \\
    &=-\frac{A_H}{4}\left(1+\frac{12\alpha_{\rm GB}}{r_h^2}\right)
      =-S_{\rm GB}.
\end{aligned}
    \label{eq:egb_tube_ghy_evaluation}
\end{equation}
Here $A_H=2\pi^2r_h^3$. The second equality uses $F(r_h)=0$, $U(r_h)=1+12\alpha_{\rm GB}/r_h^2$, and cap regularity, $\alpha\beta F'(r_h)=4\pi$. The resulting contribution is minus the Wald entropy, in agreement with Eq.~\eqref{eq:egb_bolt_contribution}. Together with the vanishing residual term $B_{\rm aux}^{(s)}$, these boundary contributions yield the constrained action in Eq.~\eqref{eq:egb_constrained_action}.

\subsection{Auxiliary Projections, Interface Sewing, and Charge Normalization in New Massive Gravity}
\label{app:nmg_auxiliary_reduction}

After imposing $\varrho_{\mu\nu\rho\sigma}=R_{\mu\nu\rho\sigma}$, the auxiliary tensor obtained from the NMG Lagrangian in Eq.~\eqref{eq:nmg_lagrangian} is
\begin{align}
    \phi_{\rm NMG}^{\mu\nu\rho\sigma}
    =-\frac{1}{8\pi}
    \Bigg\{
    &g^{\mu[\rho}g^{\sigma]\nu}
    -\frac{3\alpha_{\rm NMG}}{4}R\,
    g^{\mu[\rho}g^{\sigma]\nu}
    \notag\\
    &+\frac{\alpha_{\rm NMG}}{2}
    \left(
    g^{\mu\rho}R^{\nu\sigma}
    -g^{\mu\sigma}R^{\nu\rho}
    -g^{\nu\rho}R^{\mu\sigma}
    +g^{\nu\sigma}R^{\mu\rho}
    \right)
    \Bigg\}.
    \label{eq:nmg_auxiliary_tensor}
\end{align}
Substituting the locally AdS$_3$ curvature in Eq.~\eqref{eq:nmg_ads_branch} gives Eq.~\eqref{eq:nmg_auxiliary_ads_reduction}. The corresponding radial double-normal projection is
\begin{equation}
    \Phi_{(n),{\rm NMG}}^{ij}
    =
    -\frac{\Xi_{\rm NMG}}{16\pi}\gamma^{ij}
    =
    \Xi_{\rm NMG}\Phi_{(n),{\rm EH}}^{ij},
    \label{eq:nmg_boundary_auxiliary_reduction}
\end{equation}
while the temporal projections are
\begin{equation}
    \Phi_{abcd}
    =
    -\frac{\Xi_{\rm NMG}}{16\pi}
    \left(h_{ac}h_{bd}-h_{ad}h_{bc}\right),
    \qquad
    \Phi_{abc}=0,
    \qquad
    \Psi_{ab}
    =
    -\frac{\Xi_{\rm NMG}}{8\pi}h_{ab}.
    \label{eq:nmg_time_slice_projections}
\end{equation}
Here $\Xi_{\rm NMG}=1+\alpha_{\rm NMG}/(2\ell^2)$, as defined in Eq.~\eqref{eq:nmg_auxiliary_ads_reduction}. The corresponding components in the radial decomposition are
\begin{equation}
\begin{aligned}
    \theta_{\phi\phi}^{\parallel}
    &=-\frac{\Xi_{\rm NMG}}{16\pi}\sigma_{\phi\phi},
    &
    \chi_\phi^{\parallel}&=0,
    \\
    \Psi_\perp
    &=-\frac{\Xi_{\rm NMG}}{8\pi},
    &
    \chi_{2}^{\parallel\phi\phi}&=0.
\end{aligned}
    \label{eq:nmg_corner_auxiliary_data}
\end{equation}
In particular, $\chi_{2}^{\parallel\phi\phi}=0$ makes the residual auxiliary boundary term vanish on this branch, as stated in Eq.~\eqref{eq:nmg_residual_auxiliary_vanishing}.

With the momenta evaluated using Eq.~\eqref{eq:fr_canonical_momenta}, the projections above reduce the unrenormalized responses in Eqs.~\eqref{eq:fr_brown_york_energy_density} and~\eqref{eq:fr_brown_york_momentum_density} to $\Xi_{\rm NMG}$ times their Einstein values. For the Euclidean continuation of Eq.~\eqref{eq:nmg_btz_metric}, $k(r)=\sqrt{F(r)}/r$ and $K_{r\phi}=ij/(2r\sqrt{F(r)})$. With $\sqrt\sigma=r$, these expressions give the densitized interface responses in Eq.~\eqref{eq:nmg_interface_response_evaluation} at $r=r_T$.

For the paired BTZ configurations of Sec.~\ref{subsec:nmg_factorization}, the common spatial geometry and the projections in Eq.~\eqref{eq:nmg_corner_auxiliary_data} give $[q_{\mathsf a}]=0$. In the common-normal convention, the explicit responses in Eq.~\eqref{eq:nmg_interface_response_evaluation} also agree across the interface. Hamilton's first equation and stationarity give $Z_{\phi\phi}^{(1)}=Z_{\phi\phi}^{(2)}=0$, completing the verification of Eq.~\eqref{eq:fr_partial_neumann_sewing_conditions} for the full NMG data in Eq.~\eqref{eq:nmg_complete_interface_data}. This explicit verification on the BTZ branch is consistent with the general stationary sewing result established in Appendix~\ref{app:fr_stationary_sewing}.

\paragraph{Charge Normalization.}
At an outer cutoff circle $r=R>r_+$, both unrenormalized responses follow from Eq.~\eqref{eq:nmg_interface_response_evaluation} with $r_T$ replaced by $R$: $\mathcal E_R^{\rm NMG}=-\Xi_{\rm NMG}\sqrt{F(R)}/(8\pi)$ and $\mathcal J_{\phi,R}^{\rm NMG}=i\Xi_{\rm NMG}j/(16\pi)$. For this prescription, the contribution $I_{\mathcal B}^{\rm ren}$ is supplied by the local AdS$_3$ volume counterterm
\begin{equation*}
    I_{\mathcal B}^{\rm ren}(R)=I_{\rm ct}(R)
    =\frac{\Xi_{\rm NMG}}{8\pi\ell}
      \int_{\mathcal T_R}\dd\tau\,\dd\phi\,N\sqrt{\sigma}.
\end{equation*}
It adds $\Xi_{\rm NMG}R/(8\pi\ell)$ to the energy response and leaves the momentum response unchanged, since it is independent of the shift. The renormalized responses are therefore
\begin{equation*}
\begin{aligned}
    \mathcal E_{R,\mathrm{ren}}^{\rm NMG}
    &=\frac{\Xi_{\rm NMG}}{8\pi}
      \left(\frac R\ell-\sqrt{F(R)}\right),
    \\
    \mathcal J_{\phi,R,\mathrm{ren}}^{\rm NMG}
    &=\frac{i\Xi_{\rm NMG}j}{16\pi}.
\end{aligned}
\end{equation*}
With this counterterm prescription, the massless BTZ solution $m=j=0$ has vanishing energy and angular momentum. With $\phi$ of period $2\pi$, the real angular momentum is
\begin{equation}
    J_{\mathcal B}^{\rm NMG}
    =-i\int_0^{2\pi}\dd\phi\,\mathcal J_{\phi,R,\mathrm{ren}}^{\rm NMG}
    =\frac{\Xi_{\rm NMG}j}{8}.
    \label{eq:nmg_angular_continuation_check}
\end{equation}
The integrated renormalized energy response $E_{{\rm BY},{\rm ren}}^{\rm NMG}(R)$ is associated with the unit time normal. The Hamiltonian energy is obtained by including the lapse factor $N_R=\sqrt{F(R)}$ before taking the asymptotic limit. Thus
\begin{equation}
\begin{aligned}
    E_{{\rm BY},{\rm ren}}^{\rm NMG}(R)
    &=\int_0^{2\pi}\dd\phi\,\mathcal E_{R,\mathrm{ren}}^{\rm NMG}
      =\frac{\Xi_{\rm NMG}}4
       \left(\frac R\ell-\sqrt{F(R)}\right),
    \\
    N_R E_{{\rm BY},{\rm ren}}^{\rm NMG}(R)
    &=\frac{\Xi_{\rm NMG}}4
      \frac{\sqrt{F(R)}}{R/\ell+\sqrt{F(R)}}
      \left(m-\frac{j^2}{4R^2}\right),
    \\
    E_{\mathcal B}^{\rm NMG}
    &=\lim_{R\to\infty}N_R E_{{\rm BY},{\rm ren}}^{\rm NMG}(R)
      =\frac{\Xi_{\rm NMG}m}{8}.
\end{aligned}
    \label{eq:nmg_cutoff_energy_normalization}
\end{equation}
These charges reproduce Eq.~\eqref{eq:nmg_charges_entropy}. To evaluate the physical boundary contribution, we combine the lapse and shift terms in the periodic frame, where $\widetilde{\mathcal N}_R^\phi=ij/(2R^2)-i\Omega_{\mathcal B}$. The renormalized boundary Hamiltonian at the cutoff is therefore
\begin{equation*}
\begin{aligned}
    H_R^{\rm ren}
    &=N_R E_{{\rm BY},{\rm ren}}^{\rm NMG}(R)
      +\left(\frac{j}{2R^2}-\Omega_{\mathcal B}\right)
       J_{\mathcal B}^{\rm NMG},
    \\
    \lim_{R\to\infty}H_R^{\rm ren}
    &=E_{\mathcal B}^{\rm NMG}
      -\Omega_{\mathcal B}J_{\mathcal B}^{\rm NMG}.
\end{aligned}
\end{equation*}
Integrating the limiting Hamiltonian over the Euclidean period gives the physical boundary action $\beta(E_{\mathcal B}^{\rm NMG}-\Omega_{\mathcal B}J_{\mathcal B}^{\rm NMG})$.

\bibliographystyle{JHEP}
\bibliography{ref}
\end{document}